\documentclass[12pt]{article}
\usepackage[dvips]{color,graphicx}
\usepackage{amssymb} 
\def \Fig#1#2#3 {
\begin{figure}
\centering
\epsfxsize=#2cm \epsfbox{#1.eps}
\caption{#3}
\label{#1}
\end{figure}
}

\def\href#1#2{#2}

\def\IP{\relax{\rm I\kern-.18em P}}

\newcommand{\nc}{\newcommand}
\newcommand{\beq}{\begin{equation}}
\newcommand{\be}{\begin{equation}}
\newcommand{\eeq}{\end{equation}}
\newcommand{\ee}{\end{equation}}
\newcommand{\beqa}{\begin{eqnarray}}
\newcommand{\ba}{\begin{eqnarray}}
\newcommand{\eeqa}{\end{eqnarray}}
\newcommand{\ea}{\end{eqnarray}}
 
\def\tq{\tilde{q}} 
\newcommand{\pa}{\partial}

\def\cH{{\cal H}}
\def\Im{{\mbox{\rm Im}}}

\def\id{{\rm id}}

\def\c{\gamma} 
\def\pl{\partial} 
\def\bpl{\bar \partial} 
\def\H3p{H_3^+}
\def\QR{\mathbb{R}} 
\def\QC{\mathbb{C}} 
\def\a{\alpha}
\def\b{\beta} 
\def\c{\gamma} 
\def\d{\delta} 
\def\e{\epsilon}

\def\ra{\rightarrow}

\def\bw{\bar w} 
\def\bW{\bar W}

\def\cW{{\cal W}}  

\def\cV{{\cal V}} 
\def\cJ{{\cal J}} 
 
\def\ap{{\alpha '}} 

\def\om{\omega}

\def\cG{{\cal G}}
\def\bT{\bar T}

\def\bJ{{\bar J}}
\def\bz{{\bar z}}
\def\pl{\partial}
\def\bpl{\bar \partial}

\def\id{{\rm id}}
\def\c{\gamma} 
 
\def\bj{{\bar \jmath}} 
\def\cA{{\cal A}}
\def\cH{{\cal H}}
\def\H{{\cal H}}
\def\cJ{{\cal J}}

\nc{\nn}{\nonumber}
\def\e{\epsilon}
\def\o{\otimes}

\def\a{\alpha}
\def\b{\beta} 
\def\s{\sigma} 
\def\IP{\relax{\rm I\kern-.18em P}}

\def\QC{\mathbb{C}}

\def\QR{\mathbb{R}}

\def\QZ{\mathbb{Z}}

\def\bi{{\bar \imath}}
\def\bj{{\bar \jmath}} 
\def\bn{{\bar n}} 

\newcommand{\SLC}{{\rm SL(2,\QC \rm)}}

\newcommand{\SU}{{\rm SU(2)}}
\newcommand{\al}{\alpha}

\newcommand{\Ga}{\Gamma}

\newcommand{\la}{\lambda}

\newcommand{\CF}{{\mathcal F}}

\def\bL{{\bar L}}

\newcommand{\SL}{{\mathsf L}}

\def\U{{U}}

\def\ppe{\hspace*{-2.5mm}}
\def\ew{\hspace*{-1mm}}
\newcommand{\Fus}[6]{F_{{\scriptstyle #1}{\scriptstyle #2}}
  \hspace*{.3mm}\displaystyle{[} \ew \begin{array}{ll} {\scriptstyle #3 }
  \ppe & {\scriptstyle #4} \ppe \\[-2mm] {\scriptstyle #5}\ppe &
  {\scriptstyle #6}\ew \end{array}\displaystyle{]}}

\def\nn{\nonumber}

\def\cH{{\cal H}}

\def\a{\alpha}

\def\ra{\rightarrow}

\newcommand{\vvert}[3]{\displaystyle{(} \ew \begin{array}{ll} 
  \ & \hspace*{-4mm} {\scriptstyle #1} \\[-2mm] {\scriptstyle #2} 
  \ppe & {\scriptstyle #3} \ew \end{array} \displaystyle{)}}

\input epsf
\newcount\figno
\def\fig#1#2#3{
\par\begingroup\parindent=0pt\leftskip=1cm\rightskip=1cm\parindent=0pt
\baselineskip=15pt
\global\advance\figno by 1
\epsfxsize=#3
\centerline{\epsfbox{#2}}
\vskip 12pt
{\bf \small Figure \the\figno:} {\small #1}\par
\endgroup\par
}
\def\figlabel#1{\xdef#1{\the\figno 
\mbox{ }}}
\def\encadremath#1{\vbox{\hrule\hbox{\vrule\kern8pt\vbox{\kern8pt
\hbox{$\displaystyle #1$}\kern8pt}
\kern8pt\vrule}\hrule}}
\def\half{\frac12}
\def\p{\partial} 
\def\SL{{\rm SL}$_2(\QR)$}
\def\SLU{{\rm SL}$_2(\QR)${\rm /U(1)}}
\def\mSLU{{\rm SL}_2(\QR){\rm /U(1)}}
\def\rar{\rightarrow}
\def\Z{\QZ}
\def\tz{\tilde z}
\def\tth{\tilde \tau}
\def\vt{\vartheta}
\def\AA{$AdS_2$}
\def\halfpi{\frac{\pi}{2}}
\def\N{{\cal N}}
\def\D{{\rm D}}
\def\Dz{{\rm D0}}
\def\pp{\,}
\def\j{2j+1}
\def\la{{\langle}}
\def\ra{{\rangle}} 
\def\cA{{\cal A}}

\begin{document}
\baselineskip=19pt
\title{\bf Non-compact String Backgrounds\\[8mm]
           and Non-rational CFT \\[1cm]}
\author {{\sc Volker Schomerus} \\[2mm]
        DESY Theory Group, Notkestrasse 85,\\
      D-22603 Hamburg, Germany\footnote{Permanent address 
     after May 2005.}\\[2mm]  
     Service de Physique Th\'eorique, CEA Saclay,\\ 
      F-91191 Gif-sur-Yvette, France\\[1mm] 
 {\tt email:volker.schomerus@desy.de}}
\vskip.8cm
\date{September, 2005}
%
\begin{titlepage}      \maketitle       \thispagestyle{empty}

\vskip1cm
\begin{abstract}
\noindent     
This is an introduction to the microscopic techniques of non-rational
bulk and boundary conformal field theory which are needed to describe 
strings moving in non-compact curved backgrounds. The latter arise 
e.g.\ in the context of AdS/CFT-like dualities and for studies of 
time-dependent processes. After a general outline of the central 
concepts, we focus on two specific but rather prototypical models: 
Liouville field theory and the 2D cigar. Rather than following the
historical path, the presentation attempts to be systematic and 
self-contained.     
\end{abstract}
\vspace*{-17.9cm}
{\tt {DESY-05-168}}\\{\tt  {SPHT-T05-139}}
\bigskip\vfill
\noindent
{Lectures presented at Spring School on Superstring Theory, 
Trieste, March 16-23, 2004, at the ESI workshop on {\em 
String theory in curved backgrounds and bCFT}, Vienna, 
April 1-June 30, 2004, and at the Summer School on Strings, 
Gravity \& Cosmology, Vancouver, August 3-13, 2004.}   
\end{titlepage}

\tableofcontents
\newpage

\section{Introduction}

The microscopic techniques of (boundary) conformal field theory 
have played a key role for our understanding of string theory 
and a lot of technology has been developed in this field over the 
last 20 years. Most of the powerful results apply to strings 
moving in compact spaces. This is partly explained by the fact 
that, for at least one decade, progress of world-sheet methods 
was mainly driven by our need to understand non-trivial string 
compactifications. In addition, compact target spaces are simply 
more easy to deal with. Note in particular that compactness renders 
the spectra of the underlying world-sheet models discrete. This 
is a crucial feature of the associated `rational conformal field 
theories' which allows to solve them using only algebraic tools
(see e.g.\ \cite{DiFrancesco,Petkova:2000dv,Schomerus:2002dc,
Angelantonj:2002ct,Stanev:2001na}). More recently, however, 
several profound problems of string theory urge us to consider 
models with continuous spectra. The aim of these lectures is 
therefore to explain how conformal field theory may be extended 
beyond the rational cases, to a description of closed and open 
strings moving in non-compact target spaces.%
\smallskip 

There are several motivations from string theory to address 
such issues. One of the main reasons for studies of non-rational 
conformal field theory comes from AdS/CFT-like dualities, i.e.\ 
from strong/weak coupling dualities between closed string 
theories in AdS-geometries and gauge theory on the boundary
(see e.g.\ \cite{Aharony:1999ti}). If we want to use 
such a dual string theory description to learn something 
about gauge theory at finite `t Hooft coupling, we have 
to solve string theory in AdS, i.e.\ in a curved and 
non-compact space. Similarly, the study of little string 
theory, i.e.\ of the string theory on NS5-branes, involves 
a dual theory of closed strings moving in a non-trivial 
non-compact target space \cite{Kutasov:2001uf}. Finally, 
all studies of time dependent processes in string theory, 
such as e.g.\ the decay of tachyons \cite{Sen:2004nf}, 
necessarily involve non-compact target space-time since 
time is not compact. Both, the application to some of the most 
interesting string dualities and to time dependent backgrounds 
should certainly provide sufficient motivation to develop 
non-rational conformal field theory. 
\smallskip 

For compact backgrounds there are many model independent 
results, i.e.\ solutions or partial solutions that apply 
to a large class of models regardless of their geometry.
The situation is quite different with non-compact target 
spaces. In fact, so far all studies have been restricted 
to just a few fundamental models. This will reflect 
itself in the content of these lectures as we will mainly 
look at two different examples. The first is known as 
{\em Liouville theory} and it describes strings moving 
in an exponential potential (with a non-constant dilaton). 
Our second example is the \SLU-coset theory. Its target 
space is a Euclidean version of the famous 2-dimensional 
black hole solution of string theory.%
\smallskip 

Fortunately, these two models have quite a few interesting 
applications already (which we shall only sketch in passing). 
From a more fundamental point of view, our examples are the 
non-compact analogues of the minimal models and the coset 
SU(2)/U(1) which have been crucial for the development of 
rational conformal field theory and its 
applications to string theory. As we shall  review below, 
Liouville theory is e.g.\ used to build an important exactly 
solvable 2D toy model of string theory and it is believed 
to have applications to the study of tachyon condensation. 
The coset \SLU, on the other hand, appears as part of 
the transverse geometry of NS5-branes. Moreover, since   
the space $AdS_3$ is the same as the group manifold of 
\SL, one expects the coset space \SLU \, to  participate 
in an interesting low dimensional version of the AdS/CFT 
correspondence.%
\medskip 

The material of these lectures will be presented in 
four parts. In the first lecture, we shall review some 
basic elements of (boundary) conformal field theory. 
Our main goal is to list the quantities that characterize 
an exact conformal field theory solution and to explain 
how they are determined. This part is entirely 
model independent and it can be skipped by readers
that have some acquaintance with boundary conformal 
field theory and that are more interested in the 
specific problems of  non-rational models. In lecture 2 
and 3 we shall then focus on the solution of Liouville 
theory. Our discussion will begin with the closed string 
background. Branes and open strings in Liouville theory 
are the subject of the third lecture. Finally, we shall 
move on to the coset \SLU. There, our presentation will 
rely in parts on the experience from Liouville theory 
and it will stress the most interesting new aspects 
of the coset model.   
\newpage

\section{2D Boundary conformal field theory} 
\setcounter{equation}{0}

The world-sheet description of strings moving in any target 
space of dimension $D$ with background metric $g_{\mu\nu}$ 
is based on the following 2-dimensional field theory for a 
d-component bosonic field $X^\mu, \mu = 1, \dots, d$, 
\be \label{action} 
S[X] \ = \ \frac{1}{4\pi \ap}\, \int_\Sigma \, d^2 z \, 
  g_{\mu\nu} \, \partial X^\mu \, \partial X^\nu + \dots 
\ \ . 
\ee  
In addition to the term we have shown, many further 
terms can appear and are necessary to describe the 
effect of non-trivial background B-fields, dilatons, 
tachyons or gauge fields when we are dealing with open 
strings. For superstring backgrounds, the world-sheet 
theory also contains fermionic fields $\Psi^\mu$. We 
shall see some of the extra terms in our examples 
below. 
\smallskip 

The world-sheet $\Sigma$ that we integrate over in eq.\ 
(\ref{action}) will be either the entire complex plane 
(in the case of closed strings) or the upper half-plane 
(in the case of open strings). In the complex coordinates 
$z,\bz$ that we use throughout these notes, lines of 
constant Euclidean time are (half-)circles around the 
origin of the complex plane. The origin itself can 
be thought of as the infinite past. 
\smallskip 

When we are given any string background, our central task
is to compute string amplitudes, e.g.\ for the joining of 
two closed strings or the absorption of a closed string 
mode by some brane etc. These quantities are directly 
related to various correlation functions in the 2D 
world-sheet theory (\ref{action}) and they may be 
computed, at least in principle, using path integral 
techniques. The remarkable success of 2D conformal field 
theory, however, was mainly based on a different approach 
that systematically exploits the representation theory of 
certain infinite dimensional symmetries which are known 
as chiral- or $\cW$-algebras. We will explain some of the 
underlying ideas and concepts momentarily. 
\smallskip

One example of such a symmetry algebra already arises for 
strings moving in flat space. In this case, the equations 
of motion for the fields $X^\mu$ require that 
$\Delta X^\mu = \pl \bpl X^\mu = 0$. Hence, the field 
$$ J^\nu(z) \ :=\  \partial X^\nu(z,\bz) \ = \ 
   \sum_n \, \a^\nu_n \, z^{-n-1} $$ 
depends holomorphically on $z$ so that we can expand it 
in terms of Fourier modes $\a^\nu_n$. The canonical 
commutation relations for the bosonic fields $X^\mu$ are 
easily seen to imply that the modes $\a^\nu_n$ obey the 
following relations, 
$$ [ \, \a^\nu_n\, , \, \a^\mu_m\, ] \ = \ 
    n \, \delta^{\nu\mu} \, \delta_{n,-m} \ \ . $$ 
This is a very simple, infinite dimensional algebra
which is known as U(1)-current algebra. There exists 
a second commuting copy of this algebra that is 
constructed from the anti-holomorphic field $\bJ^\nu 
= \bpl X^\nu(z,\bz)$. Even though the operators $\a^\nu_n$ 
are certainly useful in describing oscillation modes 
of closed and open strings, their algebraic structure is not 
really needed to solve the 2D free field theory underlying 
string theory in flat space. 
\smallskip

Things change drastically when the background in curved, i.e.\ 
when the background fields depend on the coordinates $X^\mu$. 
In fact, whenever this happens, the action (\ref{action}) 
ceases to be quadratic in the fields and hence its integration 
can no longer be reduced to the computation of Gaussian integrals. 
In such more intricate situations, it becomes crucial to find and 
exploit generalizations of the U(1)-current algebra. We will see 
this in more detail below after a few introductory comments on 
chiral algebras.    
  
\subsection{Chiral algebras}  

Chiral algebras can be considered as symmetries of 2D conformal 
field theory. Since they play such a crucial role for all exact
solutions, we shall briefly go through the most important 
notions in the representation theory of chiral algebras. These
include the set $\cJ$ of representations, modular transformations, 
the fusion of representations and the fusing matrix $F$. The 
general concepts are illustrated in the case of the U(1)-current 
algebra. Readers feeling familiar with the aforementioned notions
may safely skip this subsection.%
   
\paragraph{Representation theory.} Chiral- or W-algebras are generated 
by the modes $W^\nu_n$ of a finite set of (anti-)holomorphic fields 
$W^\nu(z)$. These algebras mimic the role played by Lie algebras in 
atomic physics. Recall that transition amplitudes in 
atomic physics can be expressed as products of Clebsch-Gordan 
coefficients and so-called reduced matrix elements. While the 
former are purely representation theoretic data which depend only on 
the symmetry of the theory, the latter contain all the information 
about the physics of the specific system. Similarly, amplitudes in 
conformal field theory are built from representation theoretic 
data of W-algebras along with structure constants of the various
operator product expansions, the latter being the reduced matrix 
elements of conformal field theory. In the conformal bootstrap, 
the structure constants are determined as solutions of certain 
algebraic equations which arise as factorization constraints 
and we will have to say a lot more about such equations as we proceed. 
Constructing the representation theoretic data, on the other hand, is 
essentially a mathematical problem which is the same for all models 
that possess the same chiral symmetry. Throughout most of the following 
text we shall not be concerned with this part of the analysis and 
simply use the known results. But it will be useful to have a 
few elementary notions in mind.   
\smallskip

We consider a finite number of bosonic chiral fields $W^\nu(z)$ with 
positive integer conformal dimension $h_\nu$ and require that there 
is one distinguished chiral field $T(z)$ of conformal dimension $h 
= 2$ whose modes $L_n$ satisfy the usual Virasoro relations for 
central charge $c$. Their commutation relations with the Laurent 
modes $W^\nu_n$ of $W^\nu(z)$ are assumed to be of the form  
\be  [L_n , W^\nu_m ] \ = \ \bigl( n(h_\nu -1) - m \bigr) \, 
W^\nu_{n+m}\ \ .  \label{LWcomm} \ee
In addition, the modes of the generating chiral fields also
possess commutation relations among each other which need not 
be linear in the modes. The algebra generated by the modes 
$W^\nu_n$ is the chiral or W-algebra $\cW$ (for a precise 
definition and examples see \cite{SchBou} and in particular 
\cite{Bonn}). We shall also demand that $\cW$ comes equipped 
with a $\ast$-operation.  
\smallskip 

Sectors $\cV_i$ of the chiral algebra are irreducible (unitary) 
representations of $\cW$ for which the spectrum of $L_0$ is bounded 
from below. Our requirement on the spectrum of $L_0$ along with 
the commutation relations (\ref{LWcomm}) implies that any $\cV_i$ 
contains a sub-space $V_i^0$ of ground states which are annihilated 
by all modes $W^\nu_n$ such that $n>0$. The spaces $V_i^0$ carry an 
irreducible representation of the zero mode algebra $\cW^0$, i.e.\ 
of the algebra that is generated by the zero modes $W^\nu_0$, and we 
can use the operators $W^\nu_n, n<0,$ to create the whole sector 
$\cV_i$ out of states in $V_i^0$. Unitarity of the sectors means 
that the space $\cV_i$ may be equipped with a non-negative bi-linear 
form which is compatible with the $\ast$-operation on $\cW$. This 
requirement imposes a constraint on the allowed representations of 
the zero mode algebra on ground states. Hence, one can associate a 
representation $V_i^0$ of the zero mode algebra to every sector 
$\cV_i$, but for most chiral algebras the converse is not true.
In other words, the sectors $\cV_i$ of $\cW$ are labeled by elements 
$i$ taken from a subset $\cJ$ within the set of all irreducible 
(unitary) representations of the zero mode algebra. 
\smallskip 

For a given sector $\cV_i$ let us denote by $h_i$ the lowest eigenvalue 
of the Virasoro mode $L_0$. Furthermore, we introduce the {\em character}
$$   \chi_i(q) \ = \ \mbox{\rm tr}_{\cV_i}\, \bigl( q^{L_0 - \frac{c}{24}}
  \bigr)  
 \ \ . $$ 
The full set of these characters $\chi_i, i \in \cJ,$ has the remarkable 
property to close under modular conjugation, i.e.\ there exists a complex 
valued matrix $S = (S_{ij})$ such that 
\be \chi_i(\tq ) \ = \ S_{ij} \, \chi_j(q)  \  \ 
\label{Smatrix}
\ee 
where $\tq = \exp (-2\pi i/\tau)$ for $q = \exp(2\pi i\tau)$, as usual. 
\smallskip

Just as in the representation theory of Lie-algebras, there exists 
a product $\circ$ of sectors which is known as the fusion product. 
Its definition is based on the following family of homomorphisms 
(see e.g.\ \cite{MooSei2}) 
\ba \delta_z ( W^\nu_n ) & := & e^{-z L_{-1}} W^\nu_n e^{zL_{-1}} \o 
   {\bf 1}  + {\bf 1} \o W^\nu_n \nn \\[2mm] & = &  
   \sum_{m=0} \, \left( \begin{array}{c} h_\nu + n - 1 \\ m \end{array} 
   \right)\, z^{n + h_\nu -1-m} \, W^\nu_{1+m-h_\nu} 
   \o {\bf 1} + {\bf 1} \o W^\nu_n 
\ea
which is defined for $n > - h_\nu$. The condition on $n$ guarantees 
that the sum on the right hand side terminates after a finite number 
of terms. Suppose now that we are given two sectors $\cV_j$ and 
$\cV_{i}$. With the help of $\delta_z$ we define an action of 
the modes $W^\nu_n, n > - h_\nu,$ on their product. This action can 
be used to search for ground states and hence for sectors $k$ in 
the fusion product $j \circ i$. To any three such labels $j,i,
k$ there is assigned an intertwiner 
$$V \vvert{\ j}{k}{i}(z): \ \cV_{j} \o \cV_{i} \ \rightarrow 
\ \cV_{k}$$ 
which intertwines between the action $\delta_z$ on the product and 
the usual action on $\cV_{k}$. If we pick an orthonormal basis 
$\{ |j,\nu \rangle \}$ of vectors in $\cV_{j}$ we can represent the 
intertwiner $V$ as an infinite set of operators 
$$V  \vvert{j,\nu}{k}{i}(z) \ := \ V  \vvert{\ j}{k}{i}
[|j,\nu\rangle;\, \cdot\  ](z)\, : \ \cV_{i}\  \rightarrow \ \cV_{k}
\ \ . $$ 
Up to normalization, these operators are uniquely determined by the
intertwining property mentioned above. The latter also restricts their 
operator product expansions to be of the form 
$$ 
V \vvert{j_1,\mu}{k}{r}(z_1)\  V \vvert{j_2,\nu}{r}
{i}(z_2) \ = \ \sum_{s,\rho}\ \Fus{r}{s}{j_2}{j_1}
{i}{k} \ V \vvert{s,\rho}{k}{i}(z_2) \ \langle s, 
\rho| V \vvert{j_2,\nu}{s}{j_1}(z_{12})|j_1,\mu\rangle \ \ , 
$$ 
where $z_{12} = z_1-z_2$. The coefficients $F$ that appear in this 
expansion form the {\em fusing matrix} of the chiral algebra $\cW$.  
Once the operators $V$ have been constructed for all ground states 
$|j,\nu\rangle$, the fusing matrix can be read off from the leading 
terms in the expansion of their products. Explicit formulas can be 
found in the literature, at least for some chiral algebras. We also 
note that the defining relation for the fusing matrix admits a nice 
pictorial presentation (see Figure 1). It presents the fusing matrix 
as a close relative of the $6J$-symbols which are known from the 
representation theory of finite dimensional Lie algebras.  
\vspace{1cm}
\fig{Graphical description of the fusing matrix. All the lines are
directed as shown in the picture. Reversal of the orientation can 
be compensated by conjugation of the label. Note that in our 
conventions, one of the external legs is oriented outwards. This 
will simplify some of the formulas below.}
{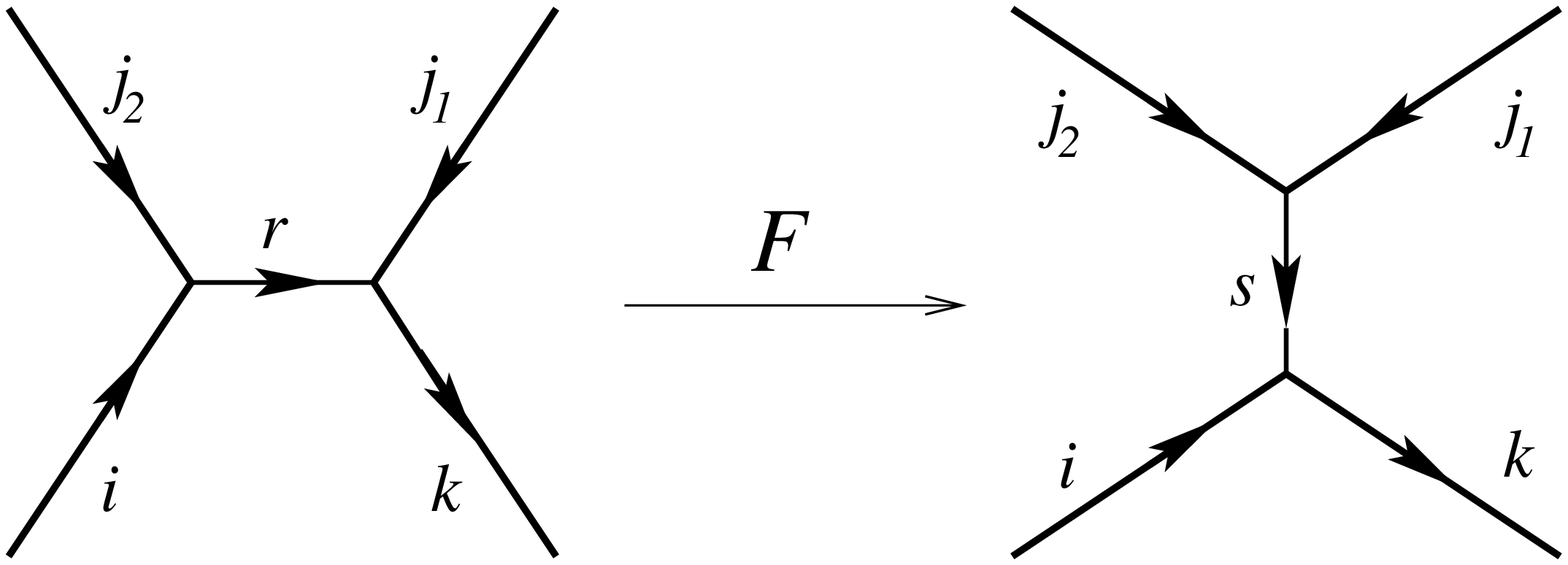}{12truecm}
\figlabel{\basic}
\vspace{5mm}

 \paragraph{Example: the U(1)-theory.} The chiral algebra of a single 
free bosonic field is known as U(1)-current algebra. It is generated by 
the modes $\a_n$ of the current $J(z)$ with the reality condition $\a_n^* 
= \a_{-n}$. There is only one real zero mode $\a_0 = \a_0^*$ so that 
the zero mode algebra $\cW^0$ is abelian. Hence, all its irreducible 
representations are 1-dimensional and there is one such representation 
for each real number $k$. The vector that spans the corresponding 
1-dimensional space $V_k^0$ is denoted by $|k\rangle$. It 
is easy to see that the space $\cV_k$ which we generate out of 
$|k\rangle$ by the creation operators $\a_{-n}$ admits a positive 
definite bilinear form 
for any choice of $k$. Hence, $\cJ = \mathbb{R}$ coincides with the 
set of irreducible representations of the zero mode algebra in this 
special case. 
\smallskip 

The character $\chi_k$ of the sector $\cV_k$ with conformal weight 
$h_k = \a' k^2/2$ is given by 
$$ \chi_k(q) \ = \ \frac{1}{\eta(q)}\ q^{\a' \frac{k^2}{2}} \ \ .$$
Along with the well known property $\eta(\tilde q) = \sqrt{-i \tau} 
\eta(q)$, the computation of a simple Gaussian integral shows that 
\be  \label{U1S}  
\chi_k(\tq) \ = \ \sqrt{\a'}\, \int dk' \, e^{2\pi i\, \a' kk'}\, 
  \chi_{k'}(q) \ = :\ \sqrt{\a'}\, \int dk' 
  \, S_{kk'} \chi_{k'}(q) \ \ .  
\ee  
This means that the entries of the S-matrix are phases, i.e.\ 
$S_{kk'} = \exp (2 \pi i \a' k k')$. Furthermore, it is not
too difficult to determine the fusion of two sectors. In fact, 
the action of $\delta_z$ on the zero mode $\a_0$ is given by
$$ \delta_z(\a_0) \ = \ \a_0 \o {\bf 1} + {\bf 1} \o \a_0 $$ 
since the current $J$ has conformal weight $h=1$. This shows that 
the fusion product amounts to adding the momenta, i.e.\ $k_1 \circ 
k_2 = k_1 + k_2$. In other words, the product of two sectors $k_1$ 
and $k_2$ contains a single sector $k_1 + k_2$.  
\smallskip 

In this case even the Fusing matrix is rather easy to compute. 
In fact, we can write down an explicit formula for the intertwining 
operators $V$. They are given by the normal ordered exponentials 
$\exp ikX$ of the chiral field $X(z) = \int^z dz' J(z')$, restricted 
to the spaces $\cV_{k'}$. When the operator product of two such 
exponentials with momenta $k_1$ and $k_2$ is expanded in the 
distance $z_1-z_2$, we find an exponential with momentum $k_1 
+ k_2$. The coefficient in front of this term is trivial, implying 
triviality for the fusing matrix.

\subsection{The bootstrap program}  

Having reviewed some elements from the representation theory of
chiral algebras we will now turn to discuss the `reduced matrix 
elements' that are most important for the exact solution of 
closed and open string backgrounds. These include the coupling 
of three closed strings and the coupling of closed strings to 
branes in the background. We shall find that these data are 
constrained by certain nonlinear (integral) equations. Two of 
these conditions, the crossing symmetry for the bulk couplings 
and the cluster property for the coupling of closed strings to 
branes will be worked out explicitly.  

\paragraph{Bulk fields and bulk OPE.} 
\def\bV{\bar V}
\def\bcV{\bar {\cal V}}
\def\bh{\bar h} 
Among the bulk fields $\Phi(z,\bz)$ of a conformal field theory 
on the complex plane we have already singled out the so-called 
{\em chiral fields} which depend on only one of the coordinates 
$z$ or $\bar z$ so that they are either holomorphic, $W = W(z)$, 
or anti-holomorphic, $\bW = \bW(\bar z)$. The most important of 
these chiral fields, the Virasoro fields $T(z)$ and $\bT(z)$, 
come with the stress tensor and hence they are present in any 
conformal field theory. But in most models there exist further 
(anti-)\-holomorphic fields $W(z)$ whose Laurent modes give 
rise to two commuting chiral algebras. 
\smallskip 

These chiral fields $W(z)$ and $\bW(\bz)$ along with all other 
fields $\Phi(z,\bz)$ that might be present in the theory can be 
considered as operators on the state space $\cH^{(P)}$ of the 
model. The latter admits a decomposition into irreducible 
representations $\cV_i$ and $\bcV_\bi$ of the two commuting 
chiral algebras, 
\be
 \cH^{(P)} \ = \ {\bigoplus}_{i,\bi}  \cV_i \o \bcV_\bi\ \ . 
\label{CSspace} 
\ee
While writing a sum over $i$ and $\bi$ we should keep in mind that
for non-compact backgrounds the `momenta' $i$ are typically continuous, 
though there can appear discrete contributions in the spectrum as well. 
Throughout this general introduction we shall stick to summations  
rather than writing integrals. Let us finally also agree to reserve 
the label $i=0$ for the vacuum representation $\cV_0$ of the chiral 
algebra. 
\smallskip 

To each state in the space $\cH^{(P)}$  we can assign a (normalizable) 
field. Fields associated with ground states of $\cH^{(P)}$ are particularly 
important. We shall denote them by $\Phi_{i,\bi}(z,\bar z)$ and refer to 
these fields as {\em primary fields}. In this way, we single out one field 
for each summand in the decomposition (\ref{CSspace}). All other fields in 
the theory can be obtained by multiplying the primary fields with chiral 
fields and their derivatives. 
\medskip%

So far, we have merely talked about the {\em space} of bulk fields. 
But more data is needed to characterize a closed string background. 
These are encoded in the short distance singularities of correlation 
functions or, equivalently, in the structure constants of the 
operator product expansions
\be \label{POPE} 
 \Phi_{i,\bi} (z_1,\bz_1) \Phi_{j,\bj}(z_2,\bz_2) \ = \ 
  \sum_{n \bn} \, {C_{i\bi;j\bj}}^{n\bn} \, z_{12}^{h_n-h_i - h_j}  
   \bz_{12}^{\bh_n - \bh_i - \bh_j} 
     \Phi_{n,\bn}(z_2,\bz_2) + \dots \ \ . 
\ee 
Here, $z_{12} = z_1-z_2$ and $h_i, \bh_\bi$ denote the conformal weights 
of the field $\Phi_{i,\bi}$, i.e.\ the values of $L_0$ and $\bL_0$ on 
$V_i^0 \o \bV_i^0$. The numbers $C$ describe the scattering amplitude for 
two closed string modes combining into a single one (``pant diagram'').  
Since all higher scattering diagrams can be cut into such 3-point vertices, 
the couplings $C$ should encode the full information about our closed string 
background. This is indeed the case. 

\paragraph{Crossing symmetry.} 
Obviously, the possible closed string couplings of a consistent string 
background must be very strongly constrained. The basic condition on 
the couplings $C$ arises from the investigation of 4-point 
amplitudes. Figure 2 encodes two ways to decompose a
diagram with four external closed string states into 3-point vertices. 
\vspace{1cm}
\fig{\label{figcross}
Graphical representation of the crossing symmetry conditions. 
The double lines represent closed string modes and remind us of the 
two commuting chiral algebras (bared and unbarred) in a bulk theory.}
{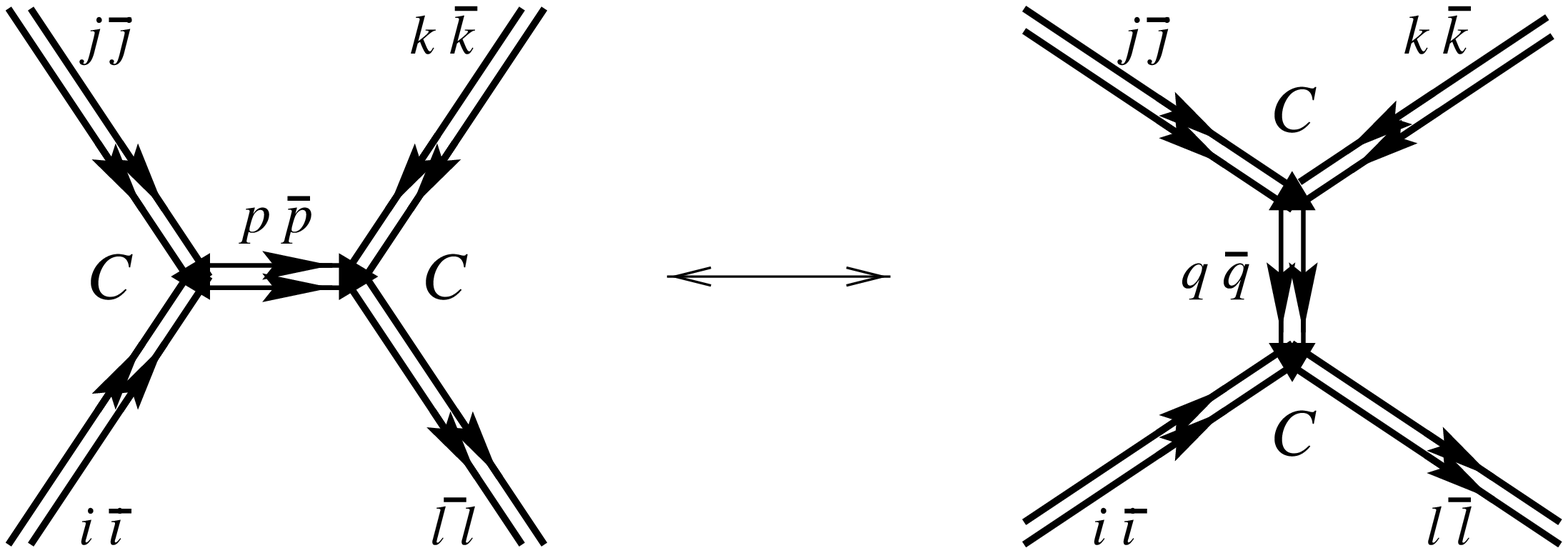}{12truecm}
\figlabel{\basic}
\vspace{5mm}
Accordingly, there exist two ways to express the amplitude through products
of couplings $C$. Since both cutting patterns must ultimately lead 
to the same answer, consistency of the 4-point amplitude gives rise
to a quadratic equation for the 3-point couplings. A more detailed 
investigation shows that the coefficients in this equation are 
determined by the Fusing matrix $F$, 
\begin{equation}\label{cross}
 \sum_{p \bar p} \  \Fus{p}{q}{j}{k}{i}{l}\, \Fus{\bar p}{\bar q}
{\bj}{\bar k}{\bi}{\bar l}\ 
{C_{i\bi,j\bj}}^{p\bar p} \ {C_{p \bar p, k\bar k}}^{l\bar l} \ = \ 
{C_{j\bj,k\bar k}}^{q\bar q} \ {C_{i\bi, q \bar q}}^{l\bar l} 
\end{equation}    
These factorization constraints on the 3-point couplings $C$ of 
closed strings are known as {\em crossing symmetry condition}. 
The construction of a consistent closed string background is 
essentially equivalent to finding a solution of eq.\ (\ref{cross}).

\paragraph{Example: The free boson.} Let us once more pause for a 
moment and illustrate the general concepts in the example of a single 
free boson. In this case, the state space of the bulk theory is given 
by 
\be \label{bulkFB}  
\cH^{(P)} \ = \ \int dk\  \cV_k \otimes \bcV_k \ \ . 
\ee
As long as we do not compactify the theory, there is a continuum of 
sectors parametrized by $i = k = \bi$. The formula (\ref{bulkFB}) 
provides a decomposition of the space of bulk fields into irreducible 
representations of the chiral algebra that is generated by the modes
$\a_n$ and $\bar \a_n$ (see above). States in $\cV_k \o \bcV_k$ are 
used to  describe all the modes of a closed string that moves with 
center of mass momentum $k$ through the flat space. 
\smallskip 

Associated with the ground states $|k\rangle \o |k\rangle$ we have 
bulk field $\Phi_{k,k}(z,\bz)$ for each momentum $k$. These fields 
are the familiar closed string vertex operators, 
$$ \Phi_{k,k}(z,\bz) \ = \ : \exp (i k X(z,\bz)) :\ \ .   $$
Their correlation functions are rather easy to compute (see e.g.\ 
\cite{Pol1Book}). From such expressions one can read off the 
following short distance expansion 
$$ \Phi_{k_1,k_1} (z_1,\bz_1) \,  \Phi_{k_2,k_2} (z_2,\bz_2) 
  \ \sim \ \int dk \, \delta(k_1+k_2 -k) \, |z_1-z_2|^{\ap 
     (k^2_1 + k^2_2 - k^2)} \, \Phi_{k,k}(z_2,\bz_2) + \dots 
\ \ . 
$$ 
Comparison with our general form (\ref{POPE}) of the operator 
products shows that the coefficients $C$ are simply given by 
\begin{equation}
{C_{k_1\bar k_1, k_2 \bar k_2}}^{k\bar k} \ = \ 
{C_{k_1 k_2}}^{k} \ = \ \delta(k_1 + k_2 - k) \ \ . 
\label{FBOPE}   
\end{equation} 
Note that the exponent and the coefficient of the short distance 
singularity are a direct consequence of the equation of motion 
$\Delta X(z,\bz) = 0$ for the free bosonic field. In fact, the 
equation implies that correlators of $X$ itself possess the 
usual logarithmic singularity when two coordinates approach 
each other. After exponentiation, this gives rise to the 
leading term in the operator product expansion of the fields  
$\Phi_{k,k}$. In this sense, the short distance singularity 
encodes the dynamics of the bulk field and hence characterizes 
the background of the model. Finally, the reader is invited to 
verify that the couplings $C$ satisfy the crossing symmetry 
condition (\ref{cross}) with a trivial fusing matrix.  
  
\subsection{The boundary bootstrap} 

\paragraph{Branes - the microscopic setup.}
With some basic notations for the (``parent'') bulk theory set up, 
we can begin our analysis of {\em associated} boundary theories 
(``open descendants''). These are conformal field theories on 
the upper half-plane $\Im z \geq 0$ which, in the interior 
$\Im z>0$, are locally equivalent (see below) to the given bulk 
theory. The state space $\cH^{(H)}$ of the boundary conformal 
field theory is equipped with the action of a Hamiltonian $H^{(H)}$ 
and of bulk fields $\Phi(z,\bar z)=
\Phi(z,\bz)^{(H)}$ which are well-defined for $\Im z >0$. While the 
space of these fields is the same as in the bulk theory, they are 
mathematically different objects since they act on different state 
spaces. Throughout most of our discussion below we shall neglect 
such subtleties and omit the extra super-script $(H)$. 
\smallskip 

Our first important condition on the boundary theory is that all the 
leading terms in the OPEs of bulk fields coincide with the OPEs 
(\ref{POPE}) in the bulk theory, i.e.\ for the fields $\Phi_{i,\bi}$ 
one has
\be \label{HOPE} 
 \Phi_{i,\bi} (z_1,\bz_1) \Phi_{j, \bj}(z_2,\bz_2) \ = \ \sum_{n \bn} 
  \, {C_{i\bi;j\bj}}^{n\bn} \, z_{12}^{h_n-h_i - h_j}  
   \bz_{12}^{\bh_n-\bh_i - \bh_j} 
     \Phi_{n,\bn}(z_2,\bz_2) + \dots 
\ee 
These relations express the condition that our brane is placed into 
the given closed string background.\footnote{In classical string theory 
the backreaction of branes on the bulk geometry is suppressed.} At the 
example of the free bosonic field we have discussed that the 
structure of the short distance expansion encodes the world-sheet 
dynamics. Having the same singularities as in the bulk theory simply 
reflects that the boundary conditions do not effect the equations 
of motion in the bulk. 
\smallskip 

In addition, we must require the boundary theory to be conformal. This 
is guaranteed if the Virasoro field obeys the following gluing condition 
\be \label{glueT} 
 T (z) \ = \ \bT (\bz) \ \ \ \ \mbox{ for } \ \ \ \
  z = \bz  \ \ . 
\ee
In the 2D field theory, this condition ensures that there is 
no momentum flow across the boundary. Note that eq.\ (\ref{glueT}) 
is indeed satisfied for the Virasoro fields $T \sim (\partial X)^2$ 
and $\bT \sim (\bar \partial X)^2$ in the flat space theory, both 
for Dirichlet and Neumann boundary conditions.%
\smallskip 

Considering {\em all} possible conformal boundary theories associated to 
a bulk theory whose chiral algebra is a true extension of the Virasoro 
algebra is, at present, too difficult a problem to be addressed 
systematically (see however \cite{GaReWa,GabRec,MaMoSe1,QueSch1} for 
some recent progress). For the 
moment, we restrict our considerations to maximally symmetric boundary 
theories, i.e.\  to the class of boundary 
conditions which leave the whole symmetry algebra $\cW$ unbroken. More 
precisely, we assume that there exists a local automorphism $\Omega$ -- 
called the {\em gluing map} -- of the chiral algebra $\cW$ such that 
\cite{RecSch1} 
\be   
   W(z) \ = \ \Omega\bW (\bar z) \ \ \mbox{ for } \ \ z = \bar z\ \ . 
\label{gluecond}
\ee
The condition (\ref{glueT}) is included in equation (\ref{gluecond}) 
if we require $\Omega$ to act trivially on the Virasoro field. The 
freedom incorporated in the choice of $\Omega$ is necessary to 
accommodate the standard Dirichlet and Neumann boundary conditions
for strings in flat space. Recall that in this case, the left and
right moving currents must satisfy $J(z) = \pm \bJ(\bz)$ all along 
the boundary. The trivial gluing automorphism $\Omega = \id$ in this 
case corresponds to Neumann boundary conditions while we have to 
choose $\Omega = -\id$ when we want to impose Dirichlet boundary 
conditions. 
\smallskip

For later use let us remark that the gluing map $\Omega$ on the 
chiral algebra induces a map $\omega$ on the set of sectors. In 
fact, since $\Omega$ acts trivially on the Virasoro modes, and 
in particular on $L_0$, it may be restricted to an automorphism
of the zero modes in the theory. If we pick any representation 
$j$ of the zero mode algebra we can obtain a new representation 
$\omega(j)$ by composition with the automorphism $\Omega$. This
construction lifts from the representations of $\cW^0$ on ground 
states to the full $\cW$-sectors. As a simple example consider 
the U(1) theory with the Dirichlet gluing map $\Omega (\a_n) = 
- \a_n$. We restrict the latter to the zero mode $\a_0$. 
As we have explained above, different sectors are labeled by the 
value $\sqrt{\a'}k$ of $\a_0$ on the ground state $|k\rangle$. If 
we compose the action of $\a_0$ with the gluing map $\Omega$, we 
find $\Omega (\a_0) |k\rangle = - \sqrt{\a'} k |k\rangle$. This 
imitates the action of $\a_0$ on $|\overline{\phantom{,, }}\, k
\rangle$. Hence, the map $\omega$ is given by $\omega(k) = - k$.

\paragraph{Ward identities.} 
As an aside, we shall discuss some more technical consequences
that our assumption on the existence of the gluing map $\Omega$ 
brings about. To begin with, it gives rise to an action of one 
chiral algebra $\cW$ on the state space $\cH\equiv \cH^{(H)}$
of the boundary theory. Explicitly, the modes $W_n = W_n^{(H)}$ 
of a chiral field $W$ dimension $h$ are given by 
$$ W_n \ := \ \frac{1}{2\pi i} \int_C \, z^{n+h-1}\, W(z) 
   \, dz + 
   \frac{1}{2\pi i} \int_C \, \bz^{n+h-1} \, \Omega \bW(\bz)  
   d \bz \ \ . $$ 
The operators $W_n$ on the state space $\cH$ are easily 
seen to obey the defining relations of the chiral algebra $\cW$.
Note that there is just one such action of $\cW$ constructed
out of the two chiral bulk fields $W(z)$ and $\Omega \bW(\bar z)$. 
\medskip

In the usual way, the representation of $\cW$ on $\cH$ leads 
to Ward identities for correlation functions of the boundary 
theory. They follow directly from the singular parts of the 
operator product expansions of the fields $W, \Omega \bW$ with 
the bulk fields $\Phi(z,\bar z)$. These expansions are fixed 
by our requirement of local equivalence between the bulk theory 
and the bulk of the boundary theory. Rather than explaining the
general form of these Ward identities, we shall only give one
special example, namely the relations that arise from the 
Virasoro field. In this case one find that 
\be \label{WardT} 
\left( T(w) \Phi(z,\bz)\right)_{\rm sing} \ = \ 
 \left[ \frac{h}{(w-z)^2} + \frac{\pl}{w-z} 
  +    \frac{\bar h}{(w-\bz)^2} + \frac{\bpl}{w-\bz}
 \right] \ \Phi(z,\bz) \ \ .  
\ee   
The subscript `sing' reminds us that we only look at 
the singular part of the operator product expansion.
Let us remark that the first two terms in the brackets
are well known from the Ward identities in the bulk 
theory. The other two terms can be interpreted as 
arising from a `mirror charge' that is located at the
point $w= \bz$ in the lower half-plane.%

\paragraph{One-point functions.} 
So far we have formalized what it means in world-sheet terms 
to place a brane in a given background (the principle of `local 
equivalence') and how to control its symmetries through gluing 
conditions (\ref{gluecond}) for chiral fields. Now 
it is time to derive some consequences and, in particular, to 
show that a rational boundary theory is fully characterized by 
just a family of numbers. 
\smallskip

Using the Ward identities described in the previous paragraph 
together with the OPE (\ref{HOPE}) in the bulk, we can reduce 
the computation of correlators involving $n$ bulk fields to the 
evaluation of 1-point functions $\langle \Phi_{i,\bi}\rangle_\a$
for the bulk primaries (see Figure 3). Here, the subscript $\a$ 
has been introduced to label different boundary theories that can 
appear for given gluing map $\Omega$.  
\vspace{1cm} 
\fig{With the help of operator product expansions in 
the bulk, the computation of $n$-point functions in a 
boundary theory can be reduced to computing 1-point 
functions on the half-plane. Consequently, the latter 
must contain all information about the boundary 
condition.} 
{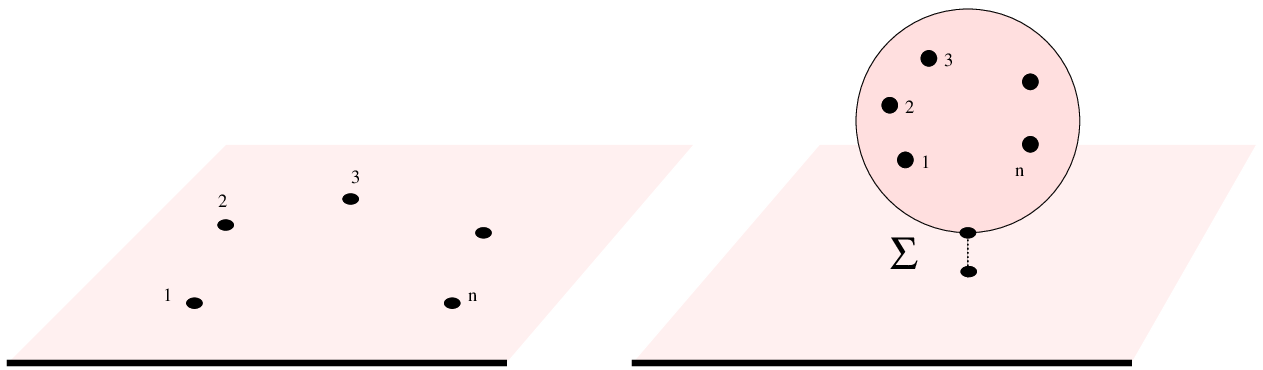}{14truecm} 
\figlabel{\basic}
\vspace{.5cm}

To control the remaining freedom, we notice that the transformation 
properties of $\Phi_{i,\bi}$ with respect to $L_n,\ n= 0, \pm 1,$ 
\ba 
 [\, L_n\, , \, \Phi_{i,\bi}(z,\bar z)\, ] & = & 
   z^n\, ( \, z \partial + h_i(n+1)\, ) \Phi_{i,\bi}(z,\bar z)   
  \nn \\[2mm] & & \hspace*{5mm} + \,  
   \bar z^n\, ( \, \bar z {\bar \partial} + \bar h_\bi(n+1)\, ) 
    \Phi_{i,\bi}  (z,\bar z) \ \   \nn
\ea
determine the 1-point functions up to scalar factors. Indeed, 
an elementary computation using the invariance of the vacuum 
state reveals that the vacuum expectation values $\langle 
\Phi_{i,\bi} \rangle_\a$ must be of the form 
\be
 \langle \Phi_{i,\bi} (z,\bar z) \rangle_\alpha  =  
 \frac{\cA^\alpha_{i\bi}}{|z - \bar z|^{h_i + h_\bi}}\ \ .
 \label{1ptfct}  
\ee
Ward identities for the Virasoro field and other chiral 
fields, should they exist, also imply $\bi = \omega(i^+)$ 
as a necessary condition for a non-vanishing 1-point 
function ($i^+$ denotes the representation conjugate to $i$), i.e. 
$$ \cA^\alpha_{i\bi} \ = \ A^\alpha_i \ 
   \delta_{\bi,\omega(i)^+}  \ \ . 
$$ 
Since $h_i = h_{i^+} = h_{\omega(i)}$ we can put  $h_i + h_\bi= 
2 h_i $ in the exponent of eq.\ (\ref{1ptfct}). Our arguments 
above have reduced the description of a boundary condition to 
a family of scalar parameters $A^\a_i$ in the 1-point functions. 
Once we know their values, we have specified the boundary theory. 
This agrees with our intuition that a brane should be completely 
characterized by its couplings to closed string modes such as 
the mass and RR charges.

\paragraph{The cluster property.}  
We are certainly not free to choose the remaining parameters $A^\a_{i}$ 
in the 1-point functions arbitrarily. In fact, there exist strong {\em 
sewing constraints} on them that have been worked out by several authors 
\cite{CarLew,Lewe1,PrSaSt1,PrSaSt2,BPPZ1}. These can be derived from the 
following {\em cluster property} of the 2-point functions 
\be \lim_{a \rightarrow \infty} \ \langle 
    \Phi_{i,\bi} (z_1,\bz_1) 
\Phi_{j,\bj} (z_2+a, \bz_2+a) \rangle_\alpha 
    \ = \  \langle \Phi_{i,\bi}(z_1,\bz_1) \rangle_\alpha \langle 
   \Phi_{j,\bj} (z_2,\bz_2)\rangle_\alpha \ \ . \label{cluster}
\ee 
Here, $a$ is a real parameter, and the field $\Phi_{j,\bj}$ 
on the right hand side can be placed at $(z_2,\bz_2)$ since the 
whole theory is invariant under translations parallel to the 
boundary. 
\smallskip

Let us now see how the cluster property restricts the choice of  
possible 1-point functions. We consider the 2-point function 
of the two bulk fields as in eq.\ (\ref{cluster}). There are two 
different ways to evaluate this function. On the one hand, we can 
go into a regime where the two bulk fields are very far from each 
other in the direction along the boundary. By the cluster property, 
the result can be expressed as a product of two 1-point functions 
and it involves the product of the couplings $A^\a_i$ and $A^\a_j$. 
Alternatively, we can pass into a regime in which the two bulk 
fields are very close to each other and then employ the operator 
product (\ref{HOPE}) to reduce their 2-point function to a sum 
over 1-point functions. Comparison of the two procedures provides 
the following important relation, 
\be A^\a_i \, A^\a_j \ = \ \sum_k \ \Xi^{k}_{ij} \ 
   A^\alpha_0 \ A^\a_k 
  \ \ .  \label{class} \ee
It follows from our derivation that the coefficient $\Xi^k_{ij}$ 
can be expressed as  a combination 
\be \Xi^k_{ij} \ = \ {C_{i\bi;j\bj}}^{k\bar k} \ \Fus{k}{0}{j\ }{\omega(j)}
{i\ }{\omega(i)^+} \label{XiCF} \ee 
of the coefficients $C$ in the 
bulk OPE and of the fusing matrix. The latter arises when we pass 
from the regime in which the bulk fields are far apart to the regime 
in which they are close together (see Figure 4).  
\vspace{1cm}
\fig{Equations (\ref{class}), (\ref{XiCF}) are derived by comparing 
two limits of the 2-point function. The dashed line represents the 
boundary of the world-sheet and we have drawn the left moving sector 
in the lower half-plane (doubling trick).}
{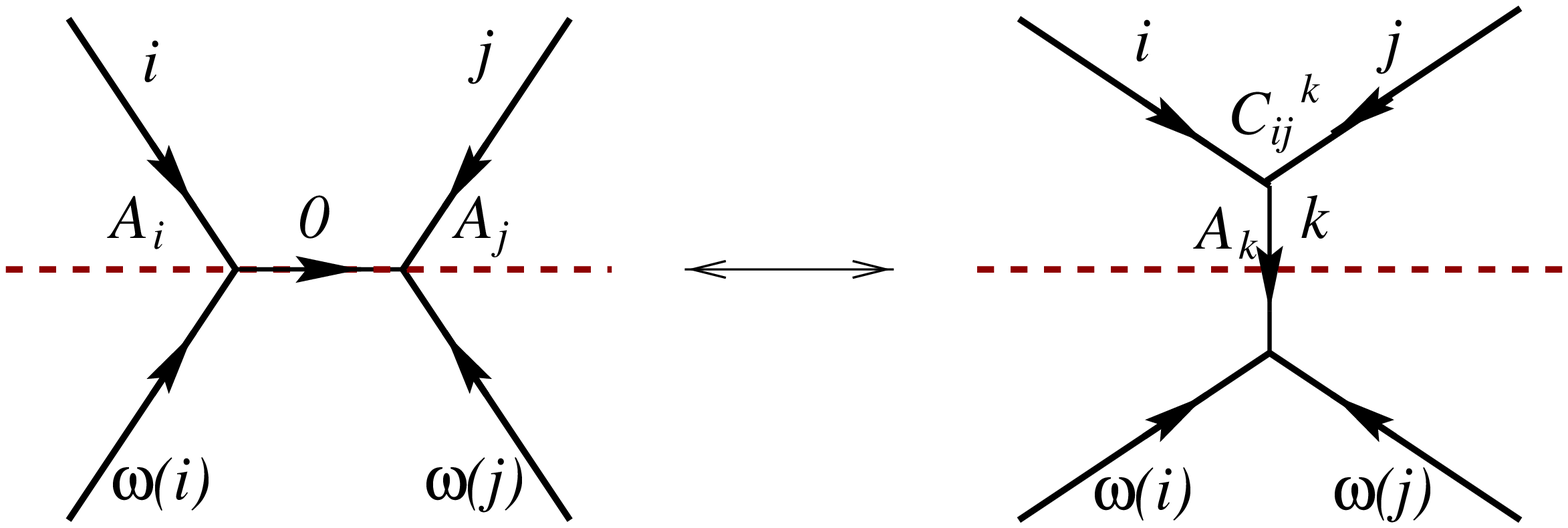}{13truecm}
\figlabel{\basic}
\vspace{5mm} 
The importance of eq.\ (\ref{class}) for a 
classification of boundary conformal field theories has been stressed 
in a number of publications \cite{FucSch1,BPPZ1,RecSch2} and 
is further supported by their close relationship with algebraic 
structures that entered the classification of bulk conformal field 
theories already some time ago (see e.g.\ \cite{Pasq,PetZub1,PetZub2}). 
\smallskip

The algebraic relations (\ref{class}) typically possess several 
solutions which are distinguished by our index $\a$. Hence, maximally 
symmetric boundary conditions are labeled by pairs $(\Omega,\alpha)$. 
The automorphism $\Omega$ is used to glue holomorphic and 
anti-holomorphic fields along the boundary and the consistent choices 
for $\Omega$ are rather easy to classify. Once $\Omega$ has been fixed, 
it determines the set of bulk fields that can have a non-vanishing 
1-point function. For each gluing automorphism $\Omega$, the non-zero 
1-point functions are constrained by algebraic equations (\ref{class}) 
with coefficients $\Xi$ which are determined by the closed string 
background. A complete list of solutions is available in a large 
number of cases with compact target spaces and we shall see a few
solutions in  our non-compact examples later on. Right now, however, 
it is most important to explain how one can reconstruct further  
information on the brane from the couplings $A^\a_i$. In particular, 
we will be able to recover the entire spectrum of open string 
excitations. Since the derivation makes use of {\em boundary states}, 
we need to introduce this concept first.        
 
\paragraph{Example: The free boson.} We wish to close this subsection 
with a short discussion of boundary conditions in the theory of a 
single free boson. The corresponding bulk model has been discussed
earlier so that we know its bulk couplings (\ref{FBOPE}). Using 
the triviality of the Fusing matrix, we are therefore able to spell 
out the cluster condition (\ref{class}) for the couplings $A$, 
$$ A^\a_{k_1} \, A^\a_{k_2} \ = \ \int dk_3 \, 
           \delta(k_1+k_2 -k_3) \, A_{k_3}^\a \ = \ 
         A^\a_{k_1+k_2}\ \ .   
$$ 
Solutions to this equation are parametrized by a single real 
parameter $\a = x_0$ and possess the form 
$$ A^{x_0}_k \ = \ e^{2 i k x_0} \ \ . $$
Obviously, the parameter $\a = x_0$ is interpreted as the transverse 
position of a point-like brane. In writing down the cluster condition
we have assumed that all fields $\Phi_k$ possess a non-vanishing 
coupling. This is possible, if the momentum label $k$ and the 
associated label $\omega (k)$ of its mirror image add up to 
$k + \omega(k) = 0$. In other words, we must choose the gluing 
automorphism $\Omega$ to act as $\Omega(J) = - J$, i.e.\ we have
to impose Dirichlet boundary conditions on the field $X$ . For 
Neumann boundary conditions, on the other hand, $\Omega(J) = J$ 
and hence only the identity field $\Phi_0$ can have a non-vanishing 
1-point coupling. The factorization constraint for the latter is
certainly trivial.

\subsection{Boundary states}
 
It is possible to store all information about the couplings $A^\a_i$ 
in a single object, the so-called {\em 
boundary state}. To some extent, such a boundary state can be 
considered as the wave function of a closed string that is sent off 
from the brane $(\Omega,\alpha)$. It is a special linear combination 
of generalized coherent states (the so-called Ishibashi states). The 
coefficients in this combination are essentially the closed string 
couplings $A^\a_i$. 
\smallskip

One way to introduce boundary states is to equate correlators of bulk 
fields on the half-plane and on the complement of the unit disk in the 
plane. With $z, \bar z$ as before, we introduce coordinates $\zeta, 
\bar\zeta$ on the complement of the unit disk by 
\be
\zeta \ =\  {\frac{1-iz}{1+i z}}\quad\quad {\rm and}  \quad\quad
\bar\zeta \ =\  \frac{1+i\bar z}{1-i \bar z}\ \ . 
\label{coortrsfzeta}\ee
If we use $\,|0\rangle$ to denote the vacuum of the bulk conformal field theory, then 
the boundary state $|\alpha\rangle = |\alpha\rangle_\Omega$ can be 
uniquely characterized by \cite{CarLew,RecSch1}
\be
 \langle\,  \Phi^{(H)}(z,\bar z) \rangle_{\alpha}
\ =\ \left(\frac{d\zeta}{dz}\right)^h \left(\frac{d\bar \zeta}{d
\bar z}\right)^{\bar h}  \cdot  
\langle 0|\, \Phi^{(P)}(\zeta,\bar\zeta) | \alpha 
\rangle\label{bstdef}
\ee
for primaries $\Phi$ with conformal weights $(h,\bar h)$.   
Note that all quantities on the right hand side are defined in the 
bulk conformal field theory (super-script P), while objects on the 
left hand side live on the half-plane (super-script H).  
\smallskip

In particular, we can apply the coordinate transformation from $(z,\bar z)$ 
to $(\zeta,\bar \zeta)$ on the gluing condition (\ref{gluecond}) to 
obtain
$$ W(\zeta) \ = \ (-1)^h \, \bar \zeta^{2h} \Omega \bW(\bar \zeta) $$ 
along the boundary at $\zeta \bar \zeta = 1$. Expanding this into modes,
we see that the gluing condition (\ref{gluecond}) for chiral fields 
translates into the following linear constraints for the boundary 
state, 
\be
\bigl[\,W_n-(-1)^{h_W}\Omega\bW_{-n}\,\bigr]\,
|\alpha\rangle_{\Omega}\ =\ 0\ .
\label{glueconplane}
\ee
These constraints posses a linear space of solutions. It is spanned 
by generalized coherent (or Ishibashi) states $|i\rangle\! \rangle$.  
Given the gluing automorphism $\Omega$, there exists one such solution 
for each pair $(i,\omega(i^+))$ of irreducibles that occur in the bulk 
Hilbert space \cite{Ishi1}. $|i\rangle\!\rangle_{\Omega}$ is unique up 
to a scalar factor which can be used to normalize the Ishibashi states 
such that 
\be
 {}_{\Omega}\langle\!\langle j |\,  \tilde q^{L_0^{(P)}-\frac{c}{24}} 
\,|i\rangle\!\rangle_{\Omega} \ = \ \delta_{i,j}\, \chi_i(\tilde q)
\ \ . \label{ishchar}
\ee
Following an idea in \cite{Ishi1}, it is easy to write down an expression 
for the generalized coherent states (see e.g.\ \cite{RecSch1}), but the 
formula is fairly abstract. Only for strings in a flat background their
constructions can be made very explicit (see below). 
\smallskip
 
Full boundary states $|\alpha\rangle_{\Omega} \equiv |(\Omega,\alpha)
\rangle$ are given as certain linear combinations of Ishibashi states, 
\be \label{bstf1} 
 |\alpha\rangle_{\Omega} \ = \ \sum_i B^i_{\alpha}\, 
|i\rangle\!\rangle_{\Omega}\ .
\ee 
With the help of (\ref{bstdef}), one can show \cite{CarLew,RecSch1} that 
the coefficients $B^i_\alpha$ are related to the 1-point functions of the 
boundary theory by 
\be
A_{i}^{\alpha} \ = \ B^{i^+}_{\alpha}  \ \ . 
\label{AeqBB}\ee
The decomposition of a boundary state into Ishibashi states contains 
the same information as  the set of 1-point functions and therefore 
specifies the ``descendant'' boundary conformal field theory of a given bulk conformal field theory completely. 
\medskip 

\paragraph{Example: The free boson.} Here we want to spell out explicit 
formulas for the boundary states in the theory of a single free boson. 
Let us first discuss this for Dirichlet boundary conditions, i.e.\ the 
case when the U(1)-currents of the model satisfy the gluing condition
(\ref{gluecond}) with $\Omega^D \bJ = - \bJ$. Since $k^+ = -k$ (recall 
that fusion of sectors is given by adding momenta) and $\omega^D (k) = 
-k$, we have $\omega^D(k)^+= k$ and so there exists a coherent state 
for each sector in the bulk theory. These states are given by 
$$ 
   |k\rangle\!\rangle_D \ = \ \exp \left(\sum_{n=1}^\infty\ \frac{1}{n}\,  
     \a_{-n} \bar \a_{-n} \right)\ |k\rangle \otimes \overline{|k\rangle} 
\ \ .    
$$ 
Using the commutation relation of $\a_n$ and $\bar \a_n$ it is easy to 
check that $|k\rangle\!\rangle_D$ is annihilated by $\a_n - \bar \a_{-n}$ 
as we required in eq.\ (\ref{glueconplane}). Since 1-point functions of 
closed string vertex operators for Dirichlet boundary conditions have the 
form (\ref{1ptfct}) with $A_k^{D,x_0} = \exp (i 2 k x_0)$, we obtain  
the following boundary state,  
$$ |x_0 \rangle_D \ = \ \sqrt{\alpha'}\, \int  dk\  
    \ e^{-i 2 k x_0} \ |k\rangle\! \rangle_D \ \ . $$ 
For Neumann boundary conditions the analysis is different. Here we 
have to use the trivial gluing map $\Omega^N = \id$ and a simple 
computation reveals that the condition $\omega^N (k) = k^+$ is only 
solved by $k = 0$. This means that we can only construct one coherent 
state, 
$$ |0\rangle\!\rangle_N \ = \ \exp(- \sum_{n=1}^\infty \frac{1}{n} 
     \a_{-n} \Omega^B\bar \a_{-n} )\ |0\rangle \otimes 
    \overline{|0\rangle} \ \ . 
$$ 
This coherent state coincides with the boundary state $|0\rangle_N = 
|0\rangle\!\rangle_N$ for Neumann boundary conditions. 

\subsection{The modular bootstrap} 

\paragraph{The boundary spectrum.}\hspace*{-2mm}%
While the 1-point functions (or boundary states) uniquely characterize
a boundary conformal field theory, there exist more quantities we are 
interested in. In particular, we shall now see how the coefficients of 
the boundary states determine the spectrum of open string vertex 
operators that can be inserted along the boundary of the world-sheet. 
\vspace{1cm}
\fig{The open string partition function $Z_{\a\b}$ can be computed 
by world-sheet duality. In the figure, the time runs upwards so that 
the left hand side is interpreted as an open string 1-loop diagram 
while the right hand side is a closed string tree diagram.}
{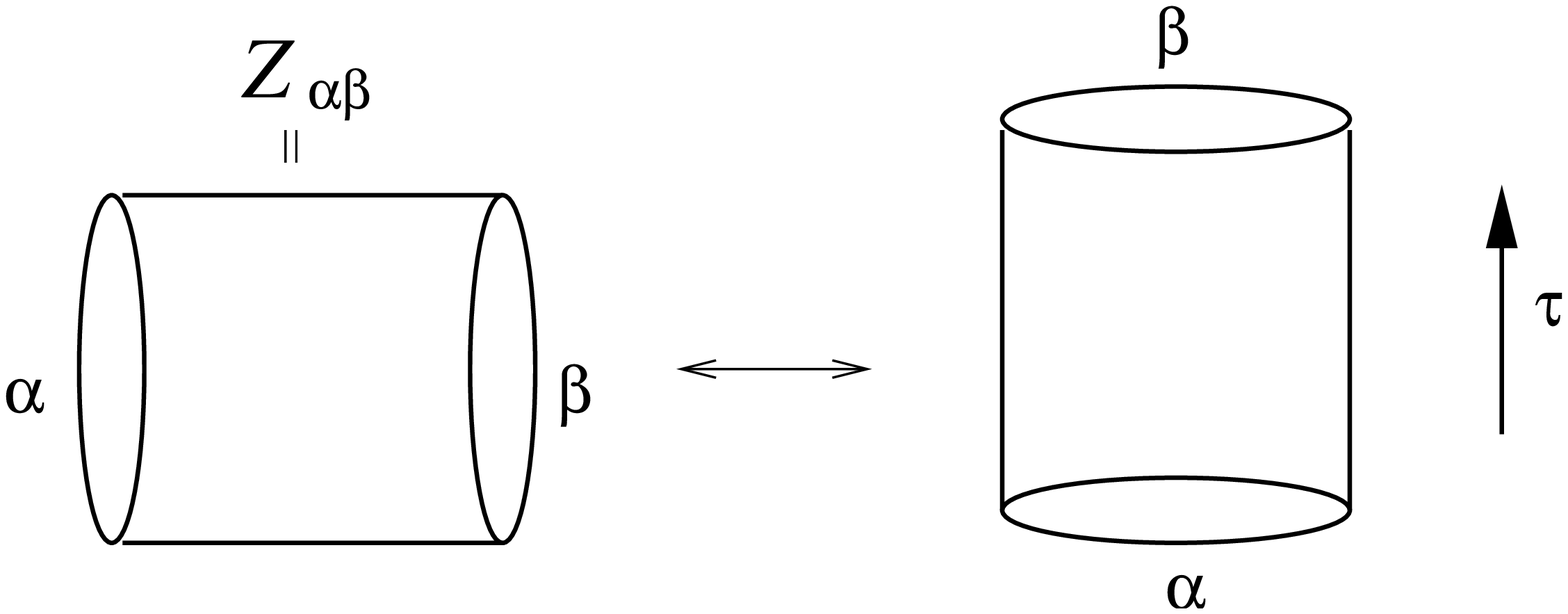}{13truecm}
\figlabel{\basic}
\vspace{1cm}

Our aim is to determine the spectrum of open string modes which can 
stretch between two branes labeled by $\a$ and $\beta$, both being 
of the same type $\Omega$. In world-sheet terms, the quantity we 
want to compute is the partition function on a strip with boundary 
conditions $\a$ and $\beta$ imposed along the two sides. This is 
depicted on the left hand side of Figure 5. The figure also 
illustrates the main idea of the calculation. In fact, world-sheet 
duality allows to exchange space and time and hence to turn the one 
loop open string diagram on the left hand side into a closed string 
tree diagram which is depicted on the right hand side. 
The latter corresponds to a process in which a closed string is created 
on the brane $\a$ and propagates until it gets absorbed by the brane 
$\beta$. Since creation and absorption are controlled by the amplitudes 
$A^\a_i$ and $A^\beta_j$, the right hand side - and hence the partition 
function on the left hand side - is determined by the 1-point functions
of bulk fields. 
\smallskip 

Let us now become a bit more precise and derive the exact relation 
between the couplings $A$ and the partition function. Reversing the 
above sketch of the calculation, we begin on the left hand side of 
Figure 5 and compute 
$$ \langle \theta \beta| {\tilde q}^{H^{(P)}} |\alpha\rangle \ = \ 
    \sum_j A^\b_{j^+} A^\a_j \langle\!\langle j^+| 
   {\tilde q}^{L_0^{(P)}-\frac{c}{24}}|j^+\rangle \! \rangle 
   \ =  \   \sum_j A^\b_{j^+} A^\a_j \chi_{j^+}(\tilde q)\ \ .  
$$       
Here we have dropped all subscripts $\Omega$ since all the boundary 
and generalized coherent states are assumed to be of the same type. 
The symbol $\theta$ denotes the world-sheet CPT operator in the bulk 
theory. It is a anti-linear map which sends sectors to their conjugate, 
i.e.\ 
$$ \theta A^\b_{j^+}\ |j\rangle\!\rangle \ = \ \left( A^\b_{j^+}
   \right)^* \ |j^+\rangle\! \rangle \ \ . $$ 
Having explained these notations, we can describe the steps we 
performed in the above short computation. To begin with we inserted  
the expansion (\ref{bstf1}),(\ref{AeqBB}) of the boundary 
states in terms of Ishibashi states and the formula $H^{(P)} = 1/2(L_0 
+ \bar L_0) - c/24$ for the Hamiltonian on the plane. With the help of 
the linear relation (\ref{glueconplane}) we then traded $\bar L_0$ for 
$L_0$ before we finally employed the formula (\ref{ishchar}). At this 
point we need to recall the property (\ref{Smatrix}) of characters to 
arrive at 
\be \label{BPF1}  
 \langle \theta \beta| {\tilde q}^{H^{(P)}} |\alpha\rangle
 \ = \ \sum_j A^\b_{j^+} A^\a_j S_{j^+ i} \chi_i(q) \ =:\  
   Z_{\a\b}(q)\ \ .  
\ee
As argued above, the quantity we have computed should be interpreted 
as a boundary partition function and hence as a trace of the operator 
$ \exp(2\pi i \tau H^{(H)})$ over some space $\cH_{\a\b}$ of states 
for the system on a strip with boundary conditions $\a$ and $\beta$ 
imposed along the boundaries. 
\smallskip

If at least one of the two branes is compact, we expect to find a
discrete open string spectrum. In this case, our computation leads 
to a powerful constraint on the numbers $A^\a_i$. In fact, since 
our boundary conditions preserve the chiral symmetry, the partition 
function is guaranteed to decompose into a sum of the associated 
characters. If this sum is discrete, i.e.\ not an integral, the 
coefficients in this expansion must be integers and so we 
conclude\vspace*{2mm}
\be \label{BPF2}  
 Z_{\a\b}(q) \ = \ \sum_i \ {n_{\a\b}}^i\, \chi_i(q) 
  \ \ \ \mbox{ where } \ \ \ {n_{\a\b}}^i \ = \  \sum_j A^{\b}_{j^+} 
 A^\a_j S_{j^+i} \ \in \ \mathbb{N} \ \ .
\ee
Although there exists no general proof, it is believed that every 
solution of the factorization constraints (\ref{class}) gives rise 
to a consistent spectrum with integer coefficients ${n_{\a\b}}^i$. 
A priori, the integrality of the numbers ${n_{\a\b}}^i$ provides 
a strong constraint, known as the {\em Cardy condition},  on the 
set of boundary states and it has often been used instead of eqs.\ 
(\ref{class}) to determine the coefficients $A^\a_i$. Note that 
the Cardy conditions are easier to write down since they only involve 
the modular S-matrix. To spell out the factorization constraints 
(\ref{class}), on the other hand, one needs explicit formulas for 
the fusing matrix and the bulk operator product expansion.  
\smallskip

There is one fundamental difference between the Cardy condition 
(\ref{BPF2}) and the factorization constraints (\ref{class}) that 
is worth pointing out. Suppose that we are given a set of solutions
of the Cardy constraint. Then every non-negative integer linear 
combination of the corresponding boundary states defines another 
Cardy-consistent boundary theory. In other words, solutions of the 
Cardy condition form a cone over the integers. The factorization 
constraints (\ref{class}) do not share this property. Geometrically, 
this is easy to understand: we know that it is possible to construct 
new brane configurations from arbitrary superpositions of branes
in the background (though they are often unstable). These brane 
configurations possess a consistent open string spectrum but they 
are not elementary. As long as we are solving the Cardy condition, 
we look for such configurations of branes. The factorization 
constraints (\ref{class}) were derived from the cluster property 
which ensures the system to be in a `pure phase'. Hence, by solving 
eqs.\ (\ref{class}) we search systematically for elementary brane 
configurations that cannot be decomposed any further. Whenever the 
coefficients $\Xi$ are known, solving the factorization constraints 
is clearly the preferable strategy, but sometimes the required 
information is just hard to come by. In such cases, one can still 
learn a lot about possible brane configurations by studying Cardy's 
conditions. Let us stress again, however, that the derivation of
the Cardy condition required compactness of at least one of the 
branes. If both branes are non-compact, the open string spectrum 
will contain continuous parts which involve an a priori unknown 
spectral density function rather than integer coefficients. We 
will come back to such issues later.  
\newpage

\section{Bulk Liouville field theory}  
\setcounter{equation}{0}

Our goal for this lecture is to present the solution of the 
Liouville bulk theory. This model describes the motion of closed 
strings in a 1-dimensional exponential potential and it 
can be considered as the minimal model of non-rational conformal 
field theory. Before we explain how to determine the bulk spectrum 
and the exact couplings, we would like to make a few more comments 
on the model and its applications.  
\smallskip 

On a 2-dimensional world-sheet with metric $\c^{ab}$ and 
curvature $R$, the action of Liouville theory takes the form 
\be \label{actLiouv} 
   S_L[X] \ = \ \int_\Sigma d^2 \sigma \sqrt \c 
     \left( \c^{ab} \pl_a X \pl_b X + R Q X 
    + \mu e^{2bX} \right) 
\ee
where $\mu$ and $b$ are two real parameters of the model. 
The second term in this action describes a linear dilaton and 
such a term would render perturbative string theory invalid if 
the strong coupling region was not screened by the third term. 
In fact, the exponential potential has the effect to keep 
closed strings away from the strong coupling region of the 
model. 
\smallskip

Liouville theory should be considered as a marginal deformation 
of the free linear dilaton theory, 
$$ S_{LD}[X] \ = \  \int_\Sigma d^2 \sigma \sqrt \c 
     ( \c^{ab} \pl_a X \pl_b X + R Q X )  \ \ . $$ 
The Virasoro field of a linear dilaton theory is given by the 
familiar expression 
$$ T \ = \ (\partial X)^2 + Q \pl^2 X \ \ . $$ 
The modes of this field form a Virasoro algebra with 
central charge $c = 1 + 6 Q^2$. Furthermore, the usual 
closed string vertex operators 
$$ \Phi_\a \ =\  :\exp 2\a X: \ \ \ \ \mbox{ have } \ \ 
  h_\a \, = \, \a (Q-\a)\,  = \, \bh_\a\ \ . $$  
Note the conformal weights $h,\bh$ are real if $\a$ is of the form 
$\a = Q/2 + iP$. We interpret the real parameter $P$ as the 
momentum of the closed string tachyon scattering state that is
created by the above vertex operator.   
\smallskip 

In order for the exponential potential in the Liouville action 
to be marginal, i.e.\ $(h_b,\bh_b) = (1,1)$, we must now also 
adjust the parameter $b$ to the choice of $Q$ in such a way that 
$$ Q = b + b^{-1} \ \ . $$ 
As one may easily check, Weyl invariance of the classical action 
$S_L$ leads to the relation $Q_{c} = b^{-1}$. Quantum corrections 
deform this correspondence such that $Q = Q_c + b$. The extra term, 
which certainly becomes small in the semi-classical limit $b 
\rightarrow 0$, has a remarkable consequence: It renders the 
parameter $Q$ (and hence the central charge) invariant under 
the replacement $ b \rightarrow b^{-1}$. We will have a lot more
to say about this interesting quantum symmetry of Liouville
theory. 
\smallskip   

After this preparation we are able to describe two interesting
application of Liouville theory. The first one comes with the 
observation that Liouville theory manages to contribute a value
$c \geq 25$ to the central charge even though it involves only 
a single dimension. Hence, in order to obtain a consistent 
string background it suffices to add one more direction with 
central charge $c=1$ or less. Geometrically, this would then 
describe a string background with $D \leq 2$. These theories
have indeed been studied extensively in the past and there
exist many results, mostly due to the existence of a dual 
matrix model description. 
\smallskip 

The second application is much more recent and also less well 
tested: it has been proposed that time-like Liouville theory 
at $c=1$ describes the homogeneous condensation of a closed 
string tachyon. A short look at classical actions makes this 
proposal seem rather plausible. In fact, the condensation of 
a closed string tachyon is described on the world-sheet by 
adding the following term to the action of some static 
background,  
\be \label{tachact} 
 \d S[X] = \int_\Sigma d^2 z \, e^{\epsilon_\Phi X^0}\, 
                  \Phi(z,\bz) \ \ . 
\ee 
Here, $X^0$ is a time-like free field and the bulk field $\Phi$ 
must be a relevant field in the conformal field theory of a 
spatial slice of the static background. $\Phi$ describes the 
profile of the tachyon. If we want the tachyon to have a 
constant profile we must choose $\Phi = \mu =${\it const}. 
The parameter $\epsilon_\Phi$ is then forced to be $\e_\mu 
= 2$ so that the interaction term (\ref{tachact}) becomes
scale invariant. If we now Wick-rotate the field $X^0 = 
i X$ the interaction term looks formally like the interaction 
term in Liouville theory, only that the parameter $b$ assumes 
the unusual value $b=i$. At this point, the central charge of
the model is $c=1$ and hence we recover the content of a 
proposal formulated in \cite{Gutperle:2003xf}: The rolling 
tachyon background is a Lorentzian $c=1$ Liouville theory.

\subsection{The minisuperspace analysis}
\setcounter{equation}{0} 
\def\w{\omega} 
\def\tw{\tilde \w}

In order to prepare for the analysis of exact conformal field 
theories it is usually a good idea to first study the particle 
limit $\a'\rightarrow 0$. Here, sending $\a'$ to zero is 
equivalent to sending $b$ to zero after rescaling both 
the coordinate $X = b^{-1} x$ and the coupling $\pi \mu = b^{-2} 
\lambda$. What we end up with is the theory of a particle 
moving in an exponential potential. The stationary Schroedinger 
equation for this system is given by 
\begin{equation}
 H_L\, \phi \ :=\ \left(- \frac{\pl^2}{\pl x^2} + \lambda \, 
     e^{2 x} \right) \, \phi(x) \ = \  4 \omega^2 \phi(x)\ \ . 
\end{equation}
This differential equation is immediately recognized 
as Bessel's equation. Hence, its solutions are linear 
combinations of the Bessel functions of first kind, 
$$ \phi^\pm_\omega(x) \ = \ J_{\pm 2 i \omega} 
  (i \sqrt \lambda e^x) \ \ . $$ 
These functions $\phi^\pm$ describe an incoming/outgoing plane 
wave in the region $x \rightarrow -\infty$, but they are 
both unbounded as can be seen from the asymptotic behavior 
at $ x \rightarrow \infty$, 
$$ \phi^\pm_{\omega}(x) \ \longrightarrow \  
   e^{-\frac{x}{2}} \, \cosh \left(\sqrt{\lambda} e^x \mp \pi \omega 
   + i \frac{\pi}{4}\right)\ \ . $$ 
There is only one particular linear combination of these two 
solutions that stays finite for $x \rightarrow \infty$. This is 
given by 
\begin{equation} \label{solin} 
 \phi_\omega(x) \ = \ (\lambda/4)^{-i\w} 
   \, \Gamma^{-1}(-2 i\omega) \, K_{-2i\omega}(\sqrt{\lambda}e^{x}) \ \ . 
\end{equation}  
where $K_\nu(z) := J_\nu(iz) - \exp(i\pi \nu ) J_{-\nu}(iz)$ is 
known as modified Bessel function. We have also fixed an the 
overall normalization such that the incoming plane wave has 
unit coefficient.   
\smallskip 

Here we are mainly interested in the 3-point function, since in the full 
conformal field theory this quantity encodes all the information about 
the exact solution. Its counterpart in the minisuperspace model can be 
evaluated through the following integral over a product of Bessel 
functions,   
\begin{eqnarray}  \label{sc3pt} 
 \langle \w_1| e_{\w_2} |\w_3\rangle & := & 
    \int_{-\infty}^\infty dx\ \phi_{\w_1}(x) \, e^{2 i \w_2 x}\,  
    \phi_{\w_3}(x) \\[2mm] & = & 
    (\lambda/4)^{-2 i \tw} \, 
   \Gamma(2i\tw)  \, \prod_{j=1}^3 
   \frac{\Gamma(1+(-1)^j 2 i\tw_j)}{\Gamma(1-(-1)^j 2 i\w_j)}\ \ , 
   \\[2mm] 
\mbox{where} & & \tw = \frac12(\w_1 + \w_2 + \w_3) \ \ \ \ , \ \ \ 
                 \tw_i = \tw - \tw_i \ \ . \label{tw} 
\end{eqnarray}
The formula that was used to compute the integral can be found in standard 
mathematical tables. Let us remark already that the exact answer in the 2D
field theory will have a very similar form, only that the $\Gamma$ functions 
get replaced by a more complicated special function (see below). Observe 
also that the result has poles at $\tw = 0 = \tw_i$ and zeroes whenever 
one of the frequencies $\w_i$ vanishes. 
\smallskip

From the 3-point function it is not hard to extract the 2-point function 
of our toy model in the limit $\omega_2 = \varepsilon \rar 0$. Note that 
the factor $P$ has poles at $\omega_1 \mp \omega_3 \pm \varepsilon = 0$ 
which, after taking the limit, produce terms of the form $\delta(\omega_1\mp
\omega_3)$. More precisely, we obtain  
\begin{eqnarray} \nonumber 
 \lim_{\omega_2 \rar 0} \langle \w_1| e_{\w_2} |\w_3\rangle & \sim & 
\delta(\w_1+\w_3) + R_0(\w_1)  
\delta(\w_1-\w_3) \,  \\[2mm]  
\mbox{with} \ \ \ \ R_0(\w) & = &  \lambda^{-2 i \w} \ 
 \frac{\Gamma(2 i\omega)}{\Gamma(-2 i\omega)} \ \ .  
 \label{R0L} \end{eqnarray} 
The quantity $R_0$ is known as the reflection amplitude. It describes
the phase shift of the wave function upon reflection in the Liouville 
potential as can be seen from the following expansion of the wave 
functions $\phi_\w$ for $x \rightarrow - \infty$, 
$$ \phi_\w(x) \ \sim \ e^{2i\w x} + \lambda^{-2i\w} 
    \frac{\Gamma(2i\w)}{\Gamma(-2i\w)} \, e^{-2i \w x}\ \ . $$ 
We conclude that the two coefficients of the $\delta-$functions in the 
2-point function encode both the normalization of the incoming plane 
wave and the phase shift that appears during its reflection. 

\subsection{The path integral approach} 

In an attempt to obtain 
formulas for the exact correlation function of closed string vertex 
operators in Liouville theory one might try to evaluate their path 
integral, 
$$ \langle e^{2\a_1 X(z_1,\bz_1)} \cdots e^{2 \a_n X(z_n,\bz_n)} 
   \rangle \ = \ 
   \int {\cal D} X \, e^{-S_L[X]} \, \prod_{r=1}^n \, 
     e^{2\a_r X(z_r,\bz_r)} \ \ . 
$$   
The first step of the evaluation is to split the field $X$ into 
its constant zero mode $x_0$ and a fluctuation field $\tilde X$ 
around the zero mode. Once this split is performed, we calculate 
the integral over the zero mode $x_0$ using a simple integral 
formula for the $\Gamma$ function 
$$ \Gamma(x) \  =\ \int_{-\infty}^\infty dt \, 
      \exp (xt - e^t) \ \ . $$ 
After executing these steps, we obtain the following expression 
for the correlator of the tachyon vertex operators, 
\beqa \langle \prod_{r=1}^n \, 
     e^{2\a_r X(z_r,\bz_r)} \rangle & = & 
     \int {\cal D} \tilde X \, e^{-S_{LD}[\tilde X]} \, 
    \prod_{r=1}^n \, e^{2\a_r X(z_r,\bz_r)}
   \frac{\Gamma(-s)}{2b} \, \left( \mu \int d^2\sigma 
   \sqrt \c e^{2b \tilde X} \right)^s \\[2mm] 
\mbox{where } & & s \ = \ Q \, \frac{1}{b} - \sum_{r=1}^n 
   \, \frac{\a_j}{b}\ \ .         \label{sal}
\eeqa 
There are quite a few observations we can make about this 
result. To begin with, we see that the expression on the left 
hand side can only be evaluated when $s$ is a non-negative 
integer since it determines the number of insertions of the
interaction term. Along with eq.\ (\ref{sal}), such a condition 
on $s$ provides a strong constraint for the parameters $\a_r$ 
that will not be satisfied for generic choices of the momentum 
labels $\a_r$. But if $s$ is integer then we can evaluate the 
integral rather easily because the action for the fluctuation 
field $\tilde X$ is the action of the linear dilaton, i.e.\ of 
a free field theory. The resulting integrals over the position 
of the field insertions are not that easy to compute, but they 
have been solved many years ago and are known to be expressible 
through $\Gamma$ functions (see \cite{Dotsenko:1984ad} and 
Appendix A). 
\smallskip

Let us furthermore observe that in the case of non-negative 
integer $s$, the coefficient $\Gamma(-s)$ diverges. Hence, the 
correlators whose computation we sketched in the previous paragraph 
all contain a divergent factor. A more mathematical interpretation 
of this observation is not hard to find. Note that even our 
semi-classical couplings (\ref{sc3pt}) possess poles e.g.\ at points 
where $2 i\tw$ is a negative integer. The exact couplings are 
not expected to behave any better, i.e.\ the correlation functions 
of Liouville theory have poles at points in momentum space at which 
the expression $s$ becomes a non-negative integer. Only the residue 
at these poles can be computed directly through the free field 
computation above. The presence of singularities in the correlators 
admits a simple explanation. In fact, in Liouville theory, the 
interaction is switched off as we send $x \rightarrow - \infty$ 
so that the theory becomes free in this region. The infinities 
that we see in the correlators are associated with the fact that 
the region of small interaction is infinite, leading to 
contributions which diverge with the volume of the space. 
Finally, with singularities arising from the weekly coupled 
region, it is no longer surprising that the coefficients of 
the residues can be computed through free field computations.     
\smallskip 

Even though our attempt to compute the path integral has
lead to some valuable insights, it certainly falls short
of providing an exact solution of the theory. After all
we want to find the 3-point couplings for all triples 
$\a_1,\a_2,\a_3$, not just for a subset thereof. To proceed 
further we need to understand a few other features of 
Liouville theory. 

\subsection{Degenerate fields and Teschner's trick}

\paragraph{Equations of motion and degenerate fields.}  
Let us first study in more detail the equations of
motion for the Liouville field. In the classical theory, 
these can be derived easily from the action, 
\be \label{cEOM} 
\pl \bpl X_c \ = \ \mu b \exp 2b X_c \ \ . \ee 
There exits an interesting way to rewrite this equation. 
As a preparation, we compute the second derivatives 
$\partial^2$ and $\bar \partial^2$ of the classical field 
$\Phi^c_{-b/2}$, e.g.\ 
\be \label{cEOM'} 
\pl^2 e^{ -b X_c(z,\bz)} \ = \ b^2 T_c(z) 
        \, e^{-b X_c(z,\bz)} \ \ .
\ee  
The fields $T_c$ and $\bT_c$ that appear on the right hand side 
are the classical analogue of the Virasoro fields. It is now 
easy to verify that the classical equation of motion (\ref{cEOM}) 
is equivalent to the (anti-)holomorphicity of $T_c (\bT_c)$. 
\smallskip
 
In its new form, it appears straightforward to come up with 
a generalization of the equations of motion to the full  
quantum theory. The proposal is to demand that
\footnote{Normal ordering instructs us to move annihilation 
modes $L_n, n > 0,$ of the Virasoro field to the right.}    
\be \label{qEOM} 
\pl^2 \Phi_{-b/2} \ = \ - b^2 : T(z) \Phi_{-b/2} :   
\ \ . 
  \ee
If we evaluate this equation at the point $z=0$ and 
apply it naively to the vacuum state, we find that 
$$ |\Psi_+\rangle \ := \ \left(L^2_{-1} + b^2 L_{-2} 
    \right) |-\frac{b}{2}\rangle \ = \ 0 \ \ . $$ 
In order to understand this equation better, let us note that 
$|\Psi_+\rangle$ is a so-called {\em singular vector} in the 
sector $\cV_{-b/2}$ of the Virasoro algebra, i.e.\ a vector that
is annihilated by all modes $L_n, n> 0,$ of the Virasoro algebra 
(but is not the ground state of the sector). Such singular vectors 
can be set to zero consistently and our arguments above suggest 
that this is what happens in Liouville theory. In other words, 
decoupling the singular vector of the sector $\cV_{-b/2}$ 
implements the quantum equation of motion of Liouville theory. 
It is worthwhile pointing out that the origin for the decoupling 
of singular vectors in Liouville theory is different from rational 
conformal field theories, where it is a consequence of unitarity. 
Here, the singular vector is not among the normalizable states of 
the model and hence it  does not belong to the physical spectrum 
anyway. In fact, the only way to reach the point $\a = - b/2$ 
from the set $Q/2 + i\QR$ is through analytic continuation. The 
presence of such unphysical singular vectors would not be in 
conflict with unitarity.  
\smallskip 

Once we have accepted the decoupling of $|\Psi_+\rangle$, 
we observe that there exists another sector of the Virasoro
algebra that has a singular vector on the second 
level. This is the sector $\cV_{-b^{-1}/2}$ and its 
singular vector has the form 
$$ |\Psi_-\rangle \ := \ \left(L^2_{-1} + b^{-2} L_{-2} 
    \right) |-\frac{1}{2b}\rangle \ \ .   $$ 
Note that this second singular vector is obtained 
from the first through the substitution $b \rightarrow
b^{-1}$. Earlier on, we noted that e.g.\ the relation 
between $Q$ and $b$ received quantum corrections which 
rendered it invariant under a replacement of $b$ 
by its inverse. It is therefore tempting to conjecture 
that in the exact Liouville theory, the second singular 
vector $|\Psi_-\rangle$ decouples as well. If we 
accept this proposal, we end up with two different 
fields in the theory that both satisfy a second order 
differential equation of the form (\ref{qEOM}), namely 
the field $\Phi_{-b/2}$ and the dual field $\Phi_{-1/2b}$.
Such fields are called {\em degenerate fields} and we  
shall see in a moment that their presence has very 
important consequences for the structure constants 
of Liouville theory.  
\smallskip 

Let us point out that a relation of the form (\ref{qEOM})
imposes strong constraints on the operator product expansion 
of the fields $\Phi_{-b^{\pm 1}/2}$ with any other field 
in the theory. Since momentum is not conserved in our 
background, the expansion of two generic bulk fields  
contains a continuum of bulk fields. But if we replace 
one of the two fields on the left hand side by one of 
our degenerate fields $\Phi_{-b^{\pm 1}/2}$, then the 
whole operator product expansion must satisfy a second 
order differential equation and hence only two terms 
can possibly arise on the left hand side, e.g.\ 
\begin{equation} \label{degOPE} 
 \Phi_\a(w,\bw) \, \Phi_{-b/2}(z,\bz)  \ = \ 
   \sum_{\pm} \ \frac{c^{\pm}_b(\a)}{|z-w|^{h_\pm}} 
   \ \Phi_{\a \mp b/2}(z,\bz) \ + \ \dots 
\end{equation} 
where $h_\pm = \mp b \a + Q (-b/2 \mp b/2)$. A similar 
expansion for the second degenerate field is obtained 
through our usual replacement $b \rightarrow b^{-1}$. 
We can even be more specific about the operator expansions 
of degenerate fields and compute the coefficients $c^\pm$. 
Note that the labels $\a_1 = -b/2$ and $\a_2 = \a$ of the 
fields on the right hand side along with the conjugate 
labels $\a_3 = (\a \mp b/2)^* = Q-\a \mp b/2$ of the 
fields that appear in the operator product obey $s_+ = 1$ 
and $s_- = 0$ (the quantity $s$ was defined in eq.\ 
(\ref{sal})). Hence, $c^\pm$ can be determined through 
a free field computation in the linear dilaton background. 
With the help of our above formulas we find that $c^+ = 1$ 
and  
\begin{eqnarray} 
 c^-_b(\a) & = & - \mu \int d^2z \, \langle \, \Phi_{-b/2}(0,0) 
 \, \Phi_{\a}(1,1)\,  \Phi_{b}(z,\bz)\, \Phi_{Q-b/2-\a}
       (\infty,\infty) \, \rangle_{{\rm LD}} \nn \\[2mm] 
& = & - \mu \pi \ \frac{\c(1+b^2) \, \c(1-2b\a)}{\c(2+b^2 - 2b\a)} 
\ \    \label{cm} 
\end{eqnarray} 
where we have introduced $\c(x) = \Gamma(x)/\Gamma(1-x)$. The 
factor $(-1)$ in front of the integral is the contribution from 
the residue of factor $\Gamma(-s)$ at $s=1$. Our result in the 
second line is obtained using the explicit integral formulas 
that were derived by Dotsenko and Fateev in \cite{Dotsenko:1984ad} 
(see also appendix A).

\paragraph{Crossing symmetry and shift equations.} 
\def\cP{{\cal P}}
\def\ta{\tilde \a} 
Now it is time to combine all our recent insights into 
Liouville theory with the general strategy we have outlined
in the first lecture and to re-address the construction of the 
exact bulk 3-point couplings. Recall that these couplings may 
be obtained as solutions to the crossing symmetry condition 
(\ref{cross}). Unfortunately, in its original form, the latter 
involves four external states with momentum labels $\a_i \in 
Q/2 + i \QR$. Consequently, there is a continuum of closed 
string modes that can be exchanged in the intermediate channels
and hence the crossing symmetry requires solving a rather 
complicated integral equation. To overcome this difficulty, 
Teschner \cite{Teschner:1995yf} suggested a continuation of one 
external label, e.g.\ the label $\a_2$, to one of the values $\a_2 
= - b^{\pm 1}/2$. The corresponding field is then degenerate 
and it possesses an operator product consisting of two terms 
only. {\em Teschner's trick} converts the crossing symmetry 
condition into a much simpler algebraic condition. Moreover, 
since we have already computed the coefficients of operator 
products with degenerate fields, the crossing symmetry equation 
is in fact linear in the unknown generic 3-point couplings. One 
component of these conditions for the degenerate field 
$\Phi_{-b/2}$ reads as follows      
\begin{eqnarray} \label{spcross}
 0 & = & C(\a_1 + \frac{b}{2},\a_3,\a_4)\, c^-_b (\a_1) \cP^{--}_{+-}
    + C(\a_1 - \frac{b}{2},\a_3,\a_4) \, c^+_b (\a_1) \cP^{++}_{+-} 
   \ \ , \\[2mm] \nonumber 
\mbox{where} & & \ \ \cP^{\pm\pm}_{+-} \ = \ 
   \Fus{\a_1\mp b/2,}{\a_3 - b/2}{- b/2\, }{\a_3}{\ \ \a_1\ }{\a_4} \ 
  \Fus{\a_1\mp b/2,}{\a_3 + b/2}{-b/2\, }{\a_3}{\ \ \a_1\ }{\a_4}\ \ .  
\end{eqnarray} 
Note that the combination on the right hand side must vanish 
because in a consistent model, the off-diagonal bulk mode 
$(\a_4-b/2,\a_4+b/2)$ does not exist and hence it cannot 
propagate in the intermediate channel. The required special entries 
of the Fusing matrix were computed in \cite{Teschner:1995yf} and 
they can be expressed through a combination of $\Gamma$ functions 
(see Appendix B for explicit formulas). Once the expressions for 
$c^\pm$ and $\cP$ are inserted (note that they only involve $\Gamma$ 
functions), the crossing symmetry condition may be written as follows, 
\be 
\label{3ptse} 
\frac{C(\a_1 + b,\a_2,\a_3)}{C(\a_1,\a_2,\a_3)} \ = \ - 
  \frac{\c(b(2\a_1 +b))\c(2b\a_1)}{\pi \mu \c(1+b^2) 
  \c(b(2\ta-Q))}   
    \frac{\c(b(2\ta_1-b)))}{\c(2b\ta_2) \c(2b\ta_3)} 
\ee
with $\c(x) = \Gamma(x)/ \Gamma(1-x)$, as before. The constraint 
takes the form of a shift equation that 
describes how the coupling changes if one 
of its arguments is shifted by $b$. Clearly, this one 
equation alone cannot fix the coupling $C$. But now we recall 
that we have a second degenerate field in the theory which is 
related to the first degenerate field by the substitution $b 
\rightarrow b^{-1}$. This second degenerate field provides 
another shift equation that encodes how the 3-point couplings 
behaves under shifts by $b^{-1}$. The equation is simply 
obtained by performing the substitution $ b \rightarrow b^{-1}$ 
in equation (\ref{3ptse}). For irrational values of $b$, the two 
shift equations determine the couplings completely, at least if 
we require that they are analytic in the momenta. Once we have 
found the unique analytic solution for irrational $b$, we shall
see that it is also analytic in the parameter $b$. Hence, the 
solution may be extended to all real values of the parameter 
$b$. 

\subsection{The exact (DOZZ) solution}    
\def\t{\tau} 
\def\Y{\Upsilon} 
It now remains to solve the shift equation (\ref{3ptse}). For this 
purpose it is useful to introduce Barnes' double $\Gamma$-function 
$\Gamma_b(y)$. It may be defined through  the following integral 
representation, 
\begin{equation}\label{BGamma}
 \ln \Gamma_b(y) \ = \ \int_0^\infty \frac{d\t}{\t}
  \left[ \frac{e^{-y\t} - e^{- Q \t/2}} 
              {(1-e^{-b \t}) (1- e^{-\t/b})} 
         - \frac{\left(\frac{Q}{2} - y\right)^2}{2} e^{-\t}
           - \frac{\frac{Q}{2} -y }{\t}\right]  
\end{equation}
for all $b \in \QR$. The integral exists when $0 < {\rm Re}(y)$ and 
it defines an analytic function which may be extended onto the 
entire complex $y$-plane. Under shifts by $b^{\pm 1}$, the function 
$\Gamma_b$ behaves according to 
\begin{equation}\label{BGshift}
\Gamma_b(y+b) \ = \  \sqrt{2\pi} \, \frac{b^{by-\frac12}}{\Gamma(by)}\, 
                      \Gamma_b(y) \ \ , \ \ 
\Gamma_b(y+b^{-1}) \ = \ \sqrt{2\pi} \, \frac{b^{-\frac{y}{b}+\frac12}}
      {\Gamma(b^{-1}y)}\, \Gamma_b(y) \ \ . 
\end{equation}
These shift equations let $\Gamma_b$ appear as an interesting 
generalization of the usual $\Gamma$ function which may also be
characterized through its behavior under shifts of the argument. 
But in contrast to the ordinary $\Gamma$ function, Barnes' double 
$\Gamma$ function satisfies two such equations which are independent 
if $b$ is not rational. We furthermore deduce from eqs.\ 
(\ref{BGshift}) that $\Gamma_b$ has poles at 
\begin{equation} \label{poles} 
 y_{n,m} \ = \ - n b - m b^{-1} \ \ \ \mbox{ for } \ \ \ 
   n,m \ = \ 0,1,2, \dots \ \ . 
\end{equation}  
From Branes' double Gamma function one may construct the 
basic building block of our exact solution,  
\begin{equation} \label{ZY} 
 \Y_b (\a) \ := \ \Gamma_2(\a|b,b^{-1})^{-1}\, 
                    \Gamma_2(Q-\a|b,b^{-1})^{-1} \ \ . 
\end{equation}  
The properties of the double $\Gamma$-function imply that $\Y$ 
possesses the following integral representation
\begin{equation}\label{ZYpsilon} 
\ln \Y_b(y) \ = \ \int_0^\infty \frac{dt}{t} 
   \left[ \left(\frac{Q}{2}-y\right)^2
     e^{-t} - \frac{\sinh^2\left(\frac{Q}{2} -y\right) \frac{t}{2}}
        {\sinh\frac{bt}{2}\, \sinh \frac{t}{2b}}\right]\ \ .  
\end{equation}
Moreover, we deduce from the two shift properties (\ref{BGshift}) 
of the double $\Gamma$-function that 
\begin{equation}\label{ZYshift} 
   \Y_b(y+b) \ = \ \gamma(by) \, b^{1-2by}\, \Y_b(y) \ \ , \ \ 
   \Y_b(y+b^{-1}) \ = \  \gamma(b^{-1}y) \, b^{-1+2b^{-1}y}\, \Y_b(y)
\ \ . 
\end{equation}
Note that the second equation can be obtained from the first with the help 
of the self-duality property $\Y_b(y) = \Y_{b^{-1}}(y)$. 
\smallskip 

Now we are prepared to solve our shift equations (\ref{3ptse}). In 
fact, it is easy to see that their solution is provided by the 
following combination of $\Y$ functions \cite{Dorn:1994xn,
Zamolodchikov:1996aa},   
\begin{equation}\label{Cc>} 
C(\a_1,\a_2,\a_3) \ := \ \left[\pi \mu \gamma(b^2) b^{2-2b^2}
\right]^{(Q-2 \ta)/b} \ \frac{\Y'(0)}{\Y(2\ta-Q)} \ \prod_{j=1}^3 
  \, \frac{\Y(2\a_j)}{\Y(2\ta_j)}
\end{equation}  
where $\ta$ and $\ta_j$ are the linear combinations of $\a_j$ which are
introduced just as in eqs.\ (\ref{tw}) of the previous subsection. The 
solution (\ref{Cc>}) was first proposed several years ago by H.\ Dorn 
and H.J.\ Otto \cite{Dorn:1994xn} and by A.\ and Al.\ Zamolodchikov 
\cite{Zamolodchikov:1996aa}, based on extensive earlier work by many 
authors (see e.g.\ the reviews \cite{Seiberg:1990eb,Teschner:2001rv} 
for references). The derivation we presented here has been proposed 
by Teschner in \cite{Teschner:1995yf}. Full crossing symmetry of the 
conjectured 3-point function (not just for the special case that involves 
one degenerate field) was then checked analytically in two steps by Ponsot 
and Teschner \cite{Ponsot:1999uf} and by Teschner \cite{Teschner:2001rv,
Teschner:2003en}.
\smallskip 

In it quite instructive to see how the minisuperspace result (\ref{sc3pt})
can be recovered from the exact answer. To this end, we chose the parameters
$\a_i$ to be of the form 
$$ \a_1 \ =  \ \frac{Q}{2} + i b \w_1 \ \ \ \ , \ \ \ \ 
   \a_2 \ = \ b\w_2  \ \ \ \ , \ \ \ \ 
   \a_3 \ = \ \frac{Q}{2} + i b \w_3 \ \ 
$$
and perform the limit $b \rar 0$ with the help of the following formula for 
the asymptotics of the function $\Y$ (see e.g.\ \cite{Thorn:2002am}) 
\begin{equation}\label{YtoG}
  \Y_b(b y) \ \sim \ b \Y'(0) \, b^{-b^2y^2+(b^2-1)y}\, \Gamma^{-1}(y) 
+ \dots \ \ .
\end{equation}  
Recall that Barnes' double Gamma function possesses a double series 
(\ref{poles}) of poles. In our limit, most of these poles move out 
to infinity and we are just left with the poles of the ordinary 
$\Gamma$ function. Not only does this observation explain formula 
(\ref{YtoG}), it also makes $\Y$ appear as the most natural 
replacement of the $\Gamma$ functions in eq. (\ref{sc3pt}) that 
is consistent with the quantum symmetry $ b \leftrightarrow 1/b$ 
of Liouville theory.%
\smallskip%

Let us furthermore stress that the expression (\ref{Cc>}) can be analytically 
continued into the entire complex $\a$-plane. Even though the corresponding 
fields $\Phi_\a$ with $\a \not \in Q/2 + i \QR$ do not correspond to 
normalizable states of the model, they may be considered as well defined 
but non-normalizable fields. It is tempting to identify the identity field 
with the limit $\lim_{\a \rar 0} \Phi_{\a}$. This identification can indeed 
be confirmed by computing the corresponding limit of the coefficients $C$ 
which is given by 
\begin{eqnarray} \nonumber   
\lim_{\a_2 \rar 0} C(\a_1,\a_2,\a_3) & = & 2\pi \, \delta(\a_1 +\a_3 - Q) 
  \, + \, R(\a_1) \, \delta(\a_1-\a_3)  \\[2mm] \mbox{where} \ \ \ \ \ \ 
 R(\a) & = &  \left(\pi \mu \c(b^2)\right)^{(Q-2\a)/b}\ 
  \frac{b^{-2} \c(2b\a - b^2)}{\c(2-2b^{-1}\a + b^{-2})}\ \  
\label{RL} 
\end{eqnarray}
for all $\a_{1,3} = Q/2 + i p_{1,3}$ with $p_i \in \QR$. The 
$\delta$-functions again emerge from the singularities of $C$, just as 
in the minisuperspace example. We would like to point out that the 
reflection amplitude $R(\a)$ may also be obtained directly from the 
3-point coupling without ever performing a limit $\a_2 \rar 0$. In fact, 
in Liouville theory the labels $\a = Q/2 +iP$ and $Q-\a=Q/2-iP$ do not 
correspond to two independent fields. This is intuitively obvious because 
there exits only one asymptotic infinity so that wave-functions are 
parametrized by the half-line $P \geq 0$. Correspondingly, the 
3-point couplings $C$ possess the following simple reflection 
property, 
$$ C(\a_1,\a_2,\a_3) \ = \ R(\a_1) \ C(Q-\a_1,\a_2,\a_3) \ \  $$
with the same function $R$ that we found in the 2-point function.  
Similar relations hold for reflections in the other two momentum 
labels. This observation implies that the fields $\Phi_\a$ and 
$\Phi_{Q-\a}$ itself can be identified up to a multiplication 
with the reflection amplitude.%
\smallskip%

As a preparation for later discussions we would finally like to point out 
that the expression (\ref{RL}) for our exact reflection amplitude may 
be rewritten in terms of its semi-classical analogue (\ref{R0L}), 
\begin{equation} \label{RR0}  
 R(Q/2 + i b \w) \ = \ R_0(\w) \  \frac{\Gamma(1+2b^2 i \w)}
                  {\Gamma(1-2b^2 i \w)} \left(\frac{\pi\c(b^2)}{b^2}
                   \right)^{-2i\w} \ \ . 
\end{equation}
Here, we have also inserted the relation $\pi \mu b^2 = \lambda$ between the 
coupling constant $\lambda$ of the minisuperspace theory and our parameter 
$\mu$. The formula shows how finite $b$ corrections to the semi-classical 
result are simply encoded in a new multiplicative factor. 
\newpage

\section{Branes in the Liouville model}  
\setcounter{equation}{0}

In this lecture we will study branes in Liouville theory. It turns out 
that there exist two different types of branes. The first consists of
branes that are localized in the strong coupling region and possess a 
discrete open string spectrum. The other class of branes is 1-dimensional
and it extends all the way to $x = - \infty$. These extended branes 
possess a continuous spectrum of open strings.

\subsection{Localized (ZZ) branes in Liouville theory.} 

\paragraph{The 1-point coupling.} 
As we have explained in the first lecture, branes are uniquely 
characterized by the 1-point couplings $A_\a$ of the bulk vertex 
operators $\Phi_\a$. These couplings are strongly constrained by  
the cluster condition (\ref{class}). Our aim therefore is to come up  
with solutions to this factorization constraint for the specific 
choice of the coefficients $\Xi$ that is determined by the 3-point 
coupling (\ref{Cc>}) of the Liouville bulk theory. In the 
most direct approach we would replace the labels $(i,\bi)$ and 
$(j,\bj)$ of the bulk fields by the parameters $\a$ and $\b$, 
respectively, and then take the latter from the set $Q/2 + iP$ 
that labels the normalizable states of the bulk model. But as we 
discussed above, this choice would leave us with a complicated 
integral equation in which we integrate over all possible closed 
string momenta $\gamma \in Q/2 + i \QR$. 
\smallskip

To avoid such an integral equation, we shall follow the same 
approach (`Teschner's trick') that we applied so successfully 
when we determined the 3-point couplings in the bulk model: 
we shall evaluate the cluster property in two cases in 
which one of the bulk fields is degenerate, i.e.\ for 
$\Phi_\b = \Phi_{-b/2}$ and $\Phi_\b = \Phi_{-1/2b}$. An 
argument similar to the one explained in the last lecture 
then gives the constraint, 
\begin{equation} \label{spclus} 
   A(-\frac{b}{2}) \, A(\a) \ = \ A(\a-\frac{b}{2}) \, c^+_b(\a) \, 
\Fus{\a-\frac{b}{2},}{0}{-\frac{b}{2}}{-\frac{b}{2}\, }{\ \, \a }{\ \, \a }  
 + A(\a + \frac{b}{2}) \, c^-_b (\a)\, 
\Fus{\a+\frac{b}{2},}{0}{-\frac{b}{2}}{-\frac{b}{2}\, }{\ \, \a }{\ \, \a }  
  \end{equation} 
and a second equation of the same type with $b \rightarrow b^{-1}$. 
The functions $c_b^\pm(\a)$ are the same that appeared in eq.\ 
(\ref{degOPE}) and we have computed them before (see eq.\ (\ref{cm})). 
Furthermore, the special elements of the Fusing matrix that appear in 
this version of the cluster condition are also easy to calculate (see 
Appendix B for explicit formulas). Note that all these quantities can be 
expressed in terms of ordinary $\Gamma$ functions. Once they are 
inserted, the condition (\ref{spclus}) becomes   
\begin{eqnarray} \label{clusterZZ} 
& & \frac{\Gamma(-b^2)}{\Gamma(-1-2b^2)} \, A(-\frac{b}{2}) 
   \,  A (\a) \\[2mm]  
   & & \hspace*{2cm} = \  \frac{\Gamma(2\a b - b^2)}{\Gamma(2\a b-
           2b^2-1)} \, A(\a-\frac{b}{2}) \, - \, 
  \frac{\pi \mu }{\c(-b^2)} \, \frac{\Gamma(2\a b - b^2 -1)}
        {\Gamma(2\a b)}\,  
     A(\a+\frac{b}{2}) \ \  \nn  
\end{eqnarray} 
and there is a dual equation with $b \rightarrow b^{-1}$. Solutions 
to these equations were found in \cite{Zamolodchikov:2001ah}. They 
are parametrized through two integers $n,m= 1,2,\dots$ and read as 
follows\footnote{In writing the cluster condition (\ref{clusterZZ}) 
we have assumed the structure constants $A$ to be normalized such 
that $A(0) = 1$. The normalization of the coefficients $A_{(n,m)}
(\a)$ is chosen differently to ensure consistency with the modular 
bootstrap below. One can show easily that the rations $A(\a) = 
A_{(n,m)}(\a)/A_{(n,m)}(0)$ satisfy relation (\ref{clusterZZ}).}  
\begin{eqnarray}
 A_{(n,m)}(Q/2+iP) &  =  & \frac{\sin \pi b Q \, \sin 2\pi n P b} 
   {\sin \pi b n Q\,  \sin 2 \pi P b}\  \frac{\sin \pi b^{-1} Q 
   \, \sin 2\pi m P b^{-1}} 
   {\sin \pi b^{-1} m Q \, \sin 2 \pi P b^{-1}} \ A_{(1,1)}(Q/2+iP) 
\nonumber\\[2mm]    \label{ZZb}
 & & \hspace*{-3cm} A_{(1,1)}(Q/2+iP) \ = \ \left(\pi \mu 
   \gamma(b^2)\right)^{-\frac{iP}{b}}\, 
   \frac{2^{1/4} 4 i \pi P}{\Gamma(1-2iPb^{-1})\Gamma(1-2iPb)} \ \ .      
\end{eqnarray}
According to our general discussion in the first lecture, we have 
thereby constructed branes in the Liouville model, though later we 
shall argue that only one of them, namely the one with $(n,m) = 
(1,1)$ is actually `physical'. 
\smallskip 

The general algebraic procedure we used to determine the closed 
string couplings $A$ to these branes remains so abstract that it is 
rather reassuring to discover that at least the branes with label 
$(n,1)$ possess a semi-classical limit and therefore some nice 
geometric interpretation. In fact, it is not difficult to see that 
these branes are localized in the strong coupling region of the model. 
We check this assertion by sending the parameter $b$ to zero after 
rescaling the momentum $P = b \omega$, 
\begin{equation} \label{ZZsc} 
  A_{(n,1)} \  \stackrel{b \rar 0}{\sim} \      
   \lambda^{-i \omega} \Gamma^{-1} (-2 i \omega) \, 
  \ = \ \lim_{x_0 \rightarrow \infty} N(x_0) \ \phi_\omega (x_0) \ \ .  
\end{equation} 
Here, the function $\phi_\w$ on the right hand side is the 
minisuperspace wave-function (\ref{solin}) of a particle that moves in 
an exponential potential and $N(x_0)$ is an appropriate normalization 
that is independent of the momentum $\omega$ and that can be read off  
from the asymptotics of the Bessel function $K_\nu(y)$ at large $y$. 
Note that the first term of such an expansion does not depend on the 
index $\nu$. 
\smallskip

We see from eq.\ (\ref{ZZsc}) that the semi-classical coupling
is entirely determined by the value of the semi-classical wave
function at one point $x_0$ in the strong coupling limit $x_0 
\rightarrow \infty$ of the theory. This means that the branes 
with labels $(n,1)$ are point-like localized. It is not so 
surprising that point-like branes prefer to sit in the string 
coupling region of the model. Recall that the mass of a brane 
is proportional to the inverse of the string coupling. Since 
we are dealing with a linear dilaton background in which the 
string coupling grows exponentially from left to right, branes
will tend to reduce their energy my moving into the region 
where the string coupling is largest, i.e.\ to the very far 
right. Let us also remark that the semi-classical coupling 
of closed strings is independent of the parameter $n$. This 
implies that we cannot interprete the $(n,1)$ branes as an 
n-fold super-position of $(1,1)$ branes. The absence of a 
good geometrical interpretation for the parameter $n$ might 
seem a bit disturbing, but it is certainly not yet sufficient 
to discard solutions with $n\neq 1$ from the list of branes
in Liouville theory.  
\smallskip 
 
As in our brief discussion of the reflection amplitude 
(\ref{RR0}),  we would like to split our expression 
for the coupling $A_{(1,1)}$ into a semi-classical factor 
$A_{(1,1)}^0$ (see eq.\ (\ref{ZZsc})) and its stringy 
correction, 
\begin{equation} 
A_{(1,1)}(Q/2+ib\w) \ = \ {\cal N}_{ZZ}(b) \, 
   \ A_{(1,1)}^0(\omega) \ \frac{1}{\Gamma(1-2i b^2\omega)}  \ 
   \left(\frac{\pi\c(b^2)}{b^2} \right)^{-i\w}\ \ . 
\end{equation} 
Here, $A_{(1,1)}^0(\w) = \lambda^{-i \omega} \Gamma^{-1} 
(-2 i \omega)$, the normalization factor ${\cal N}$ is 
independent of the momentum $\omega$ and of $\lambda = \pi b^2 \mu$.

\paragraph{The open string spectrum.} 
We emphasized before that the closed string couplings $A$ 
contain all the information about the corresponding branes. 
In particular, it should now be possible to determine the 
spectrum of open string modes that live on the ZZ branes. 
We shall achieve this in the same way that we sketched in 
the first lecture, using world-sheet duality (modular 
bootstrap). In view of our previous remarks and in order 
to simplify our task a bit, let us restrict to the case 
in which the boundary conditions on both sides of the 
cylinder are taken to be $\sigma = (1,1) = \rho$, i.e.\ 
to open string excitations on a single $(1,1)$-brane.   
On the closed string side, the amplitude reads
\beqa  Z_{(1,1)} (q) & = & \int_{0}^\infty dP 
    \, \chi_P (\tilde q) \, \sinh 2 \pi P b 
    \sinh 2\pi P b^{-1}  \\[2mm] 
\mbox{where} \ \ \ \ \ \chi_P(q) & = & \eta^{-1}(q) \, q^{P^2}\ \ . 
\eeqa  
In order to calculate the right hand side we had to 
evaluate the product $A_{(1,1)}(P) A_{(1,1)}^\ast(P)$ 
with the help of the standard relation $\Gamma(x) \Gamma(1-x) 
= \pi /\sin \pi x$. Now we can rewrite the partition sum 
using a simple trigonometric identity, 
$$  
Z_{(1,1)} (q) \ = \ \int_{0}^\infty dP 
    \, \chi_P (\tilde q) \, \left( \cosh 2 \pi P Q  - 
    \cosh 2\pi P (b-b^{-1})\right)   \ \ 
$$
and then insert the usual formula for the modular transformation 
of the characters $\chi_P$, 
$$ \chi_P(q) = 2^{3/2} \int_{-\infty}^\infty dP' \cosh 4 \pi P P' 
                \chi_{P'}(\tilde q)\ \ . $$  
The final result of this short computation is that 
$$ Z_{(1,1)}(q)  \ = \ \eta^{-1}(q)  
            \left( q^{-Q^2/4 }- q^{-Q^2/4 +1} \right) 
    \ = \ q^{-\frac{c}{24}} \, \left( 
            1 + q^2 + q^3 + \dots \right) \ \  . 
$$ 
There are a few comments one can make about this answer. To 
begin with, the spectrum is clearly discrete, i.e.\ it does 
not involve any continuous open string momentum in target space. 
This is certainly consistent with our interpretation of the 
$(1,1)$-brane as a localized object in the strong coupling 
region. We can appreciate another important feature of our 
result by contrasting it with the corresponding expression 
for point-like branes in a flat 1-dimensional background,  
$$ Z^{\rm free}_{D0}(q) \ = \ q^{-1/24} \left( 1 + q + 2 q^2 + 3 q^3 
   \dots \right) $$
Here, the first term in brackets signals the presence of an 
open string tachyon on the point-like brane while the second 
term corresponds to the massless scalar field. All higher terms 
are associated with massive modes in the spectrum of brane
excitations. The presence of a scalar field on the point-like 
brane is directly linked to the modulus that describes the 
transverse position of such a brane in flat space. Note now 
that the corresponding term is missing in the spectrum on 
the ZZ brane. We conclude that our ZZ brane does
not possess moduli, i.e.\ that it is pinned down at $x_0 
\rightarrow \infty$, in perfect agreement with our 
geometric arguments above.  

\paragraph{Application to 2D string theory.} 
Let us conclude this subsection on the ZZ branes with a few 
remarks on some recent applications. As we pointed out earlier, 
Liouville theory is a building block for the 2D string theory. 
The latter is dual to the following model of matrix quantum 
mechanics, 
$$ S_{\rm MQM} \ \sim\ - \beta\int dt \left[ 
     \frac12 (\partial_t M(t))^2 + V(M(t))\right] 
$$ 
where $M(t)$ are hermitian $N\times N$ matrices and $V$ 
is a cubic potential. To be more precise, the duality 
involves taking $N$ and $\beta$ to infinity while keeping 
their ratio $\kappa = N/\beta$ fixed and close to some 
critical value $\kappa_c$. In this double scaling
limit, the matrix model can be mapped to a system of 
non-interacting fermions moving through an inverse 
oscillator potential, one side of which has been 
filled up to a Fermi level at $\Delta \kappa = \kappa - 
\kappa_c \sim g_s^{-1}$. With a quick glance at Figure 6, 
we conclude that the model must be non-perturbatively 
unstable against tunneling of Fermions from the left to 
the right. This instability is reflected in the asymptotic 
expansion of the partition sum and even quantitative 
predictions for the mass $m \sim a/g_s$ of the instantons 
were obtained. The general dependence of brane masses on 
the string coupling $g_s$ along with the specific form of 
the coupling (\ref{ZZb}) have been used recently to 
identify the instanton of matrix quantum mechanics with 
the localized brane in the Liouville model \cite{McGreevy:2003kb,%
Martinec:2003ka,Klebanov:2003km,Alexandrov:2003nn}. In this 
sense, branes had been seen through investigations of matrix 
quantum mechanics more than ten years ago, i.e.\ long 
before their central role for string theory was fully
appreciated. 
\vspace{1cm} 
\fig{In the double scaling limit, the hermitian matrix model can 
be mapped to a system of non-interacting fermions moving 
through an inverse oscillator potential, one side of which 
has been filled up to a Fermi level at $\Delta \kappa = 
\kappa - \kappa_c \sim g_s^{-1}$.} 
{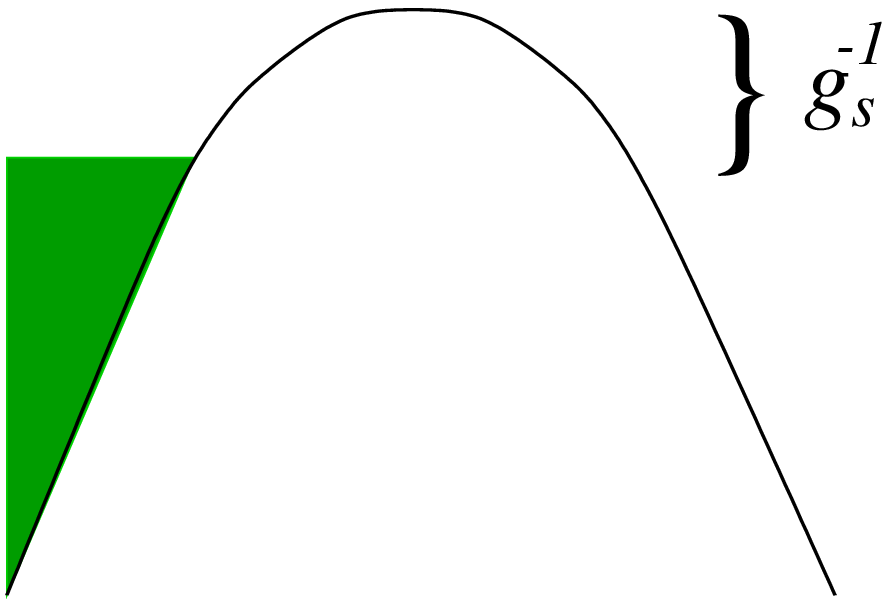}{7truecm} 
\figlabel{\basic}
\vspace{.5cm}

\subsection{Extended (FZZT) branes in Liouville theory}

\paragraph{The 1-point coupling.}
In a flat 1-dimensional space one can impose Dirichlet 
and Neumann boundary conditions on $X$ and thereby describe 
both point-like and extended branes. It is therefore natural to 
expect that Liouville theory admits extended brane solutions
as well. When equipped with an exponential potential for the 
ends of open strings, the action of extended branes should 
take the form 
$$ S_{BL}[X] \ = \ S_L[X] + 
\int_{\pl \Sigma} du \, \mu_B  \, e^{b X(u)} \ \ . 
$$ 
In the case of free field theory, we switch from point-like to 
extended branes by changing the way in which we glue left and 
right moving chiral currents $J$ and $\bJ$. But unlike for flat 
space, the symmetry of Liouville theory is generated by the 
Virasoro field alone and the gluing between left and right 
moving components of the stress-energy tensor is fixed to 
be trivial. Hence, we cannot switch between different brane 
geometries simply by changing a gluing map $\Omega$, as is 
flat space. At this point one may wonder whether our 
previous analysis has been complete and we should conclude 
that there are only localized branes. We will see in a moment, 
however, that this conclusion is incorrect and that at one 
point in our analysis of branes above we have implicitly 
assumed that they possessed a discrete open string spectrum. 
\smallskip 

To understand the issue, let us review our usual derivation 
of the cluster condition. We should always start from the 
cluster condition for two fields within the spectrum of 
normalizable states in the theory, i.e. with $\a,\b$ that 
belong to the set $Q/2 + i \QR$. As we have seen in the 
previous lecture, operator product expansions of such 
fields involve a continuum of modes. This applies not only 
to the modes that emerge when two closed strings collide, 
but also to the absorption of closed strings by a brane. 
More precisely, a closed string mode with label $\b = Q/2 
+ iP_\beta$ can excite a continuum of open string states 
on the brane, provided such a continuum of states exists, 
i.e.\ that the brane is non-compact. In formal terms, the 
process is captured by the following bulk-boundary operator 
product expansion\begin{equation} \label{bBOPE} 
 \Phi_\beta (z,\bar z) \ \stackrel{z \rightarrow x} 
    {\sim} \ 
   \int_{\frac{Q}{2}+i\QR} d\c\  B(\beta,\gamma) \, \Psi_\gamma (x) 
  \ \ . 
\end{equation} 
Here, $\Psi_\c$ with  $\c \in \frac{Q}{2} + i \QR$ are open 
string vertex operators and the operator product coefficients 
$B$ may depend on the choice of boundary condition. 
\smallskip

In evaluating all our factorization constraints, we try to 
avoid any integrals over intermediate closed or open string 
modes. This is why we employ ``Teschner's  trick'' (see 
lecture 2). It instructs us to analytically continue the 
variable $\b$ to the values $\b = - b^{\pm 1}/2$, i.e.\ to 
the labels of degenerate fields. Representation theory of 
the Virasoro algebra then ensures that only two open string vertex 
operators $\Psi_\c$ can occur on the right hand side of the 
operator product expansion (\ref{bBOPE}), namely the terms 
with $\c = 0$ and $\c = - b^{\pm 1}$. Hence, we are tempted 
to conclude that the coefficients $A(-b^{\pm 1}/2)$ that appear 
on the left hand side of the cluster condition are simply given 
by the unknown quantity $B(-b^{\pm 1}/2,0)$. Such a conclusion, 
however, would be a bit too naive. As we shall argue in a moment, 
$B(\beta,0)$ is actually singular at $\beta = - b^{\pm 1}/2$.  
 
In order to see this we have to understand how exactly the 
theory manages to pass from an expansion of the form 
(\ref{bBOPE}) to a discrete bulk-boundary operator 
product expansion at $\beta = - b^{\pm 1}$. In the following 
argument we consider $B(\b,\c)$ as a family of functions in 
the parameter $\c$. If there exist branes that extend all 
all the way to the weak coupling region, the corresponding  
functions $B_\b(\c) = B(\b,\c)$ are expected to possess 
singularities, for the same reasons that e.g.\ the bulk 
3-point function displays poles (see lecture 2). As we 
change the parameter $\b$ to reach 
$\b = - b^{\pm 1}/2$, the position of the poles of $B$ 
will change and some of these poles can actually cross 
the contour of integration in eq.\ (\ref{bBOPE}), thereby 
producing discrete contributions on the right hand side of 
the bulk-boundary operator product expansion. At generic 
points in the $\c$-plane, these discrete parts are accompanied 
by a continuous contribution. But when we reach the degenerate 
fields, the continuous parts must vanish and we remain with the 
two discrete terms that are consistent with the fusion 
of degenerate representations. There is one crucial observation 
we can take out of this discussion: The coefficients in front 
of the discrete fields in the bulk boundary operator product 
of degenerate bulk fields are not given through an evaluation 
of the function $B$ at some special points but rather through 
the residues of $B$ at certain poles. Since we understand the 
origin of such singularities as coming from the infinite region 
with weak interaction, we know that we can compute the coefficients 
using free field calculations. In the case at hand we find that 
\begin{eqnarray} \nonumber 
 {\mbox{\it res}}_{\beta = - b/2} \left(B(\beta,0)\right)
& = & - \mu_B\, \int_{-\infty}^\infty du\ \langle
\, \Phi_{-b/2}({\scriptstyle \frac{i}{2}},-{\scriptstyle \frac{i}{2}})\, 
    \Psi_Q (\infty) \, \Psi_b(u)\, 
\rangle_{\rm LD}
\\[2mm] 
& = & - \frac{2\pi \mu_B \Gamma(-1-2b^2)}{\Gamma^2(-b^2)} 
\label{res}  
\end{eqnarray} 
It is this quantity rather than the value $B(-b/2,0)\sim 
A(-b/2)$ that appears in the cluster condition for extended 
branes. In other words, we obtain the cluster condition 
for extended branes in Liouville theory by replacing 
the quantity $A(-b/2)$ in eq.\ (\ref{clusterZZ}) through 
the result of the computation (\ref{res}),   
\begin{equation} \label{clusterFZZT} 
- \frac{2\pi \mu_B}{\Gamma(-b^2)} 
   \, A (\a)  
   \, =\,  \frac{\Gamma(2\a b - b^2)}{\Gamma(2\a b-
           2b^2-1)} \, A(\a-\frac{b}{2}) \, - \, 
  \frac{\pi \mu }{\c(-b^2)} \, \frac{\Gamma(2\a b - b^2 -1)}
        {\Gamma(2\a b)}\,  
     A(\a+\frac{b}{2})\ . 
\end{equation} 
As usual, there is a dual equation with $b \rightarrow b^{-1}$.
Observe that the cluster conditions for extended branes are linear 
rather than quadratic in the desired couplings. Solutions to 
these equations were found in \cite{Fateev:2000ik}. They are 
given by the following expression 
\begin{eqnarray} \label{AFZZT} 
 A_{s}(Q/2+iP) & = & 2^{1/4} \left(\pi \mu \c(b^2)\right)^{- \frac{iP}{b}} 
             \, \frac{\cos 2\pi s P}{-2\pi i P}   
        \  \Gamma(1+2ib^{-1}P) \Gamma(1+2ibP) \\[2mm] 
\mbox{where} \ & &  \sqrt{\mu} \cosh \pi s b \ = \ \mu_B 
   \  \sqrt{\sin\pi b^2}\ \ . 
\label{musrel} \end{eqnarray}   
Here we parametrize the couplings $A$ through the new parameter 
$s$ instead of the coupling constant $\mu_B$ which appears in 
the boundary term of the action. Both parameters are related 
through the equation (\ref{musrel}). 
\smallskip 
 
Let us again compare this answer to the semi-classical 
expectation. Evaluation of the coupling $A$ in the limit 
$b \rightarrow 0$ gives 
\begin{eqnarray}\label{A0s} 
 A_s (Q/2+ib\omega) & \stackrel{b \rightarrow 0}{\sim} & 
      A^0_s(\omega)  \ :=\  (\lambda/4)^{-i\omega}\, \Gamma(2 i\omega)
              \,   \cos(2\pi \rho \omega )  \\[2mm]   
  & & \ \ \ = \ \int_{-\infty}^{\infty} dx \  e^{-\lambda_B e^x} \,  
                  \phi_\omega(P)\ \ . \nonumber 
\end{eqnarray} 
Before taking the limit, we have rescaled the momentum $P = b 
\omega$, the bulk (boundary) couplings $\pi \mu b^2 = \lambda \  
(\pi \mu_B b^2 = \lambda_B)$ and the boundary parameter $s = b^{-1} 
\rho$ so that the relation (\ref{musrel}) becomes $\lambda 
\cosh^2 \pi \rho = \lambda_B^2$. In the second line we have 
rewritten the answer\footnote{Use e.g.\ formula 6.62 (3) of 
\cite{GradRys} to express the integral in the second line 
through a hypergeometric function. Then apply formula 15.1.19 
of \cite{Abramowitz} to evaluate the latter.} to show that it 
reproduces the semi-classical limit for the coupling of 
closed strings to an extended brane with boundary potential 
$V_B(x) = \lambda_B \exp(x)$. 
\smallskip 

Once more, we can split the coupling to FZZT branes into the
semi-classical part $A^0_s$ and a stringy correction factor,
\begin{equation} \label{FZZTAS} 
A_s(Q/2+ib\w) \ = \ {\cal N}_{FZZT} \, A^0_s(\w) \,  
                    \Gamma(1+2ib^2\w) \left(\frac{\pi\c(b^2)}{b^2} 
                    \right)^{-i\w}\ \ . 
\end{equation}  
As before, we have used that $\lambda = \pi \mu b^2$ and we inserted 
$\rho = b^{-1} s$. The quantity $A^0_s$ has been defined in eq.\ 
(\ref{A0s}) above.   
 \smallskip 

Let us conclude this subsection with a short comment on the 
brane parameter $s$. It was introduced above as a convenient way 
to encode the dependence of $A$ on the boundary coupling 
$\mu_B$. There is, however, much more one can say about 
the reparametrization of FZZT branes in terms of $s$. In 
particular, it is rather tempting to extend $s$ 
beyond the real line and to allow for imaginary values. 
But in the complex plane, each value of $\mu_B$ can 
be represented by infinitely many values of $s$ and one may 
wonder about possible relations between branes whose $s$ 
parameters differ by multiples of $\Delta_s = 2i/b$. With 
the help of our explicit formulas it is not difficult to 
verify e.g.\ that 
\begin{equation} \label{relZZFZZT} 
A_{iQ}(\a) \ = \ A_{iQ-2i/b}(\a) + A_{(1,1)}(\a)\ \ .   
\end{equation}  
This interesting relation between FZZT and ZZ branes was first 
observed in \cite{Martinec:2003ka} and then beautifully 
interpreted in the context of minimal string theory \cite{SeiS}. 
Geometrically, we may picture eq.\ (\ref{relZZFZZT}) as 
follows: Let us begin by considering large real values of 
$s$. These correspond to large values of $\mu_B$ and hence 
to a brane whose density decreases fast toward the strong 
coupling regime, as can be seen from the semi-classical 
limit above. While we lower $s$, mass is moved further 
to the right. This process continues as we move along the 
imaginary axis and reach the value $s = iQ$. At this point, 
part of the brane's mass is sucked into the strong coupling 
regime where it forms a ZZ brane. The shift in the parameter 
$s$ to back to $s=iQ-2i/b$ (we assume $b>1$) may be visualized 
as a retraction of the remaining extended brane.   

\paragraph{The open string spectrum.} 
Our analysis of the open string spectrum on extended branes 
requires a few introductory remarks. For concreteness, let 
us consider a 1-dimen\-sional quantum system with a positive 
potential $V(x)$ which vanishes at $x \rightarrow -\infty$ 
and diverges as we approach $x =\infty$. Such a system has 
a continuous spectrum which is bounded from below by $E=0$ 
and, under some mild assumptions, the set of possible 
eigenvalues does not depend at all on details of the 
potential $V$ (see e.g.\ \cite{Mess}). There is much more 
dynamical information stored in the reflection amplitude 
of the system, i.e.\ in the phase shift $R(p)$ that plane 
waves undergo upon reflection at the potential $V$. $R(p)$ 
is a functional of the potential which is very sensitive to 
small changes of $V$. In fact, it even encodes enough data 
to reconstruct the entire potential.  
\smallskip 

From the reflection amplitude $R(p)$ we can extract a spectral 
density function $\rho$. To this end, let us regularize the 
system by placing a reflecting wall at $x=-L$, with large positive 
$L$. Later we will remove the cutoff $L$, i.e.\ send it to infinity. 
But as long as $L$ is finite, our system has a discrete spectrum 
so that we can count the number of energy or momentum levels in 
each interval of some fixed size and thereby we define a density 
of the spectrum. Its expansion around $L = \infty$ starts with the 
following two terms 
\begin{equation}\label{dos_asym}
\rho^{L}(p) \ =\ \frac{L}{\pi}+\frac{1}{2\pi i}\frac{\pa}{\pa p}
\ln R(p)\ + \ \dots \ \ 
\end{equation} 
where the first one diverges for $L\rar \infty$. This divergence is 
associated with the infinite region of large $x$ in which the whole 
system approximates a free theory and consequently it is universal, 
i.e.\  independent of the potential $V(x)$. The sub-leading term, 
however, is much more interesting. We can extract it from the 
regularized theory e.g.\ by computing relative spectral densities 
before taking the limit $L \rightarrow \infty$. 
\smallskip 

It is not difficult to transfer these observations from quantum 
mechanics to the investigation of non-compact branes. We are 
thereby lead to expect that the annulus amplitude $Z$ diverges 
for open strings stretching between two non-compact branes. This 
divergence, however, must be universal and there should also 
appear an interesting sub-leading contribution which is related 
to the phase shift that arises when open strings are reflected 
by the Liouville potential. All these features of the annulus
amplitude can be confirmed by explicit computation. 
\smallskip

Once more, we try to compute the annulus amplitude from the 
couplings (\ref{AFZZT})  of closed strings to extended branes. 
For real $s$, the result is\footnote{When $s$ is becomes complex, 
there exist issues with the convergence of integrals. These can 
ultimately lead to extra discrete contributions in the open string 
spectrum \cite{Teschner:2000md}. Note that such discrete terms are 
in perfect agreement with the observation (\ref{relZZFZZT}).}  
\begin{eqnarray} 
Z_{ss}(q)  & = & \int_{-\infty}^{\infty} dP' \, 
                  \chi_{P'}(\tilde q) \ \frac{\cos 2\pi s P'}
                 {\sinh 2\pi P' b \sinh 2\pi P' b^{-1}} 
          \ = \  \int_{-\infty}^{\infty} dP \, \rho_{ss}(P) 
                \, \chi_P(q)    \nonumber \\[2mm]      
& &  \mbox{where} \ \ \ \ \   \rho_{ss}(P) \ = \  
             \int_{-\infty}^{\infty} \frac{dt}{2\pi} \, 
         \frac{\cos^2 st}{\sinh tb \sinh tb^{-1}}\ \ .    
\end{eqnarray}  
As we have predicted before, the spectral density $\rho_{ss}(P)$ 
diverges. The divergence arises from the double pole at $t=0$ 
in the integral representation of $\rho_{ss}$. The coefficient 
of the double pole, and hence of the divergent term, does not 
depend on the boundary parameter $s$, i.e.\ it is universal. If 
we consider annulus 
amplitudes relative to some fixed reference brane with 
parameter $s^*$, however, we obtain an interesting finite 
answer and hence, according to our introductory remarks, a
prediction for the reflection amplitude of open strings. The
latter appears in the 2-point function of open string vertex 
operators,  
\begin{equation} \label{intro3} 
\langle \Psi_{\c_1} (u_1) \, \Psi_{\c_2}(u_2) \rangle_s    
\ \sim \ \left( 2 \pi \delta(\c_1 + \c_2 - Q) + 
   \delta(\c_1 - \c_1) \ R(\c_1|s)\right)  \, 
        \frac{1}{|u_1 - u_2|^{2h_{\c_1}}} \ \ . 
\end{equation}
In order to turn our computation of the annulus amplitude 
into an independent test of the couplings (\ref{AFZZT}), 
we are therefore left with the problem of finding an 
expression for the boundary 2-point function, or, 
more generally, the couplings for open strings on 
extended branes in Liouville theory. Formulas for the 
2-point couplings have indeed been found using 
factorization constraints \cite{Fateev:2000ik} and 
they are consistent with the modular bootstrap. 
\smallskip 

Let us briefly mention that even general expressions for 3-point 
couplings of open strings on extended Liouville branes are known 
\cite{Ponsot:2001ng}. The same is true for the exact bulk-boundary 
structure constants $B$ (see eq.\ (\ref{bBOPE})), both for extended 
branes \cite{Hosomichi:2001xc} and for ZZ branes \cite{Ponsot:2003ss}.    
The methods that are used to obtain such additional data are 
essentially the same that we have used several times  
throughout our analysis. The interested reader is referred to 
the original literature (see also \cite{Nakayama:2004vk} for a 
very extensive list of references).  
\smallskip 

While the modular bootstrap on non-compact branes alone does not 
lead to constraints on the 1-point couplings $A_s$, at least not
without further analysis of open string data, one may test our 
formulas (\ref{FZZTAS}) by studying the annulus amplitude for 
open strings that stretch between the discrete and extended 
branes of Liouville theory. We shall simply quote the final result 
of this straightforward computation, 
\begin{equation} \label{Zmns} 
    Z_{(m,n),s}(q) \  = \  {{\sum}'}_{k=1-m}^{m-1}\,  
                     {{\sum}'}_{l=1-n}^{n-1} \ 
                  \chi_{(s+i(k/b+lb))/2}(q) 
\end{equation} 
Here, $\Sigma'$ denotes a summation in steps of two. We observe 
that the spectrum of open strings is discrete, just as one would 
expect for the setup we consider. But whenever $(n,m) \neq (1,1)$, 
we encounter complex exponents of $q$ on the right hand side. 
This is inconsistent with our interpretation of the quantity $Z$ 
as a partition function and therefore it suggests that ZZ branes 
with $(n,m) \neq (1,1)$ are unphysical. Before we accept such a 
conclusion we might ask ourselves why the couplings with $(n,m) 
\neq (1,1)$ did show up when we solved the cluster condition 
(\ref{clusterZZ}). It turns out that there is a good reason. 
Let us observe that the coefficients of our cluster condition 
are analytic in $b$ and there is formally no problem to 
continue these equations to arbitrary complex values of $b$. 
When $b$ becomes purely imaginary, the central charge $c$ 
assumes values $c\leq 1$ which are realized in minimal 
models. For the latter, the existence of a two-parameter set 
of non-trivial and physical discrete branes is well established.  
These solutions had to show up in our analysis simply because 
the constraints we analyzed were analytic in $b$. But while 
such branes are consistent for $c \leq 1$, there is no reason 
for them to remain so after continuation back to $c \geq 25$. 
And indeed we have seen in the modular bootstrap that they 
are not! Needless to stress that the problem with complex 
exponents in eq.\ (\ref{Zmns}) disappears for imaginary $b$. 
In the corresponding models with $c\leq 1$ the brane parameters 
$m$ and $n$ also possess a nice geometric interpretation 
related to a position and extension along a 1-dimensional 
line \cite{FreSchcos}. When we try to continue back into 
Liouville theory, these parameters become imaginary. All 
this clearly supports our proposal to discard solutions
with $n\neq 1$ or $m\neq 1$.

\paragraph{The $c=1$ limit \& tachyon condensation.} 
Before we conclude our discussion of boundary Liouville theory, 
we would like to briefly comment on its possible applications  
to the condensation of tachyons. Let us recall from our introductory 
remarks in the second lecture that we need to take the central 
charge to $c=1$ or, equivalently, our parameter $b$ to $b=i$, 
\begin{equation}\label{BLact}  
 S^{c=1}_{BL}[X]  \ = \ \left( \frac{1}{4\pi} \int_\Sigma d^2 z\,  
    \partial X \bar \partial X + \mu \exp 2b X(z,\bz) + 
     \int_{\partial \Sigma} du \, \mu_B \exp b X(u)\right)_{b=i} 
    \ \ .  
\end{equation} 
Here we have allowed for an additional boundary term so as to 
capture the condensation of both open and closed string tachyons. 
It is important to keep in mind that any application of Liouville
theory to time-dependent processes also requires a Wick rotation, 
i.e.\ we need to consider correlation functions with imaginary rather 
than real momenta $P$. We shall argue below that the two steps 
of this programme, the limit $c\rightarrow 1$ and the Wick 
rotation, meet quite significant technical difficulties. 
\smallskip 

Nevertheless, there exists at least one quantity that we can compute 
easily from Liouville theory \cite{Gutperle:2003xf} and that we can 
even compare with results from a more direct calculation in the 
rolling tachyon background. It concerns the case in which merely 
open strings condense, i.e.\ in which $\mu = 0$. Because of relation 
(\ref{musrel}), switching off the bulk coupling $\mu$ is equivalent 
to considering the limit $s \rightarrow \infty$. The corresponding 
limit of the 1-point coupling (\ref{AFZZT}) is straightforward to 
compute and it is analytic in both $b$ and $P$ so that neither 
the continuation to $b=i$ nor the Wick rotation pose any problem. 
The resulting expression for the 1-point coupling is 
$$  \langle\, \exp \left(i E X^0(z,\bz)\right)\, \rangle  
  \ \sim \  \left(\pi \mu\right)^{i E}\, 
   \frac{1}{\sinh \pi E}\ \ .    $$
This answer from Liouville theory may be checked directly 
\cite{Sen:2002nu,Sen:2002vv,Larsen:2002wc} through perturbative 
computations in free field theory \cite{Callan:1994ub,
Polchinski:1994my,Fendley:1994rh,Recknagel:1998ih}.  
\smallskip

Unfortunately, other quantities in the rolling tachyon background 
have a much more singular behavior at $b=i$. Barnes' double 
$\Gamma$-function, which appears as a building block for many 
couplings in Liouville theory (see e.g.\ eq.\ (\ref{Cc>})),  is a 
well defined analytic function as long as ${\rm Re} b \neq 0$. If 
we send $b \rightarrow i$, on the other hand, $\Gamma_2$ becomes singular 
as one may infer e.g.\ from the integral formula (\ref{BGamma}). 
In fact, the integrand has double poles along the integration 
contour whenever $b$ becomes imaginary. A careful analysis 
reveals that the limit may still be well defined, but it 
is a distribution and not an analytic function.  
\smallskip 

Rather than discussing any of the mathematical details of the 
limit procedure (see \cite{SchL}) we would like to 
sketch a more physical argument that provides some insight into the
origin of the problem and the structure of the solution. For
simplicity, let us begin with the pure bulk theory. Recall from 
ordinary Liouville theory that it has a trivial dependence on the 
coupling constant $\mu$. Since any changes in the coupling can be 
absorbed in a shift of the zero mode, one cannot vary the strength 
of the interaction. This feature of Liouville theory persists when 
the parameter $b$ moves away from the real axis into the complex 
plane. As we reach the point $b=i$, our model seems to change 
quite drastically: at this point, the `Liouville wall' disappears 
and the potential becomes periodic. Standard intuition therefore
suggests that the spectrum of closed string modes develops gaps 
at $b=i$. Since the strength of the interaction cannot be tuned 
in the bulk theory, the band gaps must be point-like. The emerging 
band gaps explain both the difficulties with the $b=i$ limit and 
the non-analyticity of the resulting couplings. 
\smallskip 

Though our argument here was based on properties of the classical 
action which we cannot fully trust, the point-like band-gaps are 
indeed a characteristic property of the $c=1$ bulk theory. It was 
shown in \cite{SchL} that the bulk 3-point couplings of Liouville 
theory possess a $b=i$ limit which is well-defined for real momenta 
of the participating closed strings. The resulting model turns out 
to coincide with the $c=1$ limit of unitary minimal models which 
was constructed by Runkel and Watts in \cite{Runkel:2001ng}. Since
the couplings cease to be analytic in the momenta, the model cannot
be Wick-rotated directly. Nevertheless, it seems possible 
to construct the Lorentzian background. To this end, the Wick rotation 
is performed before sending the central charge to $c=1$. The 
corresponding couplings with imaginary momenta were constructed in 
\cite{SchL}, correcting an earlier proposal of \cite{Strominger:2003fn}. 
On the other hand, this Lorentzian $c=1$ limit depends on the path 
along which $b$ is sent to $b=i$. It is tempting to relate this 
non-uniqueness to a choice of boundary conditions at $x_0= \infty$ 
(see \cite{Fredenhagen:2003ut} a related minisuperspace toy model), 
but this issue certainly deserves further study. 
\smallskip 

A similar investigation of the $c=1$ boundary model (\ref{BLact}) 
was recently carried out in \cite{Fredenhagen:2004cj}, at least 
for Euclidean signature. The properties of this model are similar
to the bulk case, only that the band gaps in the boundary spectrum
can now have finite width. In the presence of a boundary, Liouville 
theory contains a second coupling constant $\mu_B$ which controls 
the strength of an exponential interaction on the boundary of the 
world-sheet. $\mu_B$ is a real parameter of the model since the 
freedom of shifting the zero mode can only be used to renormalize 
one of the couplings $\mu$ or $\mu_B$. Once more, the boundary 
potential becomes periodic at $b=i$ and hence the open string 
spectrum develops gaps, as in the case of the bulk model. But 
this time, the width of these gaps can be tuned by changes of 
the parameter $\mu_B$. All these rather non-trivial properties
were confirmed in \cite{Fredenhagen:2003ut} through an exact 
constructions of the spectrum and various couplings of these 
novel conformal field theories. So far, the Wick-rotated model 
has not been obtained from Liouville theory, thought there 
exist recent predictions for some of its structure constants 
\cite{Balasubramanian:2004fz} (see also \cite{Kristjansson:2004mf,
Gaberdiel:2004na,Kristjansson:2004ny} for related studies). 
\newpage

\section{Strings in the semi-infinite cigar} 
\setcounter{equation}{0}
\def\half{\frac12}
\def\p{\partial} 
\def\SL{{\rm SL}$_2(\QR)$}
\def\SLC{{\rm SL}$_2(\QC)$}
\def\SU{{\rm{SU(2)}}}
\def\mSL{{\rm SL}_2(\QR)} 
\def\SLU{{\rm SL}$_2(\QR)${\rm /U(1)}}
\def\mSLU{{\rm SL}_2(\QR){\rm /U(1)}}
\def\rar{\rightarrow}
\def\Z{\QZ}
\def\tz{\tilde z}
\def\tth{\tilde \tau}
\def\vt{\vartheta}
\def\AA{$AdS_2$}
\def\halfpi{\frac{\pi}{2}}
\def\N{{\cal N}}
\def\D{{\rm D}}
\def\Dz{{\rm D0}}
\def\pp{\,}
\def\j{2j+1}
\def\la{{\langle}}
\def\ra{{\rangle}} 
\def\cA{{\cal A}}
\def\o{{_{\rm os}}}
\def\U{{\rm U(1)}}
 
In the previous lectures we have analyzed Liouville theory mainly 
because it is the simplest non-trivial example of a model with 
non-compact target space. As we have reviewed in the introduction, 
however, many interesting applications of non-rational conformal 
field theory, in particular those that arise 
from the usual AdS/CFT correspondence, employ higher 
dimensional curved backgrounds such as $AdS_3$ or $AdS_5$. The 
aim of this final lecture is to provide some overview over 
results in this direction. As we proceed, we shall start to  
appreciate how valuable the lessons are that we have learned 
from Liouville theory. 
\smallskip 

Ultimately, one would certainly like to address strings moving 
in $AdS_5$. But unfortunately, this goes far beyond our present 
technology, mainly because consistency of the $AdS_5$ background 
requires to turn on a RR 5-form field. The 
situation is somewhat better for $AdS_3$. In this case, 
consistency may be achieved by switching on a NSNS 3-form $H$. 
For reasons that we shall not explain here, such pure NSNS 
backgrounds are much easier to deal with in boundary conformal 
field theory. In cylindrical coordinates $(\tau,\rho,\theta)$,
the non-trivial background fields of this geometry read 
\begin{eqnarray} \label{AdSg}
ds^2 & = & \frac{k}{2}\, \left( d\rho^2 - \cosh^2\rho \, d\tau^2 
               + \sinh^2 \rho \, d\theta^2 \right) \\[2mm] 
 H   & = & \frac{k}{2}\, \sinh 2 \rho \ d\theta \wedge 
           d \rho \wedge d\tau \ \ . \label{AdSH} 
\end{eqnarray}  
We wish to point out that these are the background fields of 
a WZW model on the universal covering space of the group manifold 
SL$_2(\QR)$. In our cylindrical coordinates, the background 
(\ref{AdSg},\ref{AdSH}) is manifestly invariant under
shifts of the time coordinate $\tau$. We can use this 
symmetry to pass to the 2-dimensional coset space SL$_2
(\QR)/$U(1),    
\begin{eqnarray} \label{cigarg}
 ds^2 & = & \frac{k}{2}\, \left( d\rho^2 + \tanh^2 \rho \ 
                         d\theta^2 \right) \\[2mm] 
 \exp \varphi  & = & \exp \varphi_0 \ \cosh \rho \ \ . \label{cigard} 
\end{eqnarray} 
Obviously, the 2-dimensional coset space cannot carry any 
NSNS 3-form $H$. Instead, it comes equipped with a non-trivial 
dilaton field $\varphi$. The latter arises because the orbits of
$\tau$-translations possess different length. Long orbits
at large values of $\rho$ correspond to regions of small 
string coupling. Let us observe that the dilaton field 
$\varphi$ becomes linear as we send $\rho \rightarrow \infty$. 
In this sense, the $\rho$-coordinate of the coset geometry
is similar to the Liouville direction. The cigar geometry 
also avoids the strong coupling problem of a linear dilaton 
background. Only the mechanism is somewhat different from 
the scenario in Liouville theory: On the cigar, the dilaton
field itself is deformed so that the string coupling stays 
finite throughout the entire target space. The background 
(\ref{cigarg},\ref{cigard}) and its Lorentzian counterpart were
first described in \cite{EFR,MSW,Wit}. 
\vspace{1cm} 
\fig{The cigar is parametrized by two coordinates $\rho \in [0,\infty]$ 
and $\theta \in [0,2\pi]$. It comes equipped with a non-constant dilaton 
that vanishes at $\rho = \infty$ and assumes its largest value $\varphi_0$
at the tip $\rho = 0$ of the cigar.} 
{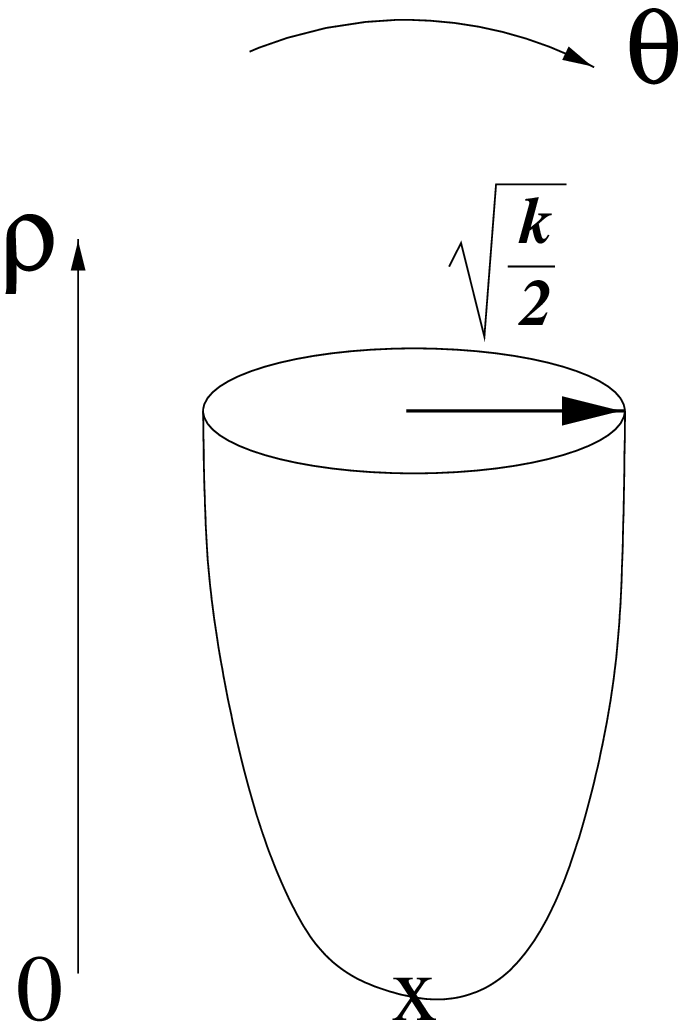}{5truecm} 
\figlabel{\basic}
\vspace{.5cm}

The close relation with $AdS_3$ makes the coset \SLU \, an 
interesting background to consider. Additional motivations
arise from the study of little string theory and the near 
horizon geometry of NS5-branes. If we place $N$ such branes
on top of each other, the near horizon geometry is given by 
\begin{equation} 
   {\cal X}^{NS5} \ \cong \ \QR^{(1|5)} \times \QR^+_Q 
   \times S^3_N  
  \ \ . \label{NS5} 
\end{equation}  
Here, the three factors are associated with the 6 directions 
along, the radial distance from and the 3-spheres surrounding 
the NS5-branes. The second factor stands for a linear dilaton 
with background charge $Q^2 = 1/N$.  The last factor represents 
a well studied rational conformal field theory, namely the 
($N=1$ supersymmetric) SU(2) WZW model at level $k' = N-2$. 
Since the dilaton field $\varphi$ in the background (\ref{NS5})
is unbounded, string perturbation theory cannot be trusted. 
To overcome this issue, it has been suggested to separate 
the $N$ NS5-branes and to place them along a circle in a 
2-dimensional plane transverse to their world-volume 
\cite{sfetsos}. Such a geometric picture is associated with 
a deformation of the corresponding world-sheet theory. In 
order to guess the appropriate deformation, we make use of 
the following well-known fact 
\begin{equation} \label{S3orb}  
  S^3_N \ =\  \SU_N \ =\ 
    \left( \U_N \times \SU_N/\U_N\right)/\QZ_N\ \ , 
\end{equation}       
where $\U_N$ denotes a compactified free boson and we orbifold   
with a group $\QZ_N$ of simple currents. After inserting eq.\ 
(\ref{S3orb}) into the NS5-brane background (\ref{NS5}), we 
can combine the linear dilaton from the latter with the $\U_N$ 
of the 3-sphere. In this way the NS5 brane background is seen 
to involve a non-compact half-infinite cylinder $\QR^+_Q \times 
\U_N$ that we can deform into our cigar geometry, thereby 
resolving our strong coupling problem, 
$$ {\cal X}^{NS5} \ \longrightarrow \ {\cal X}^{NS5}_{\rm def} 
 \ = \ \QR^{(1|5)} \times \left(\mSL_N/\U_N \times 
      \SU_N/\U_N \right)/\QZ_N \ \ ,  
$$    
with $k = N+2$.  
Hence, the interest in NS5-branes provides strong motivation 
to investigate string theory on the cigar. Let us also note that 
there exists a rich and interesting class of compactifications of 
NS5-branes on Calabi-Yau spaces that, by the same line of 
reasoning, involve the cigar as a central building block 
\cite{Ooguri:1995wj} (see also \cite{Kutasov:2001uf}). We 
should finally note that these backgrounds 
certainly involve some amount of supersymmetry which we 
suppress in our discussion here. As in the case of the 
analogous compact coset theories, adding supersymmetry has
relatively minor effects on the world-sheet theory. Since
we are more interested in the qualitative features of our 
non-compact coset model, we shall neglect the corrections 
that supersymmetry brings about, even though they are 
certainly important in concrete applications. For treatments 
of the supersymmetric models, we can refer the reader to a 
number of interesting recent publications \cite{Ahn:2003wy}-
\cite{Eguchi:2004ik}.

\subsection{Remarks on the bulk theory}

\paragraph{The minisuperspace model.} 
As in the case of Liouville theory we can get some intuition 
into the spectrum of closed string modes and their couplings from the 
minisuperspace approximation. To this end, we are looking for eigen-%
functions of the Laplacian on the cigar, 
\beqa
\Delta  & = &  
- \frac{1}{e ^{-2\varphi}\sqrt{\det
  g}}\ \partial _\mu e ^{-2\varphi}\sqrt{\det g}\ 
  g^{\mu \nu}\partial_\nu \nonumber \\[2mm] & = & 
-  \frac{2}{k}\left[  \partial_\rho^2 + (\coth \rho +
  \tanh \rho)\, \p_\rho+\coth^2\rho\, \p_\theta ^2 \right]\ \ . 
\eeqa  
The $\delta$-function normalizable eigen-functions of this operator 
can be expressed in terms of hypergeometric functions through 
\beqa
\phi^j_{n0}(\rho,\theta) & = & -\frac{\Gamma(-j+\frac{|n|}{2})^2}
  {\Gamma(|n|+1)\Gamma(-2j-1)} \ e^{in\theta}\sinh ^{|n|}\rho\
  \times \nonumber \\[2mm] & & \hspace*{2cm} \times \ 
  F\left(j+1+\frac{|n|}{2},-j+\frac{|n|}{2}, |n|+1;-\sinh^2\rho\right)
\label{wavecig}
\eeqa 
where $j \in -1/2 + i \mathbb{R}$ describes the momentum along 
the $\rho$-direction of the cigar and $n \in \mathbb{Z}$ is the 
angular momentum under rotations around the tip. For the associated 
eigenvalues one finds 
\beqa \label{EV}
\Delta ^j_{n0}\ =\ -\frac{2j(j+1)}{k}+\frac{n^2}{2k}\ \ \ .
\eeqa   
In the symbols $\phi^j_{n0}$ and $\Delta ^j_{n0}$  we have inserted 
an index `0' without any further comment. Our motivation will become 
clear once we start to look into the spectrum of the bulk conformal field theory. Let us
also stress that there are no $L^2$-normalizable eigenfunction of the 
Laplacian on the cigar. Such eigenfunctions would correspond to {\em 
discrete states} living near the tip, but in the minisuperspace 
approximation one finds exclusively {\em continuous states} which 
behave like plane waves at $\rho\rightarrow \infty$.
\smallskip 

From our explicit formula for wave-functions and general properties
of hypergeometric functions it is not hard to read off the reflection 
amplitude of the particle toy model. It is given by 
\begin{equation}\label{screfl}  
    R_0(j,n) \ = \ \frac{\phi^j_{n0}(\rho,\theta)}
                          {\phi^{-j-1}_{n0}(\rho,\theta)} 
     \ = \ \frac{\Gamma(2j+1)\Gamma^2(-j+\frac{n}{2})}
      {\Gamma(-2j-1)\Gamma^2(j+1 + \frac{n}{2})} \ \ .    
\end{equation} 
One may also compute a particle analogue of the 3-point 
couplings. Since the result is very similar to the corresponding
formula in Liouville theory, we do not want to present more details
here. 

\paragraph{The stringy corrections.}
Instead, let us now try to guess how the results of the minisuperspace 
toy model get modified in the full field theory. Each of the wave 
functions (\ref{wavecig}) in the minisuperspace theory lifts to a 
{\em primary field} in the conformal field theory. But the full story 
must be a bit more complicated. In fact, at $\rho \rightarrow \infty$, 
the cigar looks like an infinite cylinder and for the latter we know 
that primary fields are labeled by momentum $n$ and winding $w$ around 
the compact circle, in addition to the continuous momentum $iP = j+\half$ 
along the uncompactified direction. Hence, we expect that the full 
conformal field theory on the cigar has primary fields $\Phi^j_{nw}
(z,\bz)$ which are labeled by three quantum numbers $j,n$ and $w$. 
The exact couplings for these fields are known. In particular, 
the bulk reflection amplitude is given by 
\begin{eqnarray}  \label{R}
 R(j,n,w)  & = &  \frac{\Gamma(2j+1)}{\Gamma(-2j-1)}\ 
  \frac{\Gamma(-j+\frac{n-kw}{2})\Gamma(-j+\frac{n+kw}{2})} 
  {\Gamma(j+1+\frac{n-kw}{2})\Gamma(j+1+\frac{n+kw}{2})}\, \times \\[2mm]
  \nonumber  & & \hspace*{4cm} \times \, (\gamma(b^2)b^2)^{-2j-1}\
 \ \frac{\Gamma(1+b^2(2j+1))}{\Gamma(1-b^2(2j+1))}\ \ 
\end{eqnarray} 
where we used $b^2 = 1/(k-2)$. Even without any more detailed analysis
of factorization constraints, this formula is rather easy to understand. 
Let us first note that, for the modes with zero winding number $w$, we have 
\begin{equation}\label{Rw0}  
  R(j,n,w=0) \ = \ R_0(j,n)\ \frac{\Gamma(1+b^2(2j+1))}
               {\Gamma(1-b^2(2j+1))}\, (\gamma(b^2)b^2)^{-2j-1}\ \ . 
\end{equation} 
Here, we inserted the semi-classical reflection amplitude (\ref{screfl}). 
Relation (\ref{Rw0}) should be compared to the analogous eq.\ (\ref{RR0}) 
in Liouville theory. Remarkably, the stringy correction factors in both 
formulas coincide  if we identify $2j+1$ with $i\w$, at least up to a 
simple renormalization of the bulk fields. Such an agreement should 
not come as a complete surprise since the reflection amplitude 
concerns the momentum in the non-compact direction of the cigar 
which -- at $\rho \rightarrow \infty$ -- approximates a linear 
dilaton background, just as the region $x \rightarrow -\infty$ 
of Liouville theory. 
\smallskip 

The dependence of the exact reflection amplitude on the winding number 
$w$ is also rather easy to argue for. Since it is associated with the angular 
direction, one would expect that the parameter $w$ enters through the same 
rules as in the theory of a single compactified boson, i.e.\ by using the 
replacement $n \rightarrow n \pm k w$. If we apply this prescription to 
the reflection amplitude $R(j,n,w=0)$ we obtain the correct formula 
(\ref{R}). Note also that $n$ enters the reflection amplitude only 
through the semi-classical factor $R_0$.  
\smallskip 

Having gained some confidence into the formula (\ref{R}), we wish to 
look briefly at some properties of $R$. It is well known that bound 
states of a scattering potential cause singularities in the reflection 
amplitude. The converse, however, is not true, i.e.\ a reflection 
amplitude can possess singularities that are not linked to any bound 
states. Our reflection amplitude (\ref{R}) has several different 
series of poles. Singularities in the $w=0$ sector, i.e.\ of the 
expression (\ref{Rw0}), are either associated with the semi-classical 
reflection amplitude (\ref{screfl}) or the stringy correction factor. 
Since we did neither find bound states in the semi-classical model 
nor in Liouville theory, we are ready to discard these poles from 
our list of possible bound states in the cigar. The situation changes 
for $w\neq 0$. These sectors of winding strings did not exist in the 
semi-classical model and hence the corresponding poles of $R$
could very well signal bound states that are localized near the tip 
of the cigar. We shall see later that this is indeed the case, at 
least for a subset thereof.  
\smallskip 

The existence of stringy bound states near the tip of the cigar 
has been suspected for a long time \cite{dvv}, but the reasoning 
and the counting of these bound states stood on shaky grounds.  
A satisfactory derivation was only given a few years ago through 
a computation of the path integral for the partition function of 
the system. It turned out that, in addition to the expected 
{\em continuous series} with 
$$ j \ \in\  -\frac12 + i\, \mathbb{R}^+_0 \ \ \ \ \ , \ \ \ \ \ 
   n \ \in \ \mathbb{Z} \ \ \ \ , \ \ \ \ w \ \in \ \mathbb{Z} \ \ , 
$$ 
there is also a {\em discrete series} of primary fields for 
with $w,n \in \QZ$ such that $|kw|>|n|$ and 
\beqa
j \, \in \, \cJ^d_{nw} \ :=\  \left[\, \frac{1-k}{2}\, ,\, 
- \frac12 \, \right] \, \cap  
\, \left( \mathbb{N} - \half |kw| + \half |n| \right)  \ . \label{rangej}
\eeqa     
The correct list of bound states was found in \cite{hpt} (see also 
\cite{moi} for a closely related results in the case of strings on 
\SL) and it differs slightly from the predictions in \cite{dvv}. 
Later we shall confirm these findings quite beautifully through 
world-sheet duality involving open strings. Let us also point out 
that for $w = 0$, the discrete series is empty, in agreement with 
the minisuperspace analysis. The primary fields of these two series 
have conformal weights given by        
\beqa \label{h}
h^j_{n w} \ = \  - \frac{j(j+1)}{k-2} + \frac{(n+kw)^2}{4k} 
   \ \ \ \mbox{ and } \ \ \ 
   {\bar h}^j_{n w} \ = \ 
    -\frac{j(j+1)}{k-2}+\frac{(n-kw)^2}{4k} 
\ \ . 
\eeqa
Notice that these conformal weights are all positive, as for 
any Euclidean unitary conformal field theory. In the limit of 
large level $k$, the sum $h^j_{n0} + \bar h^j_{n0}$ of the left 
and right conformal weights with $w=0$ reproduces the spectrum 
(\ref{EV}) of the minisuperspace Laplacian.

\subsection{From branes to bulk - D0 branes} 

In the first lecture we have outlined a general procedure that 
allows us to construct closed and open string backgrounds. The
first step of this general recipe was to find a consistent 
spectrum of closed string modes and their 3-point couplings $C$. 
Consistency required that the latter obey the crossing symmetry 
condition (\ref{cross}). In our second step we searched for possible 
D-branes by solving the cluster condition (\ref{class},\ref{XiCF}) 
for the couplings $A$ of closed string modes to the brane. Finally, 
we showed how to recover the spectrum of open string modes from 
the couplings $A$ through the so-called modular bootstrap 
(see eq.\ (\ref{BPF1})).  
\smallskip

Looking back at the central equations (\ref{cross},%
\ref{class},\ref{XiCF},\ref{BPF1}) of our program one may have the 
idea to reverse the entire procedure and to start from the end, 
i.e.\ from the spectrum of open string modes on some brane. Suppose 
we were able to guess somehow an annulus amplitude $Z$. Then we 
could try to recover the couplings $A$ from it through formula 
(\ref{BPF1}). If this was successful, it could also teach us 
about possible closed string modes since equation (\ref{BPF1}) 
involves a sum (or integral) over closed string modes from the 
spectrum of the model.  Moreover, using eqs.\ (\ref{class},%
\ref{XiCF}), we could even recover closed string couplings $C$, 
provided we knew the fusing matrix of the chiral algebra. 
\smallskip

In this simple form, the reversed program we have sketched 
here has many problems and we shall comment on some of them 
later on. Nevertheless, there exist a few good reasons to 
believe that ultimately some refined version of this 
procedure could be quite successful. In fact, knowing the
entire open string spectrum on some brane, including all 
the massive modes, provides us with a lot of information 
not only on small fluctuations of open string, and hence 
on the dimension of the background, but also e.g.\ on the 
non-trivial cycles through the associated open string 
winding modes.  
\smallskip 

We can illustrate some of these very general remarks with two   
rather simple examples before we come back to the study of
the 2D cigar geometry. Let us begin with a point-like (D0)
brane in a flat 1-dimensional target space $\QR$. Its 
boundary partition function is given by 
$$ Z_{D0}^{\QR} (q) \ = \ \eta^{-1}(q) \ = \ 
    \int_{-\infty}^{\infty} dp \ \frac{\tilde q^{p^2}}
                                       {\eta(\tilde q)}
\ \ . $$ 
The $\eta$-function on the left hand contains all the 
oscillation modes of open strings on our point-like 
brane and there are no zero modes. When transformed to
the closed string picture we recover a continuum of
modes that are parametrized by the momentum $p$ along 
with the usual tower of string oscillations. Hence, we 
have recovered all the closed string modes that exits in 
a 1-dimensional flat space. Obviously, this example is
quite trivial and it may not be sufficient to support 
our reversed program. 
\smallskip 

But after some small modification, our analysis starts 
to look a bit more interesting. In fact, let us now 
consider a point-like (D0) brane that sits at the 
singular point of the half-line $\QR/\QZ_2$. Here, 
the non-trivial element of $\QZ_2$ acts by reflection 
$x \rightarrow -x$ at the origin $x=0$. It is rather 
easy to find the annulus amplitude for such a brane.
All we have to do is to count the number of states 
that are created by an even number of bosonic 
oscillators $a_n$ from the ground state $|0\rangle$. 
The result of this simple combinatorial problem is 
given by
$$
Z^{\QR/\QZ_2}_{D0} \ =  \frac{1}{\eta(q)} \ 
     \sum_{m\geq 0} \ \left( q^{(2m)^2}-q^{(2m+1)^2}
   \right) \ = \ \frac{1}{2\eta(q)}\, \left(1+
   \vartheta_4(q^2)\right) \ \ .  
$$
Here, $\vartheta_4$ is one of Jacobi's $\vartheta$-functions. 
After modular transformation to the closed string picture, 
we obtain 
$$ 
Z^{\QR/\QZ_2}_{D0} \ = \ 
  \int_{0}^{\infty} dp \ \frac{\tilde q^{p^2}}
                                       {\eta(\tilde q)}
  + \frac{1}{\sqrt{2}} \, \frac{\tilde q^{1/48}}
    {\prod_{n\geq 1} \left(1-\tilde q^{n-1/2}\right)} 
\ \ . 
$$ 
The interpretation of the first term is essentially 
the same as in our first example. It arises from 
closed string modes that move along the half-line. 
Of course, the origin of the second term is also well 
known: It comes with closed string states from the 
twisted sector that is characteristic for any orbifold 
geometry. In this sense, the modular bootstrap has 
predicted the existence of a twisted closed string 
sector. Even though our second example is still rather 
trivial, we now begin to grasp that sometimes it may 
be much easier to guess an annulus amplitude for open 
strings on a brane than the spectrum of closed strings. 
In a moment we shall see this idea at work is a much less 
trivial case: For open strings on a point-like brane 
in our cigar background the whole issue of bound states
does not arise simply because there is no continuous 
open string spectrum. Hence, it should be rather 
straightforward to come up with a good Ansatz for an 
open string partition function $Z$ on such branes. 
According to our general ideas, this $Z$ contains 
information on the bulk spectrum, in particular on 
the predicted closed string bound states, that we are 
able to decipher through modular transformation.     

\paragraph{The inverse procedure - some preparation.} 
We have argued  above that the reconstruction of an entire 
model from no more than the theory of open strings on a 
single brane has quite realistic chances. At least for 
rational models (i.e.\ for compact targets) the program can 
be made much more precise \cite{Petkova:2000mt,Fuchs:2001am}
(see also \cite{Fuchs:2002cm,Fuchs:2003id,Fuchs:2004dz,
Fuchs:2004xi}). The input it requires is two-fold. First 
of all, we need complete 
knowledge about the symmetry algebra and its representation 
theory, including the fusion rules, the modular S-matrix and 
the Fusing matrix. Throughout this text we have always 
assumed that this information is provided somehow. The 
second important input into the reconstruction program 
consists of data on the open strings. What we need to 
know is the spectrum of open strings on the brane and the 
operator products of the associated open string vertex 
operators. The latter is often uniquely fixed by the    
former. 
\smallskip 

In this subsection we would like to address the first part 
of the input by listing relevant facts on the representation 
theory of the symmetry algebra of cigar background. It will 
suffice to list its unitary representations and explicit 
formulas for the corresponding characters. 
\smallskip 

The chiral algebra in question is the so-called coset algebra 
\SLU . Its description is a bit indirect through the \SL\  
current algebra. The latter is generated by the modes of three 
chiral currents $J^a(z), a=0,1,2,$ with the standard relations, 
$$ [\, J^a_n\, ,\, J^b_m\, ] \ = \  {f^{ab}}_c \, J^c_{n+m} 
   - \frac12 k n \delta^{ab} \, \delta_{n,-m}\ \ , $$ 
where $f$ denotes the structure constants and of the Lie algebra 
sl$_2$. Each of the currents $J^a$ by itself generates a U(1) 
current algebra. We select one of these U(1) current algebras, 
say the one that is generated by $J^0$, and then form the coset 
chiral algebra from all the fields that may be constructed out of 
\SL\  currents and that commute with $J^0_n$. Note that 
this algebra does not contain fields of dimension $h=1$ 
since there is no \SL\  current whose modes commute with 
all $J^0_n$. But there exists one field of dimension $h=2$, 
namely the Virasoro field 
$$ T^{\rm cos}(z)\ = \ T^{\rm SL}(z) - T^{\rm U}(z) $$ 
where the two Virasoro fields on the right hand side denote 
the Sugawara bilinears in the \SL\  and the U(1) current 
algebra, respectively. It is not difficult to check that 
the field $T^{\rm cos}$ commutes with $J^0_n$ and that it
generates a Virasoro algebra with central charge 
$$ c \ =\ c^{\rm cos} \ = \ c^{\rm SL} - c^{\rm U} \ = \ 
   \frac{3k}{k-2} - 1  \ \ . $$     
Representations of the coset chiral algebra can be obtained 
through decomposition from representations of the \SL\  current 
algebra. The latter has three different types of irreducible 
unitary representations. Representation spaces $\cV^{c}_{(j,\a)}$  
of the continuous series representations are labeled by a spin 
$j \in Q/2 + i\QR$ and a real parameter $\a\in [0,1[$. For the 
discrete series, we denote the representation spaces by 
$\cV^{d}_{j}$ with $j$ real. Finally, the vacuum representation 
is $\cV^0_0$. It is the only unitary representation with a finite 
dimensional space of ground states. 
\smallskip 
     
From each of these representations we may prepare an infinite 
number of \SLU\  representations. Vectors in the corresponding 
representation spaces possess a fixed U(1) charge, i.e.\ the same 
eigenvalues of the zero mode $J^0_0$, and they are annihilated by 
all the modes $J^0_n$ with $n>0$. With the help of this simple 
characterization one can work out formulas for the characters 
of all the sectors of the coset chiral algebra. Here we only 
reproduce the results (see e.g.\ \cite{Ribault:2003ss} for a 
derivation). For the continuous series, the chiral characters 
read as follows
\beqa \label{carcont}
\chi^c_{(j,\om)}(q)\ =\ {\rm Tr}_{(j,\om)} \ q^{L_0-\frac{c}{24}}
      \ = \ \frac{q^{-\frac{(j+\frac{1}{2})^2}{k-2}+
     \frac{\om^2}{k}}}{\eta(q)^2}\ \ , 
\eeqa
where $\omega$ is some real number parametrizing the \SLU\ chiral
algebra representations descending from a continuous representation 
of \SL\ with spin $j=-\half +iP$. In the case of the discrete series, 
the expressions for chiral characters are a bit more complicated, 
\beqa\label{cardisc}
\chi_{(j,\ell-j)}^{d}(q) & = & \frac{q^{-\frac{(j+\frac12)^2}{k-2}+
 \frac{(j-\ell)^2}{k}}}{ \eta(q)^2}\ 
 \left[\, \epsilon_\ell\, 
\sum_{s=0}^\infty \ (-1)^s q^{\half
  s(s+2|\ell+\half|)} -\frac{\epsilon_\ell-1}{2}\,
  \right]\ \ \\[3mm]
& & \hspace*{.5cm}\mbox{ where } \ \ \ \epsilon_\ell \ =\ 
\left\{ \begin{array}{rcl}
1 & {\rm if } & \ell \geq \ 0 \\[1mm]
-1  & {\rm if } & \ell \leq -1
\end{array}\right. \ \  \nonumber 
\eeqa
and $l$ is an integer that labels different \SLU\ sectors that 
derive from one of the two discrete series representation of 
\SL\ with spin $j$. Finally, we obtain one series of sectors 
from the vacuum representation of the \SL\ current algebra. 
Its characters are given by  
\begin{equation}
\chi^0_{(0,n)}(q) \ = \ \chi^d_{(0,n)}(q) - \chi^d_{(-1,n+1)}(q)
\ \ .  
\end{equation} 
The integer $n \in QZ$ that parametrizes sectors in this series
refers to the U(1) charge of the corresponding states in the 
$j=0$ sector of the parent \SL\ theory.

\paragraph{D0-branes in the 2D cigar.} 
Our first aim is to construct the annulus amplitude of some
brane in the cigar. Geometrical intuition suggests that there
should exist a point-like brane at the tip of the cigar, i.e.\ 
at the point where the string coupling assumes its largest 
value. It is not too hard to guess the annulus amplitude for
such a brane. To this end, we recall that our cigar background
may be considered as a coset \SLU. Open string modes on a 
point-like brane in \SL\ are rather easy to count. They are 
generated from a ground state $|0\rangle$ by the modes 
$J^a_n, a = 0,1,2,$
$$ \cH^{\rm SL}_{D0} \ = \ {\rm span}\  J^{a_1}_{n_1} \dots 
       J^{a_s}_{n_s} |0\rangle\ \ \mbox{ with } \ \ 
       n_r < 0 \ \ .  
$$  
States $|\psi\rangle$ of open strings in the coset geometry 
form a subspace of $\cH^{\rm}_{D0}$ which may be characterized 
by the condition 
$$ J^0_n |\psi\rangle \ = \ 0 \ \ \ \mbox{ for all } \ \ \ 
   n \, \geq 1 \ \ . $$ 
It is in principle straightforward to count the number of solutions
to these equations for each eigenvalue of $L_0 = L_0^{\rm SL} -
 L_0^{\rm U}$. This counting problems leads directly to the following 
partition sum for open strings on our point-like brane (we shall 
explain the subscript $1,1$ at the end of this section) 
\begin{equation} \label{D0annulus} 
 Z_{1,1}^{\mSLU} (q) \ = \ q^{-\frac{c}{24}} \, \left( 
 1+ 2q^{1 + \frac{1}{k}} + q^2 + \dots \right) \ = \ 
  \sum_{n = -\infty}^\infty  \chi^0_{(0,n)}(q) \ \ . 
\end{equation}
The first few terms which we displayed explicitly are very easy to 
check. The full annulus amplitude is composed from an infinite 
sum of characters rather than the vacuum character $\chi^0_{(0,0)}$
of our coset chiral algebra. This may seem a bit unusual at first, 
but there are good reasons for such a behavior. In fact, when we 
expand the character $\chi^0_{(0,0)}(q)$ in powers of $q$ we find 
the following first few terms, 
$$ \chi^0_{(0,0)}(q)\ = \ q^{-\frac{c}{24}} \left( 1+ q^2 + \dots 
 \right)\ \ . $$ 
Hence, we are e.g.\ missing the term $2q^{1+ \frac{1}{k}}$ from the
correct answer in eq.\ (\ref{D0annulus}). But a quick 
look at the geometry tells us that such a term must be present in 
the annulus amplitude we consider. Recall that point-like branes 
in a flat 2-dimensional space possess two moduli of transverse 
displacement. This means that the annulus amplitude for such 
branes contains a term $2q$ which is indeed present in the flat 
space limit $k \rightarrow \infty$ of our annulus amplitude
(\ref{D0annulus}), but not in $\chi^0_{(0,0)}$. Transverse 
displacements cease to be moduli of the curved space model 
because of the varying string coupling which pins the D0 
brane down at the tip of the cigar. Correspondingly, the 
exponent of the second term in the expansion of $Z$ must 
be deformed at finite $k$. Our term $2q^{1+ \frac{1}{k}}$ 
in the proposed annulus amplitude has exactly the expected
properties. It can only come from the characters $\chi_{(0,
\pm 1)}(q)$. But once we have added these two characters to 
$\chi_{(0,0)}$, we cannot possibly stop but are forced to 
sum over all the U(1) charges $n \in \QZ$. Our argument here 
provides very strong evidence for the annulus amplitude 
(\ref{D0annulus}) in addition to the derivation we gave 
above. 
\smallskip 

Our next aim is to modular transform the expression 
(\ref{D0annulus}). The result of a short and rather direct 
computation is given by 
\begin{equation}
Z_{1,1}^{\mSLU} (q)\, =\, -2 \sqrt{b^2 k}  \ \int dP\  
  \sinh 2\pi b^2 P \ \frac{\tanh \pi P}{\eta(\tq)^2}
 \sum_{n\in \Z} \ (-1)^n\,  
\tq^{b^2(P - \frac{i}{2b^2} n)^2 + \frac{k}{4} n^2} \  .
\label{resim}
\end{equation} 
On the right hand side we have some series containing powers of 
the parameter $\tq$, but the exponents are complex. This spoils
a direct interpretation of these exponents as energies of closed
string states. To cure the problem, we exchange the summation of 
$n$ with the integration over $P$ and we substitute $P$ by the 
new variable 
$$ P_n =  P - \frac{i}{2b^2} \, n \ \  $$
in each of the summands. $P_n$ is integrated along the line 
$\mathbb{R}-in/2b^2$. The crucial idea now is to shift all
the different integration contours back to the real line. 
This will give contributions associated with the continuous 
part of the boundary state. But while we shift the contours, 
we pick up residues from the singularities. The latter lead 
to terms associated with the discrete series. To work out the 
details, we split the partition function into a continuous and 
a discrete piece,   
$$ 
 Z_{1,1}(q) \ = \ Z^c_{1,1}(q) + Z^d_{1,1}(q) \ \ . 
$$ 
Note that this split is defined with respect to the closed 
string modes. In terms of open string modes, our partition 
function contains only discrete contributions. According to 
our description above, the continuous part of the partition 
function reads as follows 
\beqa
Z_{1,1}^{c}(q) \ = \ -2\sqrt {b^2k} 
\int dP\ \sum_{w \in \mathbb{Z}} 
\frac{\tq^{b^2 P^2+\frac{k}{4}w^2}}{\eta(\tq)^2} 
\, \frac{\sinh 2\pi b^2 P 
  \sinh 2\pi P}{\cosh 2\pi P +\cos \pi kw }\  .
\label{specd0cont}
\eeqa
In writing this formula we have renamed the summation index 
from $n$ to $w$. It is not difficult to see that this part 
of our partition function can be expressed as follows 
\begin{eqnarray}
Z_{1,1}^{c}(q) &  = &   
\int dP\ \sum_{w\in\Z} \ \chi^c_{(j,\frac{kw}{2})}(\tq) \ 
  A_1(j,n,w) \, A_{1}(j,n,w)^* \ \  
 \nonumber \ \ \mbox{ where } \\[2mm] 
 A_1(j,n,w) & = & \delta_{n,0}\, \left(\frac{kb^2}
{\pi^2} \right)^{\frac{1}{4}} \ \frac{\Gamma(-j+\frac{kw}{2})
  \Gamma(-j-\frac{kw}{2})}{(-1)^w\Gamma (-2j-1)} \,  
   \frac{\Gamma(1+b^2) \nu_b^{j+1} \sqrt{\sin \pi b^2}}
     {2\Gamma(1-b^2(2j+1))}\ .
\label{cyld0cont}
\end{eqnarray}
We claim that the coefficients $A_1$ that we have introduced 
here are the couplings of closed strings to the point-like brane 
on the cigar. In the last step of our short computation, however, 
we had to take the square root of the coefficients that appear in 
front of the characters. The choice we made here is difficult to 
justify as long as we try to stay within the modular bootstrap 
of the cigar (this was also remarked in \cite{Eguchi:2003ik}). 
But once we allow our experience from Liouville theory to enter, 
the remaining freedom is essentially removed.  
\smallskip 

In order to present the argument, we compare the proposed couplings
(\ref{cyld0cont}) with their semi-classical analogue. In the 
minisuperspace model we only have wave-functions $\phi^j_{n0}$ 
corresponding to closed string modes with vanishing winding number 
$w=0$. For these particular fields, the semi-classical limit of 
our couplings reads 
\begin{equation} 
\left(\langle \Phi^j_{n0}\rangle^{\rm D0}\right)_{k \rightarrow \infty} 
 \ = \   - \frac{\Gamma(-j)^2}{\Gamma(-2j-1)} 
   \ \delta_{n,0}\ = \  \phi^j_{n0}(\rho=0) \ ,  
\label{1ptd0limit}
\end{equation} 
i.e.\ it is given by the value of the function $\phi^j_{n0}$ at the 
point $\rho = 0$, in perfect agreement with our geometric picture. 
Note that due to the rotational symmetry of the point-like brane, 
only the modes with angular momentum $n=0$ have a non-vanishing 
coupling.The semi-classical result suggest to introduce 
$$ A^0_1(j,n,w) \ := \  - (-1)^w\, \frac{\Gamma(-j+\frac{kw}{2})
  \Gamma(-j-\frac{kw}{2})}{\Gamma (-2j-1)} \, \delta_{n,0} 
$$
and to rewrite the exact answer in the form 
\begin{equation} \label{exactAscA} 
 A_1(j,n,w) \ = \ {\cal N}_{D0} \, A^0_1(j,n,w) 
   \,  \frac{(\c(b^2) b^2)^{-j-1/2}}{\Gamma(1-b^2(2j+1))} \ \ .  
\end{equation} 
Hence, the stringy improvement factor is exactly the same as for  
ZZ branes in Liouville theory. As in our discussion of the bulk 
reflection amplitude, we have incorporated the rather simple string 
effects (winding) for the $\theta$-direction into the definition of 
the `semi-classical' amplitude $A_0$. The formula (\ref{exactAscA}) 
strongly supports our expression (\ref{cyld0cont}) for the couplings 
$A_1$, and in particular the choice of square root we have made.      
\smallskip 

After this first success we now turn to the calculation of the 
discrete piece of the amplitude. It is clear from (\ref{resim}) 
that the residues we pick up while shifting the contours give the
following discrete contribution to the partition function  
\beqa
\half Z^d_{1,1}(q) & =&  -2\sqrt{b^2k}\, 
 \sum_J 
\sum_{n=1}^\infty \ (-1)^n \tq^{\frac{k}{4}n^2}\, \times
\nonumber  \\[2mm] & & \hspace*{0cm} \times \,    
\sum_{m=0}^{E(\frac{k-2}{2}n-\half)}\  \sin \left(
\frac{2\pi (m+\half)}{k-2}\right)\
\frac{\tq^{-\frac{1}{k-2}\left(
    m+\half-\frac{k-2}{2}n \right)^2}}{\eta(\tq)^2}
\ \ .
\label{specd0disc}
\eeqa
Here, $E(x)$  denotes the integer part of $x$. A careful study of the 
energies which appear in this partition sum shows that the contributing
states can be mapped to discrete closed string states with zero momentum. 
The latter are parametrized by their winding number $w$ and by their spin 
$j$ or, equivalently, by the label $\ell=\frac{kw}{2}+j \in \Z$, and the 
level number $s$ as in the character (\ref{cardisc}). The map between the 
parameters $(w,\ell,s)$ and the summation indices $(m,n)$ of formula 
(\ref{specd0disc}) is
\beqa
m\ =\ \ell-w \ \ \ \ , \ \ \ \ n\ =\ s+w\ \ .
\label{ident}
\eeqa
It can easily be inverted to compute the labels $(w,\ell,s)$ in terms
of $(m,n)$, 
\beqa
w\ = \ E\left(\frac{2m+1}{k-2}\right) 
\ \ \ \ , \ \ \ \ \ell \ = \ m+E\left(\frac{2m+1}{k-2}\right) \ \
\ \ , \ \ \ \ s\ =\ n-E\left(\frac{2m+1}{k-2}\right)\ \ .
\label{identrecip}
\eeqa
In terms of $j$ and $w$, the partition function (\ref{specd0disc}) may 
now be rewritten as follows, 
\beqa
 Z_{1,1}^{d}(q)\ = \ -2\sqrt{b^2k} \, 
 \sum_{w\in \Z}\, \sum_{j \in \cJ^d_{0w}}\, 
\sin\left(\frac{\pi (2j+1)}{k-2}\right)\, 
\chi_{(j,\frac{kw}{2})}(\tq)\ \ .
\label{cyld0disc2}
\eeqa
It remains to verify that the coefficients of the characters coincide 
with those derived from the boundary state (\ref{cyld0cont}). In our 
normalization, the boundary coefficients $A_1(j,n,w)$ are given 
through the same expression (\ref{cyld0cont}) as for the continuous 
series but we have to divide each term in the annulus amplitude by 
the non-trivial value of the bulk 2-point function  
of discrete closed string states, i.e.\  
\beqa
 {\cal Z}_{1,1}^{d}(q) & = &   
\sum_{w \in \Z} \, \sum_{j \in \cJ^d_{0w}}\ \frac{A_1(j,n=0,w)
A_{1}(j,n=0,w)^*} {\la \Phi^{j}_{0w}\Phi^{j}_{0w}\ra} \  
\chi^d_{(j,\frac{kw}{2})}(\tq)\\[2mm]  
& = &  \sum_{w \in \Z} \, \sum_{j \in \cJ^d_{0w}}\ {\rm Res}_{x=j}
 \left(\frac{A_1(x,0,w)\, A_1(x,0,w)^*}
  {R(x,n=0,w)}\right) \ \chi^d_{(x,\frac{kw}{2})}(\tq)\ .
\label{cyld0disc}
\eeqa
The second line provides a more precise version of what we mean in 
the first line using the reflection amplitude (\ref{R}) rather than 
the 2-point function. Recall that the bulk 2-point correlator 
contains a $\delta$-function which arises because of the infinite 
volume divergence. If we drop this $\delta$-function in the 
denominator by passing to the reflection amplitude, the quotient 
has poles and the physical quantities are to read off from their 
residues. 
\smallskip 

A short explicit computation shows that the argument of the 
Res-operation in eq.\ (\ref{cyld0disc}) indeed has simple poles 
at $ x \in \cJ^d_{0w}$ and that the residues agree exactly 
with the coefficients in formula (\ref{cyld0disc2}), just as it 
is required by world-sheet duality. Our calculation therefore 
provides clear evidence for the existence of closed string bound 
states, as we have anticipated. They are labeled by elements of
the set $\cJ^d$, in agreement with the findings of \cite{hpt}. 
Our input, however, was no more than a well motivated and 
simple Ansatz (\ref{D0annulus}) for the partition function of
a D0 brane at the tip of the cigar.        
\smallskip 

Before we conclude, we would like to add a few comments on further 
localized branes in the cigar geometry. Recall that the ZZ branes 
in Liouville theory were labeled by two discrete parameters $n,m 
\geq 1$, though we have argued that branes with $(n,m) \neq (1,1)$
do in some sense not belong into Liouville theory. In complete 
analogy, one can find a discrete family of point-like branes on 
the cigar which is parametrized by one integer $n \geq 1$. The 
form of the corresponding couplings $A_n$ and their annulus 
amplitudes $Z_{n,n'}$ can be found in \cite{Ribault:2003ss}. These 
branes descent from a similar discrete family of compact branes 
in the Euclidean $AdS_3$ (see \cite{pst}). The latter have been 
interpreted as objects which are localized along spheres with an 
imaginary radius. We have argued in \cite{Ribault:2003ss} that localized 
branes with $n \neq 1$ are unphysical in the cigar background.

\subsection{D1 and D2 branes in the cigar} 

Besides the compact branes that we have studied in great detail in the 
last section, there exist two families of non-compact branes on the cigar. 
One of them consists of branes which are localized along lines (D1-branes), 
members of the other family are volume filling (D2-branes). We would like 
to discuss briefly at least some of their properties.  

\paragraph{D1-branes in the 2D cigar.} 
D1-branes in the cigar are most easily studied using a new coordinate 
$u=\sinh\rho$ along with the usual angle $\theta$. We have $u\geq 0$ and 
$u=0$ corresponds to the point at the tip of the cigar. In the new 
coordinate system, the background fields read 
\begin{equation}
ds^2 \ =\ \frac{k}{2}\, \frac{du^2+u^2d\theta^2}{1+u^2}\ \ \ \ , 
 \ \ \ \ 
e^\varphi\ =\ \frac{e^{\varphi_0}}{(1+u^2)^{\frac{1}{2}}}\ \ .  
\end{equation}
When we insert these background data into the Born-Infeld action for 
1-dimensional branes we obtain 
\begin{equation} 
S_{\rm BI} \ \propto \ \int dy\, \sqrt{u'{}^2+u^2\theta'{}^2}\ \ ,
\end{equation}
where the primes denote derivatives with respect to the world-volume 
coordinate $y$ on the D1-brane. 
It is now easy to read off that D1-branes are straight lines in the plane 
$(u=\sinh\rho,\theta)$. These are parametrized by two quantities, one being 
their slope, the other the transverse distance from the origin. In our 
original coordinates $\rho,\theta$, this 2-parameter family of 1-dimensional 
branes is characterized by the equations 
\begin{equation}
\sinh\rho\ \sin(\theta-\theta_0) \ =\ \sinh r \  .
\label{eqd1}
\end{equation}
Note that the brane passes through the tip if we fix the parameter $r$ to 
$r=0$. All branes reach the circle at infinity ($\rho= \infty$) at two 
opposite points. The positions $\theta_0$ and $\theta_0+\pi$ of the latter 
depend on the second parameter $\theta_0$ (see Figure 8).
\vspace{1cm}
\fig{
D1-branes on the cigar extend all the way to two opposite points on the 
circle at $\rho = \infty$. The position of these points is parametrized 
by $\theta_0$. In the $\rho$-direction they cover all values $\rho \geq 
r$ down to some parameter $r$.}
{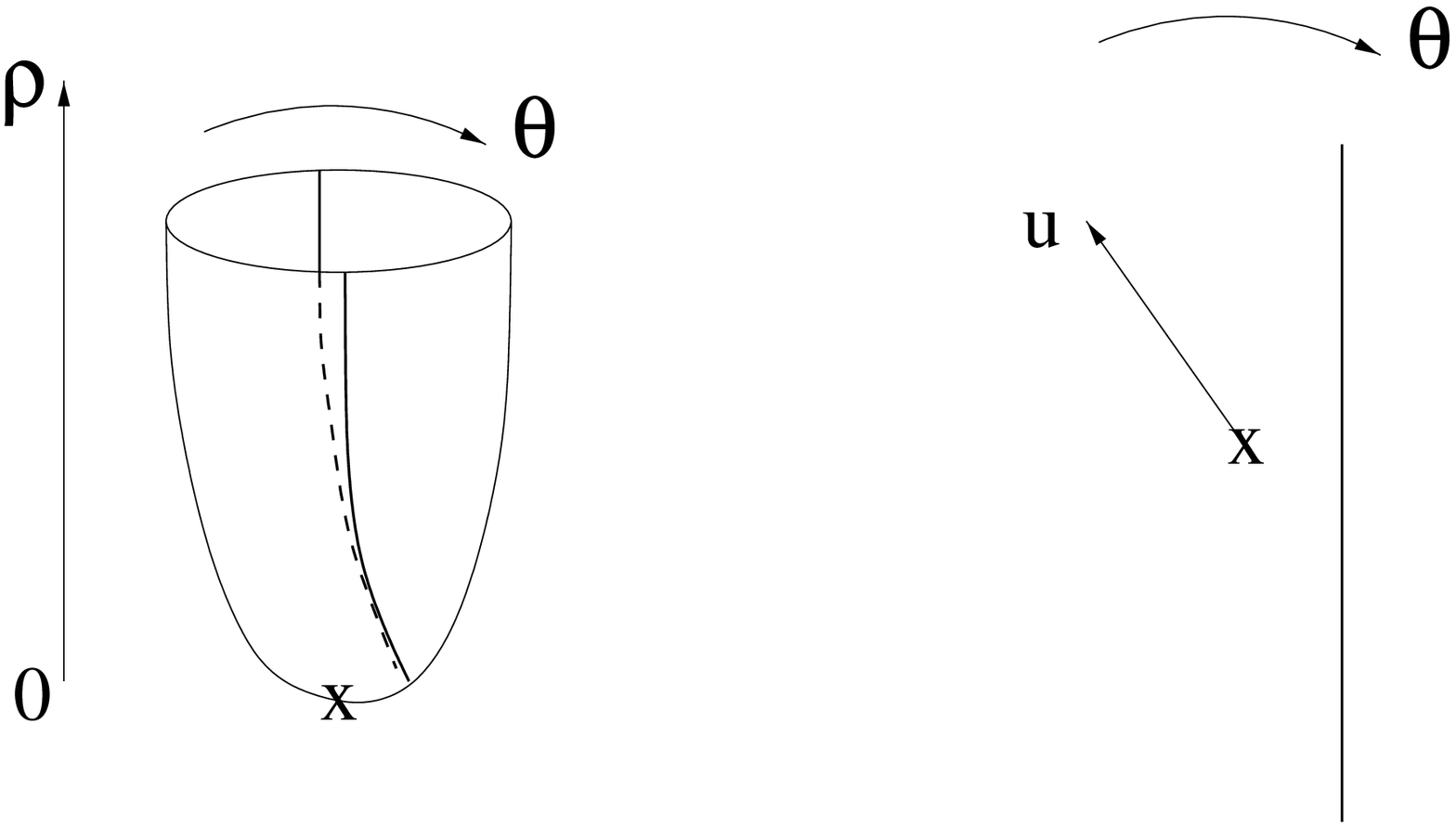}{6truecm}
\figlabel{\basic}
\vspace{1cm}

Now that we have some idea about the surfaces along which our branes
are localized we can calculate their coupling to closed string modes
in the semi-classical limit. This is done by integrating the 
minisuperspace wave functions (\ref{wavecig}) of closed string modes
over the 1-dimensional surfaces (\ref{eqd1}). The result of this 
straightforward computation provides a prediction for the 
semi-classical limit of the exact 1-point couplings $A^{D1}$,  
\begin{eqnarray}
\left(\langle \Phi^j_{n0}\rangle^{\rm D1}_r\right)_{k \rightarrow \infty} 
  & = & A^{0;\rm D1}_{(r,\theta_0)}(j,n,w=0) 
   \ = \  \label{1ptd1limit} \\[2mm] \nn  
  & = & \delta_{w,0} \ e^{ i n\theta_0}\, 
\frac{\Gamma(2j+1)}{\Gamma(1+j+\frac{n}{2})
\Gamma(1+j-\frac{n}{2})}
\left( e^{-r(2j+1)}+(-1)^n e^{r(2j+1)}\right)\ .
\end{eqnarray}
The minisuperspace model of the cigar does not include any states
associated with closed string modes of non-vanishing winding number. 
But in the case of the D1-branes, experience from the analysis of 
branes on a 1-dimensional infinite cylinder teaches us that closed 
string modes with $w\neq 0$ do not couple at all. Since the 
discrete closed string modes only appear at $w \neq 0$, they 
are like-wise not expected to couple to the D1-branes. Consequently,
our formula (\ref{1ptd1limit}) predicts the semi-classical limit of
all non-vanishing couplings in the theory.  
\medskip 

The exact 1-point couplings of the D1 branes are rather straightforward
to extra\-polate from the semi-classical result (\ref{1ptd1limit}) and the 
formula (\ref{FZZTAS}) for FZZT branes in Liouville theory. We claim that 
the exact solution is parametrized by two continuous parameters $r$ and 
$\theta_0$, just as in the semi-classical limit, and that the associated 
couplings are given by  
\begin{eqnarray}
 A^{\rm D1}_{(r,\theta_0)}(j,n,w) \  = \  
  {\cal N}_{\rm D1} \  A^{0;\rm D1}_{(r,\theta_0)}(j,n,w)   
\ \Gamma(1+b^2(2j+1))\ \left(\c(b^2)b^2 \right)^{-j-1/2} \ \ .  
\label{1ptd1}
\end{eqnarray}
These couplings were first proposed in \cite{pst}. Their consistency 
with world-sheet duality was established in \cite{Ribault:2003ss}. Let us 
also remark that our D1-branes are very close relatives of the so-called 
hairpin brane \cite{Lukyanov:2003nj}. The latter is a curved brane in a 
flat 2-dimensional target. It is localized along two parallel lines at 
infinity and then bends away from these lines into a smooth curve
(semi-infinite hairpin). Brane dynamics attempts to straighten all 
curved branes. But the shape of the hairpin brane is chosen in such 
a way that it stays invariant so that dynamical effects on the brane 
merely cause a rigid translation of the entire brane. Such rigid 
translations of the background can be compensated by introducing 
a linear dilaton. Hence, we can alternatively think of the hairpin 
brane as a 1-dimensional brane which is pending between two points  
at infinity, bending deeply into the 2-dimensional plane in order 
to reduce its mass. This alternative description of the hairpin brane 
shows the close relation with our D1 branes. Needless to say that 
the structure of the boundary states is essentially identical. 
Following \cite{Kutasov:2004dj}, a Lorentzian version of this 
boundary state has been studied to describe a time dependent 
process in which a D-brane falls into an NS5 branes (see e.g.\ 
\cite{Nakayama:2004yx,Sahakyan:2004cq,Nakayama:2004ge,
Nakayama:2005pk}).

\paragraph{D2-branes in the 2D cigar.} 
The main new feature that distinguishes the D2-branes from the branes 
we have discussed above is that they can carry a world-volume 2-form 
gauge field $F = F_{\rho\theta} d\rho \wedge d\theta$. In the presence 
of the latter, the Born-Infeld action for a D2-brane on the cigar becomes  
\begin{equation}
S_{\rm BI} \ \propto \ \int d\rho d \theta \cosh \rho 
  \sqrt{\tanh^2\rho + F_{\rho\theta}^2}\ \ . 
\end{equation}
We shall choose a gauge in which the component $A_\rho$ of the gauge 
field vanishes so that we can write $F_{\rho \theta} = \partial_\rho 
A_\theta$. A short computation shows that the equation of motion for 
the one-form gauge potential $A$ is equivalent to  
\begin{equation}
F_{\rho\theta}^2 \ = \ \frac{\beta ^2\tanh^2\rho}{\cosh^2\rho-\beta
    ^2}\ \ . 
\label{fd2}
\end{equation}
If the integration constant $\beta$ is greater than one, then the
D2-brane is localized in the region $\cosh\rho\geq \beta$, i.e.\ 
it does not reach the tip of the cigar. We will exclude this case 
in our semi-classical discussion and assume that $\beta =\sin \s 
\leq 1$. The corresponding D2-brane cover the whole cigar. 
Integrating eq.\ (\ref{fd2}) for the field strength $F_{\rho \theta} 
= \partial_\rho A_\theta$ furnishes the following expression for 
the gauge potential    
\begin{equation}
 A_\theta(\rho)\ =\ \s-\arctan\left(\frac{\tan
    \s}{\sqrt{1+\frac{\sinh^2\rho}{\cos^2 \s}}}\right)\ \ .
\end{equation}
In our normalization $A_\theta(\rho =0) = 0$, the parameter $\s$ 
is the value of the gauge potential $A_\theta$ at infinity. When 
this parameter $\sigma$ tends to $\sigma = \halfpi$, the F-field 
on the brane blows up. We should thus consider $\sigma = \halfpi$ 
as a physical bound for $\sigma$. 
\smallskip 

Let us point out that the F-field we found here vanishes at $\rho = 
\infty$. In other words, it is concentrated near the tip of the cigar. 
By the usual arguments, the presence of a non-vanishing F-field implies 
that our D2-branes carry a D0-brane charge which is given by the 
integral of the F-field. Like the F-field itself, the D0-brane 
charge is localized near the tip of the cigar, i.e.\ in a compact 
subset of the 2-dimensional background. Hence, one expects the D0-brane 
charge, and therefore the parameter $\sigma$ of the D2-brane, to be 
quantized. We shall explain below how such a quantization of the 
brane parameter $\sigma$ emerges from the conformal field theory 
treatment of D2 branes. 
\smallskip 
 
The exact one point couplings for these D2 branes were first 
proposed in \cite{Ribault:2003ss}. They are parametrized by a quantity 
\begin{equation}
 \s \ \in \ [\, 0\, ,\, \frac{\pi}{2}(1+b^2)\, [\ \ . 
\label{d2inter} 
\end{equation}
Note that  this interval shrinks to its semi-classical analogue
as we send $b$ to zero. For the associated 1-point functions 
of closed string modes we found    
\begin{eqnarray}\label{1ptd2}
A_\s^{\rm D2}(j,n,w) & = & 
   \N_{\rm D2} \ A_\s^{0;{\rm D2}}(j,n,w) \  
  \Gamma(1+b^2(2j+1))\, 
 \left(\c(b^2) b^2\right)^{-j-\half}\\[3mm]
 A_\s^{0;{\rm D2}}(j,n,w) \ = \ \delta_{n,0} & &\hspace*{-10mm} 
   \Gamma(2j+1)
  \left(\frac{\Gamma(-j+\frac{kw}{2})}{\Gamma(j+1+\frac{kw}{2})}
 \ e^{i\s(2j+1)}+\frac{\Gamma(-j-\frac{kw}{2})}
  {\Gamma(j+1-\frac{kw}{2})}\ e^{-i\s(2j+1)}\right)\ .  \nn
\end{eqnarray} 
The semi-classical coupling $A^0$ can be derived from our geometric
description of the D2 branes. In the full field theory, it receives
the same correction as the FZZT branes in the Liouville model.    
Formula (\ref{1ptd2}) holds for closed string modes from the continuous 
series. It also encodes all information about the couplings of discrete 
modes, but they have to be read off carefully because of the infinite 
factors (see the discussion in the case of D0-branes).
\smallskip

Following our usual routine, we could now compute the spectrum of 
open strings that stretch between two D2 branes with parameters $\s$ 
and $\s'$. We do not want to present the calculation here (it can 
be found in \cite{Ribault:2003ss}). But some qualitative aspects are quite
interesting. Our experience with closed string modes on the cigar 
suggests that open strings on D2 branes could possess bound states 
near the tip of the cigar. World-sheet duality confirms this 
expectation. In other words, the annulus amplitudes computed 
from eq.\ (\ref{1ptd2}) contain both continuous and discrete 
contributions. While the continuous parts involve some complicated 
spectral density, the discrete parts must be expressible as a 
sum of coset characters with integer coefficients. The latter 
will depend smoothly on the choice of branes. Integrality of
the coefficients then provides a condition on the brane labels
$\s$ and $\s'$ which reads    
\begin{equation}
\s-\s'\ =\ 2\pi\, \frac{m}{k-2}\ \ , \ \ m\in \Z\ \ .
\label{qd2quant}
\end{equation}
This is the quantization of the brane label $\s$ that we have 
argued for above. In addition, the computation of the partition 
function $Z$ also shows that the density of continuous open string 
states diverges when $\frac{\s+\s'}{2}$ reaches the upper bound 
$\halfpi (1+b^2)$. The classical version of this bound on $\s$ 
also appeared in our discussion of the geometry.  
\smallskip 

This concludes our discussion of branes in the cigar geometry. 
Let us mention that a few additional boundary states have been 
suggested in the literature, including a 2-dimensional 
non-compact brane which does not reach the tip of the cigar
\cite{Fotopoulos:2004ut} and a compact brane with non-vanishing 
couplings to closed string modes of momentum $n\neq 0$
\cite{Ahn:2004qb}. Even in the absence of a fully satisfactory 
conformal field theory analysis\footnote{Such an analysis might 
require the use of factorization constraints similar to the 
ones we discussed in the context of Liouville theory (see 
\cite{pst,Hosomichi:2004ph} for some steps in this 
direction).} it seems very plausible that such branes do 
indeed exist. In the case of the D2 branes, evidence comes 
simply from a semiclassical treatment. For the additional 
point-like branes, such an approximation is insufficient.
On the other hand, certain ring-shaped branes in flat 
space are known to collapse into point-like objects which
cannot be identified with D0 branes \cite{Lukyanov:2005nr} 
and there is no reason to doubt that similar processes can 
appear on the cigar. Furthermore, there exists a dual 
matrix model \cite{KaKoKu} with a non-perturbative 
instability that does not seem to be associated with 
the D0 branes we studied above, thereby also pointing 
towards the existence of new localized brane solutions.

\section{Conclusions and Outlook} 

In these notes we have explained the main techniques that 
are involved in solving non-rational conformal field theories. 
At least in our discussion of Liouville theory we have tried 
to follow as closely as possible the usual {\em conformal 
bootstrap} that was developed mainly in the context of 
rational conformal field theory. Among the new features 
we highlighted Teschner's trick and the use of free field 
computations. 
\smallskip 

Let us recall that {\em Teschner's trick} exploited the 
existence of so-called degenerate fields that are not part 
of the physical spectrum but can be obtained by analytic 
continuation. It turned out that the resulting bootstrap 
equations were sufficiently restrictive to determine the 
solution uniquely. Let us also recall that these equations 
are typically linear (`shift equations') in the couplings 
of physical fields (see e.g.\ eqs.\ (\ref{spcross}),
(\ref{clusterFZZT})).   
\smallskip 

The coefficients of such special bootstrap equations involve 
couplings of degenerate fields which we have been able to 
calculate through free field computations. Intuitively, we
understood the relevance of an associated free field theory 
(the linear dilaton in the case of Liouville theory) from 
the fact that the interaction of the investigated models 
is falling off at infinity. Let us mention that one can 
avoid all free field calculations and obtain the 
required couplings of degenerate fields through a `degenerate 
bootstrap'. Since degenerate fields in a non-rational model 
behave very much like fields of a rational theory, a bootstrap 
for couplings of degenerate fields is similar to the usual 
bootstrap in rational models (see \cite{Teschner:1995yf, 
Teschner:1997ft} for more details).  
\smallskip 

In the final lecture, we have attempted to reverse the procedure 
and to place the modular bootstrap in the center of the programme. 
Even though it remains to be seen whether such an approach can 
be developed into a systematic technique for solving non-rational 
models, its application to the cigar conformal field theory was 
quite successful. Nevertheless, it is important to keep in mind 
that our success heavily relied on our experience with Liouville 
theory. Actually, Liouville theory models the radial direction 
of the cigar background so perfectly that only the semiclassical 
factors in the various couplings have to be replaced when passing 
to the cigar.   
\medskip

Once the cigar conformal field theory is well understood., it is 
not difficult to lift the results to the group manifold \SL or its
Euclidean counterpart $H_3^+ \cong $ \SLC/\SU. Historically, the 
latter was addressed more directly, following step by step the  
bootstrap programme we carried out for Liouville theory in the 
second and third lecture. Minisuperspace computations for the 
various bulk and boundary spectra and couplings can be found 
at several places in the literature \cite{Gawedzki:1991yu,
Teschner:1997fv,Bachas:2000fr,Lee:2001xe,Rajaraman:2001cr,
Parnachev:2001gw,pst}. 
The shift equations for the bulk bootstrap were derived and 
solved in \cite{Teschner:1997ft,Teschner:1999ug} building on 
prior work on the spectrum of the theory \cite{Gawedzki:1991yu}. 
Full consistency has been established through an interesting 
relation of the bulk correlators with those of Liouville theory 
\cite{Ribault:2005wp} (see also \cite{Teschner:2001gi} for an 
earlier and more technical proof). As indicated above, Teschner 
did not use any free field computations and relied entirely on 
bootstrap procedures. Nevertheless, it is certainly possible 
to employ  free field techniques (see \cite{Ishibashi:2000fn,
Hosomichi:2000bm}). 
The boundary bootstrap was carried out in \cite{pst} (see also 
\cite{Giveon:2001uq} for a previous attempt and \cite{Lee:2001gh} 
for a partial discussion). The Wick rotation from the Euclidean 
to the Lorentzian bulk model was worked out in \cite{moiii}
(see also \cite{Giribet:1999ft}-\cite{Hofman:2004ny} for free 
field computations) and aspects of the boundary theory were 
addressed more recently in \cite{Israel:2005ek}.   
\smallskip

Even though the solution of Liouville theory and closely related 
models has certainly been a major 
success in non-rational model building, there remain many challenging  
problems to address. In these lectures, we have used potential 
applications to AdS/CFT like dualities as our main motivation. 
In spite of the progress we have described, the constructing 
string theory on $AdS_5$ still appears as a rather distant goal 
for now. Note that the string equations of motions require a 
non-vanishing RR background when we are dealing with the 
metric of $AdS_5$. Switching on RR fields tends to reduce 
the chiral symmetry algebras of the involved world-sheet 
theories \cite{Bershadsky:1999hk} and hence it makes 
such backgrounds very difficult to tackle. In this context, 
the example of $AdS_3$ might turn out to provide an interesting
intermediate step. All the developments we sketched in the 
previous paragraph concern the special case in which the string 
equations of motion are satisfied through a non-vanishing NSNS 
3-form $H$. Beyond this point, there exists a whole family of 
models with non-zero RR 3-form flux (see \cite{BWV,%
Bershadsky:1999hk} for more explanations). It seems likely that 
at least some of these models may be solved using tools of 
non-rational conformal field theory. 
\smallskip 

Concerning the basic mathematical structures of non-rational 
conformal field theory, the whole field is still in its infancy. 
The study of strings in compact backgrounds can draw on a rich 
pool of formulas which hold regardless of the the concrete 
geometry. In fact, for large classes of models, solutions 
of the factorization constraints may be constructed from the 
representation theoretic quantities (modular S-matrix, Fusing 
matrix etc.\ ) of the underlying symmetry. Similar results in 
non-rational models are not known. We hope that these lectures 
may contribute drawing some attention to this vast and 
interesting field which remains to be explored.  
\bigskip 
\bigskip
  
\noindent 
{\bf Acknowledgment:} I wish to thank the participants of the 
three events at which I delivered these lectures for many good
questions, their interest and feedback. Special thanks are also
due to the organizers of the schools in Trieste and Vancouver 
and to my co-organizers at the ESI in Vienna for (co-)organizing 
such stimulating meetings. This work was 
partially supported by the EU Research Training Network grants 
``Euclid'', contract number HPRN-CT-2002-00325, ``Superstring 
Theory", contract number MRTN-CT-2004-512194, and 
``ForcesUniverse'', contract number MRTN-CT-2004-005104.

\newpage
\appendix
\section{Appendix: Dotsenko-Fateev Integrals} 
\setcounter{equation}{0} 

We have seen above that residues of correlation functions in 
non-compact backgrounds can be computed through free field 
theory calculations. The latter involve integrations over 
the insertion points of special bulk or boundary fields. 
In the case of bulk fields, such integrations can be 
carried out with the help of the following formula
\begin{eqnarray} 
\int \prod_{j=1}^k \  d^2 z_j \, |z_j|^{2\a} \, 
|1-z_j|^{2\beta} \ \prod_{j<j'}\ |z_j-z_{j'}|^{4\rho} 
 &  = & \\[2mm] 
 \nonumber & & \hspace*{-4cm} = \  
k! \, \pi^k \, \prod_{j=0}^{k-1} \ \frac{\c((j+1)\rho)}
{\c(\rho)} \frac{\c(1+\a+j\rho)\c(1+\beta + j \rho)}
{\c(2 + \a + \beta + (k-1+j\rho))}\ \  .  
\end{eqnarray}
This is a complex version of the original Dotsenko-Fateev 
integral formulas \cite{Dotsenko:1984ad}, 
\begin{eqnarray} 
\int_0^1dx_1\, \int_0^{x_1}dx_2 \dots \int_0^{x_{k-1}}dx_k \ 
  \prod_{j=1}^k \,  x_j^{\a} \, 
(1-x_j)^{\beta} \ \prod_{j<j'}\ (x_j-x_{j'})^{2\rho} 
 &  = & \\[2mm] 
 \nonumber & & \hspace*{-6cm} = \  
 \prod_{j=0}^{k-1} \ \frac{\Gamma((j+1)\rho)}
{\Gamma(\rho)} \frac{\Gamma(1+\a+j\rho)\Gamma(1+\beta + j \rho)}
{\Gamma(2 + \a + \beta + (k-1+j\rho))}\ \ .  
\end{eqnarray} 
In Liouville theory, Dotsenko-Fateev integrals emerge after the
evaluation of correlators in a linear dilaton background (see 
e.g.\ eqs.\ (\ref{cm}) or (\ref{res})), either on a full plane 
(P) or on a half-plane (H). These are given by 
\begin{equation} \nonumber
\langle \, \Phi_{\a_1}(z_1,\bz_1) \dots \Phi_{\a_n}(z_n,\bz_n)\, 
    \, \rangle^{(P)}_{\rm LD} 
     \ = \ 
\frac{1}{\prod_{i>j}^n |z_i-z_j|^{8\a_i\a_j}}
\end{equation} 
and
\begin{eqnarray*} 
\langle \, \Phi_{\a_1}(z_1,\bz_1) \dots \Phi_{\a_n}(z_n,\bz_n)\, 
           \Psi_{\b_1}(x_1) \dots \Psi_{\b_m}(x_m) \, \rangle^{(H)}_{\rm LD} 
 & =  &  \\[3mm]   & &  \hspace*{-6cm} = \   
\frac{\prod_{i=1}^{n} |z_i-\bz_i|^{-2\a_i^2} \ \prod_{i,r} 
    |z_i-x_r|^{-4\a_i \b_r}}{\prod_{r<s}^m |x_r-x_s|^{2\b_r\b_s} 
   \ \prod_{i>j}^n |(z_i-z_j)(z_i-\bz_j)|^{4\a_i\a_j}}\ \   
\end{eqnarray*} 
where $\Psi_\b(x) = :\exp \b X(x): $ are the boundary vertex operators
of the linear dilaton theory. 
\smallskip 

From the above formulas we can in particular compute the non-trivial 
constant $c^-$ that appears in the operator product expansions of 
section 3.2.,   
$$ c_b^-(\a) \ = \  \int d^2z\ |z|^{2b^2}\ |1-z|^{-4b\a} \ = 
\ \pi \, \frac{\c(1+b^2) \, \c(1- 2b\a)}{\c(2+b^2-2b\a)}\ \ . 
$$ 
A similar free field theory computation was used in section 4.2. to 
determine the residue of the bulk boundary structure constants $B(\b,\c)$ 
at $\beta = -b/2$, 
$$  {\mbox{\it res}}_{\beta = - b/2} \left(B(\beta,0)\right)
\ = \ - \mu_B \, \int_{-\infty}^\infty du\ |\frac{i}{2}-u|^{2b^2}
\ = \ \frac{-2\pi \mu_B \, \Gamma(-1-2b^2)}{\Gamma^2 (-b^2)} \ \ . 
$$    
Free field theory calculations of this type can be performed for 
other singular points of correlation functions and they provide 
a rather non-trivial test for the proposed exact couplings.

\section{Appendix: Elements of the Fusing matrix}

{\def\D{h} 
\def\CF{{\cal F}} 
\def\al{\alpha}

In the first lecture we encountered the construction of the 
so-called Fusing matrix as a problem in the representation 
theory of chiral algebras. When we deal with Liouville theory, 
the relevant algebraic structure is the Virasoro algebra with 
central charge $c \geq 25$. Given the importance of the Virasoro
field for 2D conformal field theory it may seem quite surprising 
that its Fusing matrix was only obtained a few years ago by 
Ponsot and Teschner. Even though the construction itself is 
rather involved, it leads to elegant expressions through an 
integral over a product of Barnes' double $\Gamma$ functions 
\cite{Ponsot:1999uf}. When specialized to cases in which at least one
external field is degenerate\footnote{Recall that these are the 
only matrix elements that we need in eqs.\ (\ref{spcross}) and 
(\ref{spclus})).}  the general formulas simplify significantly. 
In fact, such special elements of the Fusing matrix can be 
written in terms of ordinary 
$\Gamma$ functions.     
\smallskip 

Rather than spelling out the general expression for the Fusing 
matrix and then specializing it to the required cases, we shall 
pursue another, more direct route which exploits the presence 
of the degenerate field from the very beginning. Since this 
procedure is similar to analogous constructions in rational 
theories, we will only give a brief sketch here. To this end, 
let us look at the following 4-point conformal block 
$$ 
  \cG(z) \ = \ \langle\, V_{\a_4} (\infty) \, V_{\a_3}(1) 
                 \,  V_{\a_2}(z)\, V_{\a_1}(0)\, \rangle\ \ .  
$$ 
Here, $V_\a$ denote chiral vertex operators for the Virasoro 
algebra where, in contrast to Section 2.1., we have not specified 
their source and target space. Since we are only interested in 
elements of the Fusing matrix with $\a_2 = - b^{\pm1}/2$, we can 
specialize to the case where $V_{\a_2}$ is degenerate of order 
two. Consequently, it satisfies an equation of the form 
(\ref{qEOM}). With the help of the intertwining properties of
chiral vertex operators (see Section 2.1) or, equivalently, a 
chiral version of the Ward identities (\ref{WardT}), one can 
then derive second order differential equations for the 
conformal blocks $\cG$, 
\begin{equation}
\left(-\frac{1}{b^2}\frac{d^2}{dz^2}+
  \left(\frac{1}{z-1}+\frac{1}{z}\right)
    \frac{d}{dz}-\frac{\D_3}{(z-1)^2}-\frac{\D_1}{z^2}+
    \frac{\D_3+\D_2+\D_1-\D_4}{z(z-1)} \right)\cG(z)\ =\ 0\ , 
\label{de}
\end{equation} 
where $h_i = \a_i (Q-\a_i)$ and $\a_2 = - b^{\pm1 }/2$. For 
definiteness, let us concentrate on $\a_2 = -b/2$. Two linearly 
independent solutions of the differential equations (\ref{de}) 
can be expressed in term of the hypergeometric function $F(A,B;
C;z)$ as follows,    
\begin{eqnarray*} 
  \cG_\pm (z)& =&  z^{\eta_\pm}(1-z)^\rho \,  F(A_\pm,B_\pm;C_\pm;z)
  \\[3mm] 
\mbox{with}  & & 
\eta_\pm \ =\ \D_{\al_1\pm b/2}-\D_2-\D_1 \ \ , \ \ 
\rho\ =\ \D_{\al_3-b/2}-\D_3-\D_2 \\[3mm] 
\mbox{and} & & 
A_\pm \ =\ \mp b(\al_1-Q/2)+b(\al_3+\al_4-b)-1/2 \\[2mm] 
& & B_\pm \ =\  A_\pm -2b\a_4+  b^2 + 1 \ \ , \ \ 
C_\pm  \ = \  1\mp b(2\al_1-Q)\ \ .
\end{eqnarray*}
A standard identity for the hypergeometric function $F$,  
\begin{eqnarray}
F(A,B;C;z) &=&
\frac{\Ga(C)\Ga(C-A-B)}{\Ga(C-B)\Ga(C-A)}\ F(A,B;A+B-C+1,1-z)
\nonumber \\[2mm] 
 &  & \hspace*{-3cm} + \ 
 \frac{\Ga(C)\Ga(A+B-C)}{\Ga(A)\Ga(B)}\ (1-z)^{C-A-B}F(C-A,C-B;C-A-B+1,1-z)
\nonumber \end{eqnarray}
allows to expand $\cG_\pm$ in terms a second basis $\tilde \cG_\pm$ 
in the space of solutions of the differential equation (\ref{de}),  
\begin{equation}
\cG_s(z)\ =\ \sum_{t=\pm} \ 
\Fus{\a_1-s\frac{b}{2},}{\alpha_3- t \frac{b}{2}}{-\frac{b}{2}\,}{\a_3}
                                              {\ \, \a_1}{\a_4} \ 
\tilde \cG_t(1-z)\ \ , \nonumber
\end{equation}
with $s = \pm$. Formulas for $\tilde \cG_\pm$ can be found in the 
literature. We only list the coefficients,  
\begin{eqnarray}
\Fus{\a_1-\frac{b}{2},}{\alpha_3-\frac{b}{2}} {-\frac{b}{2}\,}{\a_3}
                                              {\ \, \a_1}{\a_4}
   &=&\frac{\Gamma(b(2\alpha_1-b))\, \Gamma(b(b-2\alpha_3)+1)}
           {\Gamma(b(\alpha_1-\alpha_3-\alpha_4+\frac{b}{2})+1)
            \, \Gamma(b(\alpha_1-\alpha_3+\alpha_4-\frac{b}{2}))} 
  \nonumber \\[2mm]
\Fus{\a_1-\frac{b}{2},}{\alpha_3+\frac{b}{2}}{-\frac{b}{2}\,}{\a_3}
                                              {\ \, \a_1}{\a_4}
&=&  \frac{\Gamma(b(2\alpha_1-b))\, \Gamma(b(2\alpha_3-b)-1)}
         {\Gamma(b(\alpha_1+\alpha_3+\alpha_4-\frac{3b}{2})-1)
          \,  \Gamma(b(\alpha_1+\alpha_3-\alpha_4-\frac{b}{2}))} 
\nonumber \\[2mm] 
\Fus{\a_1+\frac{b}{2},}{\alpha_3-\frac{b}{2}}{-\frac{b}{2}\,}{\a_3}
                                              {\ \, \a_1}{\a_4}
&=& \frac{\Gamma(2-b(2\alpha_1-b))\, \Gamma(b(b-2\alpha_3)+1)}
          {\Gamma(2-b(\alpha_1+\alpha_3+\alpha_4-\frac{3b}{2}))
      \Gamma(1-b(\alpha_1+\alpha_3-\alpha_4-\frac{b}{2}))} 
 \nonumber \\[2mm] 
\Fus{\a_1+\frac{b}{2},}{\alpha_3+\frac{b}{2}}{-\frac{b}{2}\,}{\a_3}
                                              {\ \, \a_1}{\a_4}
&=&  \frac{\Gamma(2-b(2\alpha_1-b))\, \Gamma(b(2\alpha_3-b)-1)}
      {\Gamma(b(-\alpha_1+\alpha_3+\alpha_4-\frac{b}{2}))
     \, \Gamma(b(-\alpha_1+\alpha_3-\alpha_4+\frac{b}{2})+1)} 
\ \ . \nonumber 
\label{FusL}
\end{eqnarray}}
Let us  stress once more that this simple construction of the fusing 
matrix does only work for elements with one degenerate external label. 
In more general cases, explicit formulas for the conformal blocks $\cG$ 
are not available so that one has to resort to more indirect methods 
of finding the Fusing matrix (see \cite{Ponsot:1999uf}).  

\baselineskip=15pt
\begingroup\raggedright\endgroup
\end{document}